\documentclass[draft]{article}
\usepackage{amsmath,amsfonts,latexsym, amssymb,  mathrsfs, eucal}

\usepackage{color}

\newcommand{\wt}{\widetilde}
\newcommand{\wh}{\widehat}

\newcommand{\beqa}{\begin{eqnarray}}
\newcommand{\eeqa}{\end{eqnarray}}
\newcommand{\e}{\varepsilon}
\newcommand{\eps}{\varepsilon}

\newcommand{\rd}{{\rm d}}
\newcommand{\bR}{{\mathbb R}}
\newcommand{\bC}{{\mathbb C}}

\newcommand{\bZ}{{\mathbb Z}}
\newcommand{\non}{\nonumber}
\newcommand{\wH}{{K}}

\newcommand{\tr}{\mbox{Tr\,}}

\newcommand{\ba}{{\bf{a}}}

\newcommand{\bx}{{\bf{x}}}

\newcommand{\bu}{{\bf{u}}}
\newcommand{\bv}{{\bf{v}}}
\newcommand{\bw}{{\bf{w}}}

\newcommand{\mg}{{m_N}}

\newcommand{\al}{\alpha}

\newcommand{\be}{\begin{equation}}
\newcommand{\ee}{\end{equation}}

\newcommand{\la}{\lambda}
\newcommand{\Om}{{\Omega}}
\newcommand{\om}{{\omega}}

\newcommand{\cL}{{\mathscr L}}

\newcommand{\cN}{{\mathcal N}}
\newcommand{\cH}{{\mathcal H}}

\newcommand{\ov}{\overline}

\newcommand{\re}{{\text {Re}  }}
\newcommand{\im}{{\text {Im} }}
\newcommand{\E}{{\mathbb E }}
\newcommand{\R}{{\mathbb R }}
\newcommand{\N}{{\mathbb N}}

\newcommand{\Ci}{{ C_{inf}}}
\newcommand{\Cs}{{ C_{sup}}}

\renewcommand{\P}{{\mathbb P}}
\newcommand{\C}{{\mathbb C}}

\newtheorem{theorem}{Theorem}
\newtheorem{corollary}[theorem]{Corollary}
\newtheorem{lemma}[theorem]{Lemma}
\newtheorem{proposition}[theorem]{Proposition}

\newtheorem{definition}{Definition}
\newcommand{\qed}{\hfill\fbox{}\par\vspace{0.3mm}}

\numberwithin{equation}{section}
\numberwithin{theorem}{section}
\numberwithin{definition}{section}
\numberwithin{remark}{section}

\title{Bulk universality for generalized Wigner matrices}

\author{
L\'aszl\'o Erd\H os${}^1$\thanks{Partially supported
by SFB-TR 12 Grant of the German Research Council}, 
Horng-Tzer Yau${}^2$\thanks{Partially supported
by NSF grants DMS-0602038, 0757425, 0804279}  \; and Jun Yin${}^2$  \thanks{Partially supported
by NSF grants  DMS-100165} \\ \\
Institute of Mathematics, University of Munich, \\
Theresienstr. 39, D-80333 Munich, Germany \\ lerdos@math.lmu.de ${}^1$ \\ \\
Department of Mathematics, Harvard University\\
Cambridge MA 02138, USA \\  htyau@math.harvard.edu,  jyin@math.harvard.edu ${}^2$ \\ \\
\\}

\begin{document}

\date{Aug 14, 2010}

\maketitle

\begin{abstract}

Consider  $N\times N$ Hermitian or symmetric random matrices $H$
where the  distribution  of the $(i,j)$ matrix element is given by
a probability  measure $\nu_{ij}$ with a subexponential decay.
  Let $\sigma_{ij}^2$ be the 
variance for the  probability  measure $\nu_{ij}$ with the normalization
property that $\sum_{i} \sigma^2_{ij} = 1$ for all $j$. 
Under essentially the only condition that  $c\le N \sigma_{ij}^2 \le c^{-1}$
for some constant $c>0$,
we prove that, in the limit $N \to \infty$,  the eigenvalue spacing 
statistics of $H$ in the bulk of the spectrum 
coincide with those of the Gaussian unitary or orthogonal
ensemble (GUE or GOE). 
We also show that for band matrices with bandwidth $M$ 
the  local semicircle law holds to the energy scale $M^{-1}$.

\end{abstract}


\medskip

\medskip


\medskip


\newpage 

\setcounter{tocdepth}{2}

\newpage
\section{Introduction} 
One key universal quantity for random matrices  is the eigenvalue gap  distribution. 
Although the density of eigenvalues may depend on the specific model, 
the gap distribution or the short distance 
correlation function are believed to depend only on the symmetry class
 of the ensembles but are otherwise  independent of the details of  the distributions.
There are two types of universality: the edge universality and the
bulk universality.  In this paper, we will focus on  the
bulk universality  concerning the interior of the spectrum. 
The bulk universality was  proved for very general classes of invariant ensembles
(see, e.g. \cite{BI, De1, De2, DKMVZ1, DKMVZ2, M, MG,  PS} and references therein).
For  non-invariant ensembles, in particular for matrices with i.i.d. entries (Wigner
matrices), the bulk universality was difficult to establish
due to the lack of an explicit expression for the joint distribution
of the eigenvalues.

The first rigorous partial result for bulk universality in the non-unitary case
was given by  Johansson  \cite{J}   (see also Ben Arous and P\'ech\'e \cite{BP}
{and the recent improvement \cite{J1} })
 stating that the bulk universality holds for Gaussian divisible 
{\it Hermitian} ensembles, i.e., Hermitian ensembles   of the form
\be
\wh H+ s V,
\label{HaV}
\ee
where $\wh H$  is a Wigner matrix, 
$V$ is an independent standard GUE matrix
and $s$ is a positive constant of order one.
 The restriction on Gaussian divisibility  turned out to be very difficult to  
remove.  In a series of papers \cite{ESY1, ESY2, ESY3,  EPRSY}, we developed a new approach 
to prove the universality. 
The first step  was to derive the local semicircle law, an estimate of the local eigenvalue 
density,   down to energy scales containing   around $\log N$ eigenvalues. 
Once such a strong form of the  local semicircle law was obtained, the result of 
\cite{J, BP}  can be extended to a Gaussian convolution
with variance only  $s^2 \asymp N^{-1 + \e}$. This tiny Gaussian component
can then be removed via a  reverse heat flow argument and this proves  \cite{EPRSY}
the bulk universality for Hermitian ensembles provided that
the distributions of the matrix elements are sufficiently differentiable.

The bulk universality for   Hermitian ensembles was also proved later on by
  Tao and Vu  \cite{TV} under the condition that 
the first four moments of the matrix elements 
match  those of GUE, but without the differentiability assumption. 
The condition on the fourth moment
was already removed in \cite{TV}  by using the result for Gaussian divisible 
ensembles  of \cite{J, BP}; the third moment condition was then removed in  \cite{ERSTVY}
by using the result of  \cite{EPRSY}. 

The four moment theorem \cite{TV} is also valid for the symmetric ensembles, but 
the restriction on the matching of the first four moments  cannot be weakened for
 the following reason.  
The key input to remove the fourth moment matching condition for the Hermitian case, 
 the universality of the Gaussian divisible ensembles  \cite{J, BP},  
relied entirely on the asymptotic  analysis of an
{\it explicit} formula, closely related to a formula in Br\'ezin-Hikami  \cite{BH, J}, for
the correlation functions of the eigenvalues for the {\it Hermitian ensembles}  $\wh H+ s V$.
Since similar formulas for symmetric matrices  are very complicated, 
the corresponding result is not available and thus the matching of the fourth moment cannot be removed
in this way. 
Although there is a proof \cite{ERSY} of universality 
for $s^2 \ge N^{-3/4}$ without using this formula, 
the main ingredient of that proof, establishing the uniqueness of  the local equilibria
 of the  Dyson Brownian motion, 
still heavily used   explicit formulas related to GUE.

In \cite{ESY4}  a  
completely different strategy was  introduced 
based on  a {\it local relaxation flow}, which locally behaves like a Dyson Brownian motion, 
but has a faster decay to equilibrium. 
This approach
entirely eliminates  explicit formulas and  it gives a unified proof for the  universality 
of {\it symmetric} and Hermitian  Wigner matrices \cite{ESY4}. 
It was further  generalized \cite{ESYY} to  quaternion self-dual 
Wigner matrices  and  sample covariance matrices. The method not only applies to all these specific 
ensembles,    but it also gives a conceptual interpretation that the occurrence of the universality 
is due to the relaxation to local equilibrium of the DBM.  We remark that very 
recently the results of \cite{TV} were also extended  to sample covariance matrices \cite{TV3}.

The main input of all these  methods \cite{EPRSY, ESY4, ESYY}  and \cite{TV, TV3} is 
an estimate of the local density of
eigenvalues, the local semicircle law.  This has been developed in the  previous work on Wigner
matrices \cite{ESY1, ESY2, ESY3}, where the matrix elements were i.i.d. random variables.
In this paper, we extend this method to random matrices
with independent, but not necessarily identically
distributed entries.
If we denote the variance of the $(i,j)$ entry of the matrix by  $\sigma_{ij}^2$, 
our main interest  is the case that $\sigma_{ij}$ 
are not a  constant but they satisfy the normalization
condition $\sum_i \sigma_{ij}^2=1$ for all $j$. We will call such matrix ensembles
{\it universal Wigner matrices}. For these ensembles Guionnet \cite{gui}
 and Anderson-Zeitouni 
\cite{AZ} proved that the density of the
 eigenvalues converges to the Wigner semi-circle law. 
 The simplest case is that of {\it generalized Wigner matrices}, where  $N\sigma_{ij}^2$  is uniformly bounded from above and below by two fixed positive numbers.  
In this case, we prove the local semicircle law down to essentially the 
smallest possible energy scale $N^{-1}$ (modulo $\log N$ factors). 
A much more difficult case is the {\it Wigner band matrices} where, roughly speaking, 
$\sigma_{ij}^2 = 0$ if $|i-j|> M$ for some 
$M < N$. In this case, we obtain the local semicircle law to the energy scale $M^{-1}$.
  We note that a certain three-dimensional version 
of Gaussian band matrices was considered by 
 Disertori, Pinson and Spencer \cite{DPS} using the supersymmetric method. 
They proved that the expectation of the density of eigenvalues 
is smooth and it  coincides with the Wigner semicircle law.

With the local semicircle law proved up to  the almost optimal scale,
applying the method of  \cite{ESY4, ESYY}  leads 
to the identification 
of the correlation functions and the gap distribution for generalized Wigner matrices
provided that the 
distribution of the matrix elements is continuous and satisfies the 
logarithmic Sobolev inequality.
These additional assumptions can be removed
if one can extend the Tao-Vu  theorem \cite{TV} 
to generalized Wigner matrices. In Section \ref{sec:proofcomp}, we will introduce
 an approach based on a 
Green's function comparison theorem,  which states that  the joint distributions of
 Green's functions of two ensembles 
at different energies with imaginary parts of order $1/N$  are identical  
provided that   the first three moments of the two ensembles coincide and the fourth moments are close.
Since local correlation
functions and the gap distribution of the eigenvalues
can be identified from Green's functions, it follows that 
the local  correlation
functions of these two ensembles are identical at the scale $1/N$. 
We can thus use this theorem to remove all continuity and   logarithmic Sobolev inequality 
restrictions in our approach. 
In particular, this leads to the bulk universality for
 generalized Wigner matrices with the subexponential decay being 
essentially the only  assumption on the  probability law. {  We note that 
one major technical difficulty in \cite{TV}, the level repulsion estimate, is not needed in 
the proof of the Green's function 
comparison theorem. It will be clear in Section 8 that, once the local semicircle law is established,  the Green's function comparison theorem 
is a simple consequence of  the standard resolvent 
 perturbation theory.
}

\section{Main results}

We now state the main results of this paper. Since all our results hold for both Hermitian
 and symmetric ensembles, 
we will state the results for Hermitian matrices only. The modifications to the 
symmetric case are straightforward and they will be omitted. 
Let $H=(h_{ij})_{i,j=1}^N$  be an $N\times N$  Hermitian matrix where the
 matrix elements $h_{ij}=\ov{h}_{ji}$, $ i \le j$, are independent 
random variables given by a probability measure $\nu_{ij}$ 
with mean zero and variance $\sigma_{ij}^2$. 
The variance of $h_{ij}$ for $i>j$ is $\sigma_{ij}^2 =\E\,  |h_{ij}|^2 = \sigma_{ji}^2$.
 For simplicity of the presentation, 
we assume that for any fixed $1\leq i<j\leq N$, ${\rm Re}\,h_{ij}$ 
and $\im \, h_{ij}$ are i.i.d.
with distribution $\om_{ij}$  
 i.e., $\nu_{ij} = \om_{ij}\otimes\om_{ij}$ 
in the sense that $\nu_{ij}(\rd h) = \om_{ij}(\rd {\rm Re}\,h)
\om_{ij}(\rd \im \, h)$, but this assumption is
not essential for the result.
The distribution $\nu_{ij}$ and its variance $\sigma_{ij}^2$ may depend on $N$,
 but we suppress this in the notation.  We assume that, for any $j$ fixed,
\be
   \sum_{i} \sigma^2_{ij} = 1 \, .
\label{sum}
\ee
Matrices with independent, zero mean  entries and with the normalization  condition
\eqref{sum} will be called {\it universal Wigner matrices.}
For a forthcoming review on this matrix class, see \cite{Spe},
where  the terminology of {\it random band matrices} was used.

Define $\Ci$ and $\Cs$ by
\be\label{defCiCs}
      \Ci:= \inf_{N, i,j}\{N\sigma^2_{ij}\}\leq \sup_{N, i,j}\{N\sigma^2_{ij}\}=:\Cs.
\ee
Note that $\Ci=\Cs$ corresponds to the standard Wigner matrices and
the condition $0< \Ci \le \Cs <\infty$ defines more general Wigner matrices
with comparable variances.

We will also consider an even more general case when $\sigma_{ij}$
for different $(i,j)$ indices are not comparable. 
The basic parameter of such matrices is the quantity
\be\label{defM}
M:= \frac{1}{\max_{ij} \sigma_{ij}^2}.
\ee
A special case is the band matrix, where  $\sigma_{ij}=0$ for $|i-j|>W$
with some parameter $W$. In this case,  $M$ and $W$ are related by $M\le CW$.

Denote by $B:=\{ \sigma^2_{ij}\}_{i,j=1}^N$ the matrix of variances 
which is symmetric and doubly stochastic by \eqref{sum}, in particular it satisfies
$-1\leq B\leq 1$. 
Let the spectrum of $B$ be supported in 
\be\label{de-de+}
\mbox{Spec}(B)\subset [-1+\delta_-, 1-\delta_+]\cup\{1\}
\ee   
with some nonnegative constants $\delta_\pm$. 
We will always have the following spectral assumption
\be\label{speccond}
\mbox{\it  1 is a simple eigenvalue of $B$ and
$\delta_-$ is a positive constant,  independent of $N$.}
\ee

The local semicircle law will be proven under this general condition, but
the precision of the estimate near the spectral edge will also depend on $\delta_+$
in an explicit way. For the orientation of the reader, we 
mention two special cases of universal Wigner matrices that provided the main motivation
for our work.

\bigskip

\noindent {\it Example 1. Generalized Wigner matrix.}  In this case we have
\be\label{VV}
	0<\Ci\leq \Cs<\infty,
\ee
and one can easily prove that $1$ is a simple eigenvalue of $B$
and \eqref{de-de+} holds with
\be\label{de-de+2}
	\delta_\pm \ge C_{inf},
\ee 
i.e., both 
 $\delta_-$ and $\delta_+$ are positive constants independent of $N$.

\bigskip

\noindent {\it Example 2. Band matrix.}  The variances are given by 
\be\label{BM}
   \sigma^2_{ij} = W^{-1} f\Big(\frac{ [i-j]_N}{W}\Big),
\ee
where $W\ge 1$, $f:\bR\to \bR_+$ is a bounded nonnegative symmetric function with 
$\int f =1$ and we defined $[i-j]_N\in\bZ$  by the property
that  $[i-j]_N\equiv i-j \; \mbox{mod}\,\,\, N$ and
$-\frac{1}{2}N < [i-j]_N \le\frac{1}{2}N $.
 Note that 
the relation \eqref{sum} holds only asymptotically as $W\to \infty$
but this can be remedied by an irrelevant rescaling.
If the bandwidth is
comparable with $N$, then we also have to assume that
 $f(x)$ is supported in $|x|\le N/(2W)$.
The quantity  $M$ defined in \eqref{defM} satisfies $M\le  W/\|f\|_\infty$.
 In Appendix \ref{sec:spec}
 we will show that \eqref{speccond} is satisfied for the choice of \eqref{BM}
if $W$ is large enough.

\bigskip

The Stieltjes transform of the empirical eigenvalue distribution of $H$ is given by 
\be\label{defmz}
 m(z) \equiv\mg (z) = \frac{1}{N} \tr\, \frac{1}{H-z}\,,\,\,\, z=E+i\eta.
 \ee 
 We define the density of the semicircle law
\begin{equation} \label{def rho sc}
\varrho_{sc}(x) \;:=\; \frac{1}{2 \pi} \sqrt{[4 - x^2]_+}\,,
\end{equation}
and, for $\im\, z > 0$, its Stieltjes transform
\begin{equation} \label{def m sc}
m_{sc}(z) \;:=\; \int_\R \frac{\varrho_{sc}(x)}{x - z} \, \rd x\,.
\end{equation}
The Stieltjes transform $m_{sc}(z) \equiv m_{sc}$ may also be characterized as the unique solution of
\begin{equation} \label{defmsc}
m_{sc} + \frac{1}{z + m_{sc}} \;=\; 0
\end{equation}
satisfying $\im\, m_{sc}(z) > 0$ for $\im\, z > 0$, i.e.,
\be\label{temp2.8}
m_{sc}(z)=\frac{-z+\sqrt{z^2-4}}{2}\,.
\ee
Here the square root function is chosen with a branch cut along the positive real axis.
This guarantees that the imaginary part of $m_{sc}$ is non-negative. The Wigner semicircle law states that  $m_N(z) \to m_{sc} (z) $
for any fixed $z$
 provided that $\eta=\im \,  z>0$ is independent of $N$. 
The local version of this result for universal Wigner
matrices is the content of the following Theorem.

\begin{theorem}[Local semicircle law] \label{mainls}{}
Let $H=(h_{ij})$ be a Hermitian  $N\times N$ random matrix where the
 matrix elements $h_{ij}=\ov{h}_{ji}$, $ i \le j$, are independent 
random variables
with $\E\, h_{ij}=0$, $1\leq i,j\leq N$,  and assume that the variances $\sigma_{ij}^2
=\E |h_{ij}|^2$  satisfy \eqref{sum}, \eqref{de-de+} and \eqref{speccond}. 
 Suppose that the distributions of the matrix elements have a uniformly 
  subexponential decay
in the sense that  there exist constants $\al$, $\beta>0$, independent 
of $N$, such that for any $x> 0$  we have 
\be\label{subexp}
\P(|h_{ij}|\geq x^\al |\sigma_{ij}|)\leq \beta e^{- x}.
\ee
Then there exist constants $C_1$, $C_2$, $C$ and $c>0$, 
depending only on $\al$, $\beta$ and $\delta_-$ in \eqref{speccond}, such that 
for any  $z=E+i\eta$ with $\eta={\mbox Im} \, z>0$, $|z|\leq 10$ and 
\be\label{fakerelkaeta}
\frac{1}{\sqrt{M\eta}}\leq \frac{\kappa^2}{(\log N)^{C_1}},
\ee 
where $\kappa : = \big| \, |E|-2 \big|$,  the Stieltjes transform of the empirical 
eigenvalue distribution of  $H $  satisfies 
\be\label{fakemainlsresult}
\P\left(|\mg(z)-m_{sc}(z)|\geq (\log N)^{C_2}
 \frac{1}{\sqrt{M\eta}\,\kappa}\right)\leq CN^{-c(\log \log  N)}
\ee
for sufficiently large $N$.
In fact, the same result holds for the individual matrix elements of
the Green's function $G_{ii}(z) = (H-z)^{-1}(i,i)$:
\be\label{Gii}
\P\left(\max_i | G_{ii}(z)-m_{sc}(z)|\geq (\log N)^{C_2}
 \frac{1}{\sqrt{M\eta}\,\kappa}\right)\leq CN^{-c(\log \log  N)}.
\ee
\end{theorem}

We remark that once a local semicircle law is obtained on a scale essentially $M^{-1}$,
it is straightforward to show that eigenvectors are delocalized 
on a scale at least of order $M$. The precise statement will be  formulated
in Corollary \ref{cor}. 
We will prove Theorem 
\ref{mainls} in Sections \ref{sec:proofloc}--\ref{proofsolvesleq}
 by extending  the approach of  \cite{ESY1, ESY2, ESY3}. The main ingredients
of this approach   consist of  i)  a derivation of a  self-consistent
 equation for the Green's function and ii) an  
induction on the scale of the imaginary part of the energy.
The key 
 novelty in this paper is that the self-consistent equation
is formulated for the array of the diagonal elements of
the Green's function $(G_{11}, G_{22}, \ldots , G_{NN})$
instead of the Stieltjes transform $m =\frac{1}{N} \tr G 
= \frac{1}{N}\sum_i G_{ii}$ itself as in \cite{ESY3}.
This yields for the first time
a strong pointwise control on the diagonal elements $G_{ii}$,
see \eqref{Gii}.

The subexponential decay condition \eqref{subexp} can be 
 weakened  if we are not aiming 
at error estimates faster than any power law of $N$. 
This can be easily carried out and we will not pursue  it in this paper.

\bigskip

Denote the eigenvalues of $H$ by $\lambda_1,  \ldots , \lambda_N$
and let $p_N(x_1,  \ldots , x_N)$ be 
their (symmetric) probability  density. 
For any
$k=1,2,\ldots, N$, the $k$-point correlation function  of the eigenvalues is defined by 
\be
 p^{(k)}_N(x_1, x_2,\ldots, x_k):=
\int_{\bR^{N-k}} p_N(x_1, x_2, \ldots , x_N)\rd x_{k+1}\ldots \rd x_N.
\label{corrfn}
\ee
We now state our main result concerning these correlation functions. 

\begin{theorem}[Universality for generalized Wigner matrices] \label{mainsk}
We consider a generalized hermitian  Wigner matrix such that \eqref{VV} holds.
Assume that the distributions $\nu_{ij}$ of the $(i,j)$ matrix elements 
have a uniformly   subexponential decay
in the sense of \eqref{subexp}. Suppose that the
real and imaginary parts of $h_{ij}$ are i.i.d., distributed
according to $\om_{ij}$, i.e.,
$\nu_{ij}(\rd h) = \om_{ij}(\rd \im \,  h)\om_{ij}(\rd \re h)$.
Let
 $m_k(i,j)=\int x^k\rd\om_{ij}(x)$,  $1\leq k\leq 4$, denote the $k$-th moment
of $\om_{ij}$
($m_1=0$). Suppose that
 \be\label{no3no4}
\inf_{N} \min_{1\leq i,j\leq N}\left\{\frac{m_4(i,j)}{(m_2(i,j))^2}-
\frac{(m_3(i,j))^2}{(m_2(i,j))^3}\right\}>1,
 \ee
then, for any $k\ge 1$ and for any compactly supported continuous test function
$O:\bR^k\to \bR$, we have
\be
\begin{split}
 \lim_{b\to0}\lim_{N\to \infty} \frac{1}{2b}
\int_{E-b}^{E+b}\rd E' \int_{\R^k} & \rd\alpha_1 
\ldots \rd\alpha_k \; O(\alpha_1,\ldots,\alpha_k)  \\
&\times
 \frac{1}{\varrho_{sc}(E)^k} \Big ( p_{N}^{(k)}  - p_{GU\! E, N} ^{(k)} \Big )
  \Big (E'+\frac{\alpha_1}{N\varrho_{sc}(E)}, 
\ldots, E'+\frac{\alpha_k}{N \varrho_{sc}(E)}\Big) =0,
\label{matrixthm}
\end{split}
\ee
where  $p_{GU\! E, N} ^{(k)}$ is the $k$-point correlation
function of the GUE ensemble. The same statement holds for
generalized symmetric Wigner matrices, with GOE replacing the GUE ensemble.
\end{theorem}

The limiting correlation functions of the GUE ensemble are given
by the sine kernel
$$
     \frac{1}{\varrho_{sc}(E)^k} p_{GU\! E, N} ^{(k)} 
\Big (E+\frac{\alpha_1}{N\varrho_{sc}(E)}, 
\ldots, E+\frac{\alpha_k}{N \varrho_{sc}(E)}\Big)
 \to \det\{ K(\al_i-\al_j)\}_{i,j=1}^k, \qquad K(x) = \frac{\sin \pi x}{\pi x},
$$
and similar universal formula is available for the limiting gap distribution.

\bigskip
 
{\it Remark}: The quantity  in the bracket in  \eqref{no3no4} is always  
greater or equal to 1 for any real distribution with mean zero,  which can be obtained by
$$
  m_3^2= \big[ \int x^3 \rd \om  \big]^2 =  \big[ \int x(x^2-m_2) \rd \om \big]^2
  \le \big[ \int x^2 \rd \om\big]\big[\int (x^2-m_2)^2 \rd \om\big] = m_2 (m_4 -  m_2^2 )
$$
and it
is exactly 1  if the distribution is supported 
on two points. For example, if $\om_{ij}$ is a rescaling of a fixed distribution
$\wt\om$ with variance $\frac{1}{2}$, 
i.e. $\om_{ij}(x)\rd x= \sigma_{ij}^{-1}\wt\om(x/\sigma_{ij})\rd x$,
then  condition \eqref{no3no4} is satisfied under \eqref{VV}, 
as long as the support of $\wt\om$
consists of at least three points.  The case of a Bernoulli-type distribution
supported on two points require a separate
argument and it will be treated in the forthcoming paper \cite{EYY}.

\bigskip

We now state our main comparison theorem for matrix elements of  Green's functions
of two Wigner ensembles.   As in the paper  \cite{TV}, 
we assume conditions on four moments. 
It will lead quickly to Theorem~\ref{com} stating that 
the correlation functions  of eigenvalues of two matrix ensembles 
are identical up to scale $1/N$ provided that the first four moments 
of all  matrix elements  of these two ensembles are almost identical.
Here we do not assume that the real and imaginary parts are i.i.d.,
hence the $k$-th moment of $h_{ij}$ is understood as the collection of numbers
$\int \bar h^s h^{k-s}\nu_{ij}(\rd  h)$, $s=0,1,2,\ldots,k$.  
The main result in \cite{TV} compares the joint distribution of individual eigenvalues ---
which is not covered by our Theorem \ref{comparison} --- 
but it does not address directly the matrix elements of 
Green's functions. 
The key input for both theorems is  the local semicircle law on
 the almost optimal scale $N^{-1+\e}$.
The eigenvalue perturbation  used in   \cite{TV}
  requires certain estimates on the eigenvalue level repulsion; 
the proof of Theorem \ref{comparison} is a straightforward  resolvent perturbation theory.

\begin{theorem}[Green's function comparison]\label{comparison}   Suppose that we have  
two generalized $N\times N$ Wigner matrices, $H^{(v)}$ 
and $H^{(w)}$, with matrix elements $h_{ij}$
given by the random variables $N^{-1/2} v_{ij}$ and 
$N^{-1/2} w_{ij}$, respectively, with $v_{ij}$ and $w_{ij}$ satisfying
the uniform subexponential decay condition
$$
   \P \big( |v_{ij}|\ge x^\al)\le \beta e^{-x},  \qquad
  \P \big( |w_{ij}|\ge x^\al)\le \beta e^{-x},
$$
with some $\al, \beta>0$.
  Fix a bijective ordering map on the index set of
the independent matrix elements,
\[
\phi: \{(i, j): 1\le i\le  j \le N \} \to \Big\{1, \ldots, \gamma(N)\Big\} , 
\qquad \gamma(N): =\frac{N(N+1)}{2},
\] 
and denote by  $H_\gamma$  the generalized Wigner matrix whose matrix 
elements $h_{ij}$ follow
the $v$-distribution if $\phi(i,j)\le \gamma$ and they follow the $w$-distribution
otherwise; in particular $H^{(v)}= H_0$ and $H^{(w)}= H_{\gamma(N)}$. 
Let $\kappa>0$ be arbitrary
 and suppose that, for any small parameter $\tau>0$ 
and for any  $y \ge N^{-1 + \tau}$,  we have 
the following estimate on the diagonal elements of the resolvent
\be\label{basic}
\P\left(\max_{0 \le \gamma \le \gamma(N)} \max_{1 \le k \le N}  
 \max_{|E|\le 2-\kappa}\left |  \left (\frac 1 {  H_{\gamma}-E- i y} \right )_{k k } 
\right |\le N^{2\tau} \right)\geq 1-CN^{-c\log\log N}
\ee
with some constants $C, c$ depending only on $\tau, \kappa$.
 Moreover, we assume that the first three moments of
 $v_{ij}$ and $w_{ij}$ are the same, i.e.
$$
    \E \bar v_{ij}^s v_{ij}^{u} =  \E \bar w_{ij}^s w_{ij}^{u},
  \qquad 0\le s+u\le 3,
$$
 and the difference between the  fourth moments of 
 $v_{ij}$ and $w_{ij}$ is much less than 1, say
\be\label{4match}
\left|\E \bar v_{ij}^s v_{ij}^{4-s}- \E \bar w_{ij}^s w_{ij}^{4-s}
\right|\leq N^{-\delta}, \qquad s=0,1,2,3,4,
\ee
for some given $\delta>0$.  Let $\e>0$ be arbitrary and choose an 
$\eta$ with $N^{-1-\e}\le \eta\le N^{-1}$.
For any  sequence of positive integers $k_1, \ldots, k_n$, set  complex parameters
$z^m_j = E^m_j \pm i \eta$,   $j = 1, \ldots k_m$, $m = 1, \ldots, n$,
 with $ |E^m_j| \le2-2\kappa $
and with an arbitrary choice of the $\pm$ signs. 
Let  $G^{(v)}(z) =  ( H^{(v)}-z)^{-1}$ denote the resolvent
and let    $F(x_1, \ldots, x_n)$ be a function such that for
any multi-index $\al=(\al_1, \ldots ,\al_n)$ with $1\le |\al|\le 5$
and for any $\e'>0$  sufficiently small, we have
\be\label{lowder}
\max \left\{|\partial^{\al}F(x_1, \ldots, x_n)|: 
\max_j|x_j|\leq N^{\e'}\right\}\leq N^{C_0\e'}
\ee
and
\be\label{highder}
\max\left\{|\partial^\al F(x_1, \ldots, x_n)|:
 \max_j|x_j|\leq N^2\right\}\leq N^{C_0}
\ee
for some constant $C_0$.

Then, there is a constant $C_1$,
depending on $\alpha, \beta$, $\sum_m k_m$ and $C_0$ such that for any $\eta$ with
$N^{-1-\e}\le \eta\le N^{-1}$
and  for any choices of the signs 
in  the imaginary part of $z^m_j$, we have 
\begin{align}\label{maincomp}
\Bigg|\E F  \left (  \frac{1}{N^{k_1}}\tr  
\left[\prod_{j=1}^{k_1} G^{(v)}(z^1_{j})\right ]  , \ldots, 
 \frac{1}{N^{k_n}} \tr \left [  \prod_{j=1}^{k_n} G^{(v)}(z^n_{j}) \right ]  \right )  &  -
  \E F\left ( G^{(v)} \to  G^{(w)}\right )  \Bigg| \non\\
\le & C_1 N^{-1/2 + C_1 \e}+C_1 N^{-\delta+ C_1 \e},
\end{align}
where the arguments of $F$ in the second term are changed from the Green's functions of $H^{(v)}$
to $H^{(w)}$ and all other parameters remain unchanged.
\end{theorem}

{\it Remark 1:}   We formulated Theorem \ref{comparison} for functions of
traces of monomials of the Green's function because this is the form
we need in the application. However, the result (and the proof we are going to present)
holds directly for
matrix elements of monomials of Green's functions as well, namely, 
 for any choice of $\ell_1, \ldots ,\ell_{2n}$, we have 
\begin{align}\label{maincomp1}
\Bigg|\E F &  \left (  \frac{1}{N^{k_1-1} } 
\left[\prod_{j=1}^{k_1} G^{(v)}(z^1_{j})\right ]_{\ell_1, \ell_2}  , \ldots, 
 \frac{1}{N^{k_n-1}}  \left [  \prod_{j=1}^{k_n} G^{(v)}(z^n_{j}) 
\right ]_{\ell_{2n-1}, \ell_{2n}}  \right )   -
  \E F\left ( G^{(v)} \to  G^{(w)}\right )  \Bigg| \non \\
& \le  C_1 N^{-1/2 + C_1 \e}+C_1 N^{-\delta+ C_1 \e}.
\end{align}
We also remark that  Theorem \ref{comparison} holds for 
generalized Wigner matrices since $C_{sup}<\infty$ in \eqref{defCiCs}.
 The positive lower
bound on the variances, $C_{inf}>0$, is not necessary for this theorem.

\bigskip 

{\it Remark 2:}  Although we state Theorem \ref{comparison} for Hermitian and symmetric ensembles, 
similar results  hold for real and complex sample covariance ensembles;  the modification of the proof, to be given in 
Section \ref{sec:proofcomp}, is  obvious and we omit the details.

\bigskip 
To summarize, our approach to prove the universality is based on the following three steps; 
a detailed  outline  will be given in Section \ref{sec:sk}. 
{\it Step 1.}  Local semicircle law, i.e.,  Theorem \ref{mainls}. This will be proved in Sections 
\ref{sec:proofloc}--\ref{proofsolvesleq}. {\it Step 2.}  Universality for 
ensembles with smooth distributions satisfying the logarithmic Sobolev inequality (LSI),
  Theorem \ref{lagaN-1}. The
key input is the general theorem, Theorem \ref{thm:main},  concerning the universality 
for the local relaxation flow. 
In Section \ref{sectlem}, by using the local semicircle law and  the LSI, we verify 
the assumptions for this theorem. {\it Step 3.} 
Green's function comparison theorem, Theorem \ref{comparison}. This removes  the
 restriction on the smoothness and the LSI, 
and it will be proved  in Section \ref{sec:proofcomp}.

{\it Convention.} We will frequently use the notation $C, c$ for
generic positive constants whose exact values are irrelevant
and may change  from line to line. For two positive quantities $A$, $B$ we also  introduce the notation  $A \asymp  B$ to indicate that there
exists a universal constant $C$ such that $C^{-1}\leq A/B\leq C$.

\section{Proof of local semicircle law}\label{sec:proofloc}

\textit{Proof of Theorem \ref{mainls} }  Recall that $G_{ij}=G_{ij}(z)$ 
denotes the  matrix element
\be
G_{ij}=\left(\frac1{H-z}\right)_{ij}
\ee
and  
  $$m(z)=m_N(z)=\sum_{i=1}^NN^{-1}G_{ii}(z).$$
 We will prove the following more
detailed stronger result.

\begin{theorem}\label{thm:detailed}
  Assume the $N\times N$ random matrix $H$ satisfies \eqref{sum}, \eqref{de-de+}, 
\eqref{speccond} and \eqref{subexp}, $\E\, h_{ij}=0$, for any $1\leq i,j\leq N$.
 Let $z=E+i\eta$ $(\eta>0)$ and let $g(z)$ be the real valued function  defined  by
\be\label{defgz}
g(z)\equiv \min\bigg\{\sqrt{\kappa+\eta},\,\,\,\max\left\{\delta_+,\,\,\,\left|{\rm Re} 
 \,[m_{sc}(z)^2]-1\right|\right\}\bigg\},
\ee
where $\kappa\equiv ||E|-2|$ and $\delta_+$ is given in \eqref{de-de+}.  Then for all 
 $z=E+i\eta$ and 
\be\label{relkaeta}
\frac{(\kappa+\eta)^{1/4}}{\sqrt{M\eta}\,\,g^{\,2}(z)}\leq (\log N)^{-13-6\al} ,\quad |z|\leq 10,
\ee 
we have \be\label{mainlsresult}
\P\left\{\max_{i}|G_{ii}(z)-m_{sc}(z)|\geq (\log N)^{11+6\al} 
\frac{(\kappa+\eta)^{1/4}}{\sqrt{M\eta}\,\,g(z)}\right\}\leq CN^{-c(\log \log  N)}
\ee
for sufficiently large N, with positive $c$ and $C>0$ depending only $\al$ and $\beta$ in 
\eqref{subexp} and $\delta_-$ in \eqref{de-de+} and \eqref{speccond}.
\end{theorem}

{\it Remark:} The condition \eqref{relkaeta} is effectively a lower bound on $\eta$.
 The  control function $g(z)$ can be estimated by 
\be\label{defgz2}
g(z)\asymp \left\{\begin{array}{cc} 
\min\Big\{\sqrt{\kappa+\eta},\,\,\,\max\left\{\delta_+,\,\,\,\eta/\sqrt{\kappa},\,\,\, 
\kappa\right\}\Big\}, &
 |E|\leq 2 {\rm\,\,\, and \,\,\,}\kappa\geq \eta, 
\\
\\\sqrt{\kappa+\eta}, &{\rm otherwise,}
\end{array}
\right.
\ee
up to some factor of order one. Note that the precise formula \eqref{defgz} for $g(z)$ is not
 important, only its asymptotic behaviour for small 
$\kappa$, $\eta$ and $\delta_+$ is relevant. The theorem remains valid if $g(z)$ is
 replaced by $\widetilde g(z)$ with $\widetilde g(z)\leq Cg(z)$. In particular, $g(z)$
 can be chosen to be order one when $E$ is not near the edges of the spectrum. If we 
are only concerned with the case of generalized Wigner matrices, \eqref{VV}, we can choose 
 $g(z)=O(\sqrt{\kappa+\eta})$ for any $z=E+i\eta$ $(\eta>0)$. Note that Theorem 
\ref{mainls} was obtained by replacing $g(z)$ with the lower bound $\kappa\leq g(z)$
 in Theorem \ref{thm:detailed}.

\bigskip

Once the local semicircle law is established on scale $\eta\asymp 1/M$ (modulo
logarithmic factors), we obtain
the following supremum bound on the eigenvectors that can be
interpreted as a lower bound of order $1/M$ on the localization length.
The proof of this result now is simpler than in \cite{ESY1, ESY2}, since
we have a pointwise control on the diagonal elements of the Green's function.
Let ${\bf u}_\al$ denote the normalized  eigenvector of $H$ belonging
to the eigenvalue $\lambda_\al$, $\al=1,2,\ldots, N$, 
i.e., $H{\bf u}_\al=\lambda_\al {\bf u}_\al$  
and $\|{\bf u}_\al\|=1$.

\begin{corollary}\label{cor} 
Let $H$ be as in Theorem \ref{thm:detailed},
 for any fixed $\kappa>0$, there exists $C_\kappa$ that  
\be\label{eigenvectorb}
\P\left\{\exists \;  \lambda_\al\in[-2+\kappa, 2-\kappa],\;
 H{\bf u}_\al=\lambda_\al {\bf u}_\al,
\; \|\bu_\al\|=1,\;\;
\|{\bf u}_\al\|_\infty \geq C_\kappa\frac{(\log N)^{13+6\al}}{M^{1/2}} 
  \right\}\leq CN^{-c \log\log N}.
\ee
For the  case of generalized Wigner matrices, \eqref{VV}, we have the following more precise bound
\be\label{evc}
\P\left\{\exists \;  \lambda_\al, \; H{\bf u}_\al=\lambda_\al {\bf u}_\al,
\; \|\bu_\al\|=1,\;\;
\|{\bf u}_\al\|_\infty \geq \frac{C(\log N)^{13+6\al}}{N^{1/2} \big[
\big| |\la_\al|-2\big| + N^{-1}\big]^{1/2}} 
  \right\}\leq CN^{-c\log  \log N}.
\ee

\end{corollary}

\noindent
{\it Proof of Corollary \ref{cor}.}  Let  $\eta=C_\kappa M^{-1} (\log N)^{26+12\al}$; $C_\kappa$ can be chosen large enough so that \eqref{relkaeta} is satisfied for all $|\kappa'|\leq \kappa$, making use of \eqref{defgz2} . 
Choose $\{E_m\}$ as a grid of 
points in $[-2+\kappa, 2-\kappa]$ such
 that the distance between any two neighbors is of order $\eta$. 
Then with \eqref{mainlsresult}, we have
\be
\P\left\{\max_{j}\max_{m}\im \,  |G_{jj}(E_m+i\eta)|\geq \im \,  
 m_{sc}(E_m+i\eta)+1\right\}\leq CN^{-c(\log \log  N)},
\ee
where we used $g(z)\leq \sqrt{\kappa+\eta}\leq C$ from \eqref{defgz2}.
 Then, with $|m_{sc}(z)|\leq C$ (see \eqref{temp2.8}) and 
\be
\im \,  G_{jj}(E_m+i\eta)=\sum_{\al}\frac{\eta |u_\al(j)|^2}{|E_m-\lambda_\al|^2+\eta^2},
\ee
where ${\bf u}_\al=(u_\al(1),\,\,\,u_\al(2)\ldots u_\al(N))$, we have 
\be\label{Pleftmax}
\P\left(\max_{j}\max_{m}\sum_{\al}\frac{\eta |u_{\al}(j)|^2}{|E_m-\lambda_\al|^2+\eta^2}
\geq C\right)\leq CN^{-c(\log \log  N)}.
\ee  
By the definition of $E_m$, for any $\lambda_\al\in [-2+\kappa, 2-\kappa]$, there exists
 $m'$ such that $|E_{m'}-\lambda_\al|$  is of the order of $\eta$. Together with \eqref{Pleftmax},
 we obtain \eqref{eigenvectorb}.

In case of the generalized Wigner matrix \eqref{VV}, we have $g(z)=\sqrt{\kappa+\eta}$
and $M\asymp N$. Let $\eta$  be the solution
to $\eta = N^{-1} (\log N)^{26+12\al}(\kappa +\eta)^{-3/2}$, then $N^{-1}\le
\eta \le CN^{-1}(\kappa+N^{-1})^{-3/2}
(\log N)^{26+12\al}$.
With this choice of $\eta$, \eqref{relkaeta}
is satisfied, and $\max_i |G_{ii}-m_{sc}| \le C(\log N)^{-2}(\kappa +\eta)^{1/2}$
holds with an overwhelming probability 
by \eqref{mainlsresult}.
Since $|\im \,  m_{sc}(z)| \le C\sqrt{\kappa+\eta}$,
so $\max_i \im \,  G_{ii} \le C(\kappa +\eta)^{1/2}$.
By the argument above, we obtain that  $\| \bu_\al\|_\infty^2\le C \eta(\kappa+\eta)^{1/2}$
 on this event.
This proves \eqref{evc}.
\qed

\bigskip

 To prove that $G_{ii}(z)$ is very close to $m_{sc}(z)$ in the sense of 
\eqref{mainlsresult}, we will also need to control the off-diagonal elements. In fact we 
will show that all $G_{ij}$ ($i\neq j$) are bounded by $O((M\eta)^{-1})$ up 
to some factor $(\log N)^C$.
To state the result  precisely, we first define  some events in the probability space. 

\par Recall that $\lambda_\al$, $\al=1,2,\ldots, N$, denote the eigenvalues
 of $H=(h_{ij})$. Denote 
by  $\Omega^0$  the subset of the probability space such that 
\be\label{BI5}
\max_{\alpha}|\lambda_\alpha|\leq 3 \; . 
\ee

Let  $\widehat\Omega^d_z$ (here the superscript $d$ means diagonal) be  the subset of $\Omega^0$
 where the following inequality on the diagonal terms hold for any $1 \leq i \leq N$
\be\label{proplsresult1}
|G_{ii}(z)-m_{sc}(z)|\leq (\log N)^{11+6\al} \frac{(\kappa+\eta)^{1/4}}{\sqrt{M\eta}\,\,g(z)} 
\ee
(recall that $m_{sc} (z)$ was  defined in \eqref{temp2.8}).
Similarly, let  $\widehat\Omega^o_z$ 
 (here the superscript $o$ means off-diagonal) be  the subset 
of $\Omega^0$ where the following inequality on the off-diagonal terms hold for 
any $1 \leq i\neq j \leq N$
\be\label{proplsresult2}
|G_{ij}(z)|\leq (\log N)^{5+4\al} \frac{(\kappa+\eta)^{1/4}}{\sqrt{M\eta}}.
\ee
Finally, denote by  $\Omega_z^d$  the set  
\be
\label{defOmzdo}
\Omega_z^{d}= \bigcap_{k =0}^{10 N^5}  \widehat\Omega^d_{z+ i k/N^5} ,
\ee
and similarly define $\Omega_z^o$. These sets depend on $N$ but we suppress this
 from the notations. 

\bigskip

{\it Proof of Theorem \ref{thm:detailed}.}  
The following proposition immediately implies Theorem \ref{thm:detailed}. 

\qed

\begin{proposition}\label{propls} Suppose that the assumptions of Theorem 
\ref{thm:detailed} hold. 
Then, for sufficiently large N, we have
\be\label{mainresultomz}
\P(\Omega_z^d \cap \Omega_z^o)\geq 1-CN^{-c\log\log N}
\ee
for some positive constants $c$ and $C$. 
\end{proposition}

\bigskip 

Following the work of \cite{ESY1}, we will
  use a continuity argument.   In  Section \ref{proofselfeq} we will derive
 a self-consistent equation of the form
\be\label{selfequation1}
G_{ii}+\frac{1}{z+\sum_{j}\sigma^2_{ij}G_{jj}+\Upsilon_i(z)
}=0,\,\,\,\,\,\,\,\,\,\,\,\, \,\,\,i=1,2\ldots , N.
\ee
Later we will give an explicit formula for
$\Upsilon_i(z)$, but for now we take \eqref{selfequation1} 
as  the definition of $\Upsilon_i$. 
 Let  $\widehat\Omega^\Upsilon_z(N)=\widehat\Omega^\Upsilon_z$  be  the subset 
of $\Omega^0$ where the following inequality holds
\be\label{temp2.28}
\Upsilon=\Upsilon(z):= \max_i |\Upsilon_i(z)|\leq  
(\log N)^{9+6\al}\frac{(\kappa+\eta)^{1/4}}{\sqrt{M\eta}} . 
\ee
We will use the following Lemmas that will be proved later in Section \ref{proofselfeq} and \ref{proofsolvesleq}.

\begin{lemma}\label{selfeq}
Let $z=E+i\eta$ be a fixed complex number satisfying \eqref{relkaeta}.  Then there are
  constants $C$ and $c$ such that for  $N\ge N_0$, with $N_0$ sufficiently  large
 independent of $E$ and $\eta$,
 the following estimates hold. 

(1)  Suppose  $3\leq \eta\leq 10$. Then  
\begin{align}\label{main310}
\P(\widehat\Omega_z^o )\geq & \; 1-CN^{-c\log\log N},\\
\label{selfequation2}
\P\left( \widehat\Omega_{z}^o \cap \widehat\Omega^\Upsilon_z\right)\geq &\; 1-CN^{-c\log\log N},
\end{align}
and for any $1\leq i\leq N$, 
\be
\label{EOmz} |\E {\bf 1}(\widehat\Omega_{z}^o)\Upsilon_i(z)|\leq (\log N)^{10+8\al}
\frac{(\kappa+\eta)^{1/2}}{M\eta}+CN^{-c(\log \log  N)}.
\ee

(2)  Suppose that $\eta\le 3$.
 Setting  $z'=z+iN^{-5}$, we have 
\begin{align}\label{main<3}
\P(\widehat\Omega_z^o \cap \Omega_{z'}^d \cap \Omega_{z'}^o)\geq & \;\P(\Omega_{z'}^d 
\cap \Omega_{z'}^o)-CN^{-c\log\log N}, \\
\label{selfequation3}
\P\left( \widehat\Omega_{z}^o\cap\Omega_{z'}^d
 \cap \Omega_{z'}^o\cap \widehat\Omega^\Upsilon_z \right)
\geq &\; \P\left( \widehat\Omega_{z}^o\cap\Omega_{z'}^d
 \cap \Omega_{z'}^o \right)- CN^{-c\log\log N},
\end{align}
and for any $1\leq i\leq N$, 
\be
\label{EOmz2}
 |\E {\bf 1}(\widehat\Omega_{z}^o\cap\Omega_{z'}^d \cap \Omega_{z'}^o) \Upsilon_i(z)|\leq
 (\log N)^{10+8\al}\frac{(\kappa+\eta)^{1/2}}{M\eta}+CN^{-c(\log \log  N)}.
\ee

\end{lemma}

\begin{lemma}\label{solvesleq}
Suppose we are on the event  $\widehat\Omega^\Upsilon_z$ for some fixed $z=E+i\eta$ satisfying \eqref{relkaeta}. Suppose  either $3\leq \eta\leq 10$ or  the following inequality hold: 

\be\label{condmaxi}\max_{i}|G_{ii}-m_{sc}(z)|\leq 2(\log N)^{-2}g(z). 
\ee
Then, for sufficiently large $N$, we have 
\be\label{resultsolvesleq}
\max_{i}|G_{ii}-m_{sc}(z)|\leq \frac{(\log N)^{2}}{g(z)}\Upsilon(z).
\ee
\end{lemma}
 
\bigskip 
{\it Proof of Proposition \ref{propls}. } 
 Recall that $\widehat \Omega_z^d$ 
is the subset of $\Omega^0$ where \eqref{proplsresult1} holds. 
Since on  $\widehat\Omega^\Upsilon_z$ \eqref{proplsresult1} follows from  \eqref{temp2.28},   the case $3\leq \eta=\im \,  z
\leq 10$ follows from Lemma \ref{selfeq}
  and  Lemma \ref{solvesleq} by taking a union bound for $0\leq k\leq 10N^5$.

Now we  prove  \eqref{mainresultomz} for the case $\eta\leq 3$ assuming that $z=E+i\eta$ 
satisfies \eqref{relkaeta}. We have shown that 
 \eqref{mainresultomz} holds for $\eta=3$, now we will  successively 
decrease $\eta$ by $N^{-5}$ in each step, and we continue this inductive procedure as long as 
\eqref{relkaeta} is still satisfied for the reduced $\eta$. More precisely, let $z'=z+iN^{-5}$ and
  assume that  \eqref{mainresultomz} holds for $z'$. Our goal is to prove that 
\be\label{2.1}
\P (\Omega_z^d \cap \Omega_z^o) \ge \P (\Omega_{z'}^d \cap \Omega_{z'}^o) -C N^{-c\log\log N}.
\ee
The number of steps we will be taking  is of order $N^5$.
Since $N^{-c\log\log N} \ll N^{-5}$, this proves  \eqref{mainresultomz} provided that we can 
establish \eqref{2.1}.

{F}rom \eqref{main<3}, the difference between the probabilities of the sets $\widehat\Omega_z^o 
\cap \Omega_{z'}^d \cap \Omega_{z'}^o$ 
and $\Omega_{z'}^d \cap \Omega_{z'}^o$ is negligible. 
With the definition of $\Omega_z^d$ and $\Omega_z^o$ in \eqref{defOmzdo}, we have 
\be 
\Omega_{z}^d \cap \Omega_{z}^o\supset
\widehat\Omega_z^d\cap \widehat\Omega_z^o \cap \Omega_{z'}^d
 \cap \Omega_{z'}^o.
\ee
Then,  to prove \eqref{2.1}, it remains to  prove
\be\label{temp3.292}
\P (\widehat\Omega_z^d\cap \widehat\Omega_z^o \cap \Omega_{z'}^d \cap \Omega_{z'}^o) \ge
 \P (\widehat\Omega_{z}^o\cap\Omega_{z'}^d \cap \Omega_{z'}^o) -C N^{-c\log\log N},
\ee
 i.e., we need to estimate the probability of the complement of $\widehat \Omega_z^d$ on the set 
 $\widehat\Omega_z^o \cap \Omega_{z'}^d \cap \Omega_{z'}^o$. On this set, using
 \eqref{selfequation3}, we can assume that  the estimate \eqref{temp2.28} holds  with a 
very high probability. We will show below that \eqref{condmaxi} holds on $\Omega_{z'}^d$. 
Then  \eqref{resultsolvesleq} together with 
\eqref{temp2.28} imply \eqref{proplsresult1}, the defining relation of $\widehat\Omega_{z}^d$. 
This will conclude \eqref{temp3.292} and complete the proof of Proposition \ref{propls}. 
Therefore, we only have  
to verify  \eqref{condmaxi}. 
\par Now we  show that \eqref{condmaxi} holds on $\Omega^d_{z'}$.  Recall $z'=z+iN^{-5}$ and
 we have the trivial estimate 
\be
|G_{ii}(z)-m_{sc}(z)|\leq  |G_{ii}(z)-G_{ii}(z')|+|m_{sc}(z)-m_{sc}(z')|+|G_{ii}(z')-m_{sc}(z')|. 
\ee
In the set $\Omega^d_{z'}$, we have 
\be
|G_{ii}(z')-m_{sc}(z')| \leq (\log N)^{11+6\al} \frac{(\kappa+\eta)^{1/4}}{\sqrt{M\eta}\,\,g(z')}
 \le (\log N)^{-2}g(z'),
\ee
where in the second inequality we used  \eqref{relkaeta}.  
By the definition of  $g(z)$ from \eqref{defgz}, we have  $g(z)\leq \sqrt{\kappa+\eta}$. Thus,
  if \eqref{relkaeta} holds, then, in particular,    
\be\label{lbeta}
\eta\geq C M^{-1}(\log N)^{26+12\al}.
\ee
This sets a lower bound on $\eta$.  Together with  $|z-z'| = 1/N^5$, 
 we have the trivial continuity bound 
\[
|G_{ii}(z)-G_{ii}(z')|+|m_{sc}(z)-m_{sc}(z')| \le N^{-2},
\]
using  $|\partial_z m_{sc}(z)|\leq |\im \,  z|^{-2}$,  
$|\partial_z G_{ii}(z)|=|[(H-z)^{-2}]_{ii}|\leq \|(H-z)^{-2}\|  \leq |\im \,   \, z|^{-2}$ and $\eta>N^{-1}$ from \eqref{lbeta}. Thus 
\be
|G_{ii}(z)-m_{sc}(z)|\leq  N^{-2}+(\log N)^{-2}g(z').
\ee
Using $|g(z)|\geq C\eta\geq CN^{-1}$ and  $|g'(z)|\leq C\eta^{-1}\leq CN$
  for $\eta\leq 3$,  we have the following estimate 
\be\label{temp2.27}
|G_{ii}(z)-m_{sc}(z)|\leq 2(\log N)^{-2}g(z) \,\,\, 
\ee
in the set  $\Omega^d_{z'}$. Thus the assumption \eqref{condmaxi} holds in the set $\Omega^d_{z'}$.  
\qed

\bigskip

Under the assumptions of Theorem \ref{thm:detailed}, 
with \eqref{mainresultomz}, \eqref{EOmz}, \eqref{EOmz2} and the  definitions in
 \eqref{defOmzdo},  all these $\Omega$'s are sets of almost full probability, i.e.,
\be\label{allomega}
\P(\widehat\Omega_z^o),\,\,\,  \P(\widehat\Omega_z^d),\,\,\, \P(\Omega_z^o),\,\,\,
 \P(\Omega_z^d),\,\,\, \P(\widehat\Omega_z^\Upsilon)\geq1-CN^{c\log\log N}
\ee
for some $c$, $C>0$.

\section{Self-consistent equation for Green's function}\label{proofselfeq}

First, we introduce some notations.
\begin{definition}\label{basicd}
\par  For any collection of $s$ different
numbers, $k_1, k_2, \ldots k_s\in \{1, 2, \ldots, N\}$,
let $H^{(k_1,k_2,\ldots,k_s)}$ denote the $N-s$ by $N-s$ submatrix of $H$ after removing the
 $k_i$-th $(1\leq i\leq s)$ rows and columns. Sometimes we use the notation  $H^{({\bf T})}$
 where  ${\bf T}$ denote the unordered set $\{ k_1, k_2, \ldots k_s\}$.   
Similarly, we define $\ba^{(\ell;\,\,{\bf T})}$ to be 
the $\ell$-th column of $H$ with $k_i$-th $(1\leq i\leq s)$
 elements removed. Sometimes, we just use the short notation $\ba^{\ell}$=$\ba^{(\ell;\,\,{\bf T})}$.
 \par 
\par For ${\bf T}=\{ k_1, k_2, \ldots k_s\}$, we define  
 \begin{align}
 G^{({\bf T})}_{ij}:= & \; [H^{({\bf T})}-z]^{-1}(i,j), \non \\
 Z^{({\bf T})}_{ij} := &\;\ba^{i}\cdot [H^{({\bf T})}-z]^{-1}\,\ba^{j}=\sum_{k,l\notin {\bf T}}
\overline{\ba^{\,i}_k} G^{({\bf T})}_{k,l}\ba^{j}_{l\,}, \non\\
\wH^{({\bf T})}_{ij} := &\; h_{ij}-z\delta_{ij}-Z^{({\bf T})}_{ij}. \non
 \end{align}
These quantities depend on $z$, but we mostly neglect this dependence in the notation. 
\end{definition}
\bigskip
\par  We start the proof with deriving some identities between the matrix
elements of $G=(H-z)^{-1}$ and 
$G^{(k_1,k_2,\ldots,k_s)}$ using  the following well known result in linear algebra that we quote
 without proof. 
\begin{lemma}\label{basicf}
Let $A$ , $B$, $C$ be $n\times n$, $m\times n$
 and $m\times m$ matrices. We define $(m+n)\times (m+n)$
 matrix $D$ as 
\be 
D=\begin{pmatrix}
    A & B^*  \\
  B& C 
\end{pmatrix}
\ee
and  $n\times n$ matrix $\widehat D$ as
\be\label{defhatD}
\widehat D=A -B^*C^{-1}B.
\ee
Then for any $1\leq i,j\leq n$, we have $$(D^{-1})_{ij}= {(\widehat D^{-1})}_{ij}$$ for the 
corresponding matrix elements. 

Furthermore, let  ${\bf T}$ denote the unordered set $\{ k_1, k_2, \ldots k_s\}$ and $1\leq k_i\leq n$,
 $1\leq i\leq s$.  We define $D^{({\bf T})}$ to be the $n+m-s$ by
 $n+m-s$ submatrix of $D$ after removing 
the $k_i$-th $(1\leq i\leq s)$ rows and columns and define
 $\widehat D^{({\bf T})}$ to be the $n-s$ 
by $n-s$ submatrix of $\widehat D$ after removing the $k_i$-th $(1\leq i\leq s)$ rows and columns.
Then for any $1\leq i,j\leq n$ and $i,j\notin {\bf T}$,  we have 
$$\left((D^{({\bf T})})^{-1}\right)_{ij}= \left((\widehat D^{({\bf T})})^{-1}\right)_{ij}$$ 
for the corresponding matrix elements. 

\end{lemma}
\bigskip

\par Using Lemma \ref{basicf} and Definition \ref{basicd},
 for $1\leq i \not = j\leq N$, we have 
\be\label{G11}
G_{ii}=(\wH^{(i)}_{ii})^{-1}
=\frac {\wH^{(ij)}_{jj} } { \wH^{(ij)}_{jj}\wH^{(ij)}_{ii}-\wH^{(ij)}_{ij}\wH^{(ij)}_{ji} }.
\ee
For the off diagonal matrix elements $G_{ij}$, $(i\neq j)$, we have
\be\label{G12}
G_{ij}= -\frac {\wH^{(ij)}_{ij} } { \wH^{(ij)}_{jj}\wH^{(ij)}_{ii}
-\wH^{(ij)}_{ij}\wH^{(ij)}_{ji} }
 = -G_{ii}\frac{\wH^{(ij)}_{ij}}{\wH^{(ij)}_{jj}}=-G_{ii}G_{jj}^{(i)}\wH^{(ij)}_{ij}.
\ee
Similarly, we have the following result

\begin{lemma}\label{basicIG} Let ${\bf T}$ be  an
unordered set $\{k_1$, $k_2$, $\ldots$, $k_s\}$ with
   $1\leq k_t\leq N$ for $(1\leq t\leq s)$  or $\bf T=\emptyset$. For  simplicity, we use the
 notation $(i \,{\bf T})$ for $\{i\}\cup {\bf T}$ and $(i j \,{\bf T})$
 for $\{i,j\}\cup {\bf T}$. 
 Then we have the following identities:
\par\begin{enumerate}    
\item For any  $i\notin {\bf T}$ 
\be\label{GiiHii} 
 G^{({\bf T})}_{ii}=(\wH^{(i\,{\bf T})}_{ii})^{-1}.
 \ee
 \item For $i\neq j$ and $i,j\notin {\bf T}$
\be\label{GijHij} 
 G^{({\bf T})}_{ij}=-G^{({\bf T})}_{jj}G_{ii}^{(j\,{\bf T})}\wH^{(ij\,\,{\bf T})}_{ij}=
-G^{({\bf T})}_{ii}G_{jj}^{(i\,{\bf T})}\wH^{(ij\,\,{\bf T})}_{ij}.
 \ee
 	\item  For $i\neq j$ and $i,j\notin {\bf T}$
  \be\label{GiiGjii}
  G^{({\bf T})}_{ii}-G^{(j\,\,{\bf T})}_{ii}=
G^{({\bf T})}_{ij}G^{({\bf T})}_{ji}(G^{({\bf T})}_{jj})^{-1}.
  \ee
  \item  For any indices  $i$, $j$ and $k$ that are different  and 
  $i,j,k \notin {\bf T}$
 \be\label{GijGkij}
G^{({\bf T})}_{ij}-G^{(k\,\,{\bf T})}_{ij}=G^{({\bf T})}_{ik}G^{({\bf T})}_{kj}
(G^{({\bf T})}_{kk})^{-1} . \ee 
 \end{enumerate}
 
\end{lemma}
{\it Proof of Lemma \ref{basicIG}.} The first two identities \eqref{GiiHii} and \eqref{GijHij}
 are obvious extensions of  \eqref{G11} and \eqref{G12}. 
To prove  \eqref{GiiGjii},  without loss of generality, we may assume that $i=1$, $ j=2$ 
and ${\bf T}=\emptyset$.  Let $D=H-z$ and $\widehat D$ defined as in \eqref{defhatD} with
 $n=2$ and $m=N-2$. 
With Lemma \ref{basicf},  we have that for $ i,j=1$ or $2$. 
\be G_{ij}= (\widehat D^{-1})_{ij}
\ee
and 
\be
G_{ii}^{(j)}= \left((\widehat D^{(j)})^{-1}\right)_{ii}   \,.
\ee
 Since $\widehat D$ is just a $2\times 2$ matrix, one can easily check that  
\eqref{GiiGjii} holds. 
With the same method, one can obtain \eqref{GijGkij} .
\qed
\bigskip

\begin{lemma}\label{selfeq1} The diagonal matrix elements of the resolvent
 satisfy the following 
self-consistent equation. 
\be\label{mainseeq}
G_{ii} = \left(-z- \sum_{j}\sigma^2_{ij}G_{jj}+\Upsilon_i\right)^{-1}
\ee
where $\Upsilon_i(z)$ is given by
\be\label{seeqerror}
\Upsilon_i(z):=\sigma^2_{ii}G_{ii}+\sum_{j\neq i}\sigma^2_{ij}G_{ij}G_{ji}[G_{ii}]^{-1}+
\left(\wH^{(i)}_{ii}-\E_{\ba^i}\wH^{(i)}_{ii}\right),
\ee
and $\E_{\ba^i}$ is the expectation over ${\ba^i}$. 
Let ${\bf T}$ denote the set $\{ k_1,k_2,\ldots k_m\}$,
 which also could be the empty set, then
\be 
\label{BI2}
|K^{(i{\bf T})}_{ii}-\E_{\ba^i}K^{(i{\bf T})}_{ii}|\leq (\log N)^{3+2\al}
\sqrt{M^{-1}+M^{-1}\max_{k}|G^{(i\bf T)}_{kk}|^2+\max_{k\neq l}|G^{(i\bf T)}_{kl}|^2}
\ee
and for $i\neq j$
\be
\label{BI3}
|K^{(ij{\bf T})}_{ij}|
 \leq(\log N)^{4+4\al}\sqrt{M^{-1}+(M\eta)^{-1}
  \max_{l}\left\{\im \,  G^{(ij\bf T)}_{ll}\right\}}, \quad i\ne j,
\ee
hold with a probability larger than $1-CN^{-c(\log \log  N)}$ for sufficiently large $N$. 
\end{lemma}
\bigskip
{\it Proof of Lemma \ref{selfeq1}. } We can write  $G_{11}$  as follows, 
\be\label{ep2G22}
   G_{11} = (\wH^{(1)}_{11})^{-1}=\frac{1}{\E_{\ba^1}\wH^{(1)}_{11}+\wH^{(1)}_{11}-
\E_{\ba^1}\wH^{(1)}_{11}}.
\ee
Using the fact $G^{(1)}=(H^{(1)}-z)^{-1}$ is  independent of $\ba^1$ and $\E_{\ba^1}
 \overline {\ba^1(i)}\ba^1(j)=\delta_{ij}\sigma^2_{1j}$, we obtain
$\E_{\ba^1}\wH^{(1)}_{11}=-z- \sum_{j\neq 1}\sigma^2_{1j}G^{(1)}_{jj} $, and thus
\be
 G_{11} = \frac{1}{-z- \sum_{j\neq 1}\sigma^2_{1j}G^{(1)}_{jj} +\left(\wH^{(1)}_{11}
-\E_{\ba_1}\wH^{(1)}_{11}\right)}.
\ee
Combining this identity with 
\eqref{GiiGjii}, we have
\be
G_{11} = \left(-z- \sum_{j\neq 1}\sigma^2_{1j}\left(G_{jj}-G_{1j}G_{j1}G^{-1}_{11}\right)
 +\left(\wH^{(1)}_{11}-\E_{\ba^1}\wH^{(1)}_{11}\right)\right)^{-1}.
\ee
Clearly $G_{11}$ can be replaced with any $G_{ii}$ and this proves
 \eqref{mainseeq} with the definition \eqref{seeqerror}. 

Now  we prove \eqref{BI2} and \eqref{BI3}. 
  Define 
\be\label{defz}
v_{ij}\equiv h_{ij}/\sigma_{ij},
\ee
hence $\E v_{ij}=0$ and $\E |v_{ij}|^2=1$.  If $\sigma_{ij}=0$,
 i.e., $h_{ij}=0$ almost surely,
then we set $v_{ij}=0$. By the definition of $K^{(i{\bf T})}_{ii}$, we  write
\be\label{temp3.20}
K^{(i{\bf T})}_{ii}=h_{ii}-z-\sum_{k,l\notin(i{\bf T})}
\overline{\ba^i_k }G^{(i{\bf T})}_{kl}\ba^i_l
=h_{ii}-z-\sum_{k,l\notin(i{\bf T})}\overline{v_{ik}}\sigma_{ik}
 G^{(i{\bf T})}_{kl}\sigma_{li}v_{li}
\ee
and 
\be
\E_{\ba^i}K^{(i{\bf T})}_{ii}=-z-\sum_{k\notin(i{\bf T})}\sigma_{ik}
 G^{(i{\bf T})}_{kk}\sigma_{ki}.
\ee
We note $h_{ii}$, $v_{ij}$ and $G^{(i{\bf T})}_{kl}$ are independent for
 $k,l\notin(i{\bf T})$.
 With the sub-exponential decay \eqref{subexp} and $\sigma_{ij}^2\leq 1/M$,
we have for any $i,j$ 
\be\label{temp3.22}
\P\left\{ |h_{ij}|\leq (\log N)^{3+2\al}M^{-1/2}\right\}\geq 1-CN^{-c(\log \log N)}.
\ee
In Corollary \ref{generalHWT} of Appendix \ref{sec:lde}
 we will prove a general large deviations result.
Applying  \eqref{diaglde} to
 the last term in \eqref{temp3.20},  with the choice 
\be
B_{kl}=\sigma_{ik}G^{(i{\bf T})}_{kl}\sigma_{li}
\ee
and with $\sum_{j}\sigma_{ij}^2=1$ and $\sigma_{ii}^2\leq 1/M$, we obtain that \be
\left|\sum_{k,l \notin(i{\bf T})}\overline{v_{ik}}\sigma_{ik} 
G^{(i{\bf T})}_{kl}\sigma_{li}v_{li}
-\sum_{k\notin(i{\bf T})}\sigma_{ik} G^{(i{\bf T})}_{kk}\sigma_{ki}\right|
\leq (\log N)^{3+2\al}\sqrt{M^{-1}\max_{k}|G^{(i\bf T)}_{kk}|^2
+\max_{k\neq l}|G^{(i\bf T)}_{kl}|^2}
\ee
 holds with a probability larger than $1-CN^{-c(\log \log N)}$.
 Together with \eqref{temp3.22}, 
we obtain that \eqref{BI2} holds with 
a probability larger than $1-CN^{-c(\log \log  N)}$ for
 sufficiently large $N$.  
\par Next we prove \eqref{BI3}. By the definition of $K^{(ij{\bf T})}_{ij}$, 
 $i\neq j$, we can write
\be\label{temp3.202}
K^{(ij{\bf T})}_{ij}=h_{ij}-\sum_{k,l\notin(ij{\bf T})}\overline{v_{ik}}\sigma_{ik}
 G^{(ij{\bf T})}_{kl}\sigma_{lj}v_{lj}.
\ee
Applying  \eqref{resgenHWTO},
\eqref{temp3.22} and $\sigma^2_{ij}\leq 1/M$, we obtain that 
 \be\label{BI23}
|K^{(ij{\bf T})}_{ij}|
 \leq(\log N)^{4+4\al}\sqrt{M^{-1}+\sum_{k,l\notin(ij{\bf T})}\left|\sigma_{ik}
G^{(ij{\bf T})}_{kl}\sigma_{lj}\right|^2}
\ee
holds with a probability larger than $1-CN^{-c(\log \log  N)}$ for
 sufficiently large $N$.
 With Schwarz's inequality, for any $i,j$, 
\be
\sum_{kl}\left|\sigma_{ik}G^{(ij{\bf T})}_{kl}\sigma_{lj}\right|^2\leq 
\left(\sum_{kl}|\sigma_{ik}|^4|G^{(ij{\bf T})}_{kl}|^2\right)^{1/2}
\left(\sum_{kl}|G^{(ij{\bf T})}_{kl}|^2|\sigma_{jl}|^4\right)^{1/2}.
\ee
\par  Denote $u^{({ij\bf T})}_\alpha$ and $\lambda_\al^{({ij\bf T})}$ ($\al=1,2,\ldots ,N-|{\bf T}|-2$)
 the $l^2$-normalized eigenvectors and eigenvalues of $H^{({ij\bf T})}$. Let $u^{({ij\bf T})}_\alpha(l)$ denote the
 $l$-th coordinate of $u^{({ij\bf T})}_\alpha$, then for any $l$
\be
\label{BI4}
\sum_{k}|G^{(i\,j\,{\bf T})}_{kl}|^2=\left(|G^{(i\,j\,{\bf T})}|^2\right)_{ll}=\sum_{\alpha}
\frac{|u^{(i\,j\,{\bf T})}_\alpha(l)|^2}{|\lambda_\alpha^{(i\,j\,{\bf T})}-z|^2}
= 
\frac{\mbox\im \,   G^{(i\,j\,{\bf T})}_{ll}(z)}{\eta}.
\ee
Here we defined $|A|^2:= A^* A$ for any matrix $A$.
 Inserting \eqref{BI4} into \eqref{BI23}
and using  the definition of $M$ in \eqref{defM}, we obtain that \eqref{BI3} holds with a 
probability larger than $1-CN^{-c(\log \log  N)}$ for sufficiently large $N$. 
\qed
\bigskip

{\it Proof of Lemma  \ref{selfeq}.}   We first  prove \eqref{main310}
in the range  $3\leq \eta\leq 10$.  Recall that $\Omega^0$ is  the 
subset of the entire probability space where  $\|H\|\leq 3$ see \eqref{BI5}.
By \eqref{resnlambda3} from  Lemma  \ref{N-1/6} (using that
$M\ge (\log N)^9$ is implied by \eqref{relkaeta}),
 we have  $\P(\Omega^0)\geq 1-N^{-c\log\log N}$. 
Denote $\lambda_\al$ and $\bu_\al$ the eigenvalues and eigenvectors of $H=(h_{ij})$. From  the identity
\be
G_{ii}=\sum_{\al}\frac{|u_\al(i)|^2}{\lambda_\al-z}
\ee
and  $\max_{\alpha}|\lambda_\alpha| \le 3$, we have that 
\be
\eta^{-1}\geq |G_{ii}|\geq |\im \,   G_{ii}| \geq\frac{\eta}{(|E|+3)^2+\eta^2}
\ee
holds in $\Omega^0$. Together with $3\leq \eta\leq 10$ and $|E|\leq 10$, we obtain 
\be\label{cGC}
c\leq |G_{ii}|
\leq C
\ee 
with some positive constants.
{F}rom the interlacing property of the eigenvalues of the matrix and its submatrices,
 we find that not only $\|H\|\leq 3$ but also $\|H^{(\bf T)}\|\leq 3$ holds on the
 set $\Omega^0$. Thus for any $j, k$
 such that $i$, $j$ and $k$ are all different, the bounds 
\be\label{cGjkC}
c\leq|G^{}_{ii}|,\,\,\, |G^{(j)}_{ii}|,\,\,\, |G^{(j k)}_{ii}|\leq C.
\ee 
hold in $\Omega^0$ by a similar argument that led to \eqref{cGC}.
Thus \eqref{GijHij} implies   
\be
{\bf1}(\Omega^0)|G^{(ij)}_{ij}|\leq C^2{\bf1}(\Omega^0)|K^{(ij)}_{ij}|  
\ee
and \eqref{main310} follows make use of \eqref{BI3} and $\eta>3$. 
\bigskip

Now we prove \eqref{selfequation2}. Recall that the self consistent equation
 \eqref{selfequation1} with the error term $\Upsilon_i(z)$ is given by 
 \eqref{seeqerror}, i.e., 
\be \label{ups}
\Upsilon_i(z)=\sigma^2_{ii}G_{ii}+\sum_{j\neq i}\sigma^2_{ij}G_{ij}G_{ji}G_{ii}^{-1}
+\left(\wH^{(i)}_{ii}-\E_{\ba^i}\wH^{(i)}_{ii}\right).
\ee

Now we bound $\Upsilon_i(z)$ in $\widehat\Omega^o_z$. Since  $\sigma^2_{ii}\leq M^{-1}$, 
with \eqref{cGjkC}, the first term of the r.h.s. of \eqref{ups} is less than $O(M^{-1})$. 
Then with \eqref{sum}, and using  the bound on $G_{ij}$ ($i\neq j$) from \eqref{proplsresult2} and the
 one on $G_{ii}$ from \eqref{cGjkC}, we obtain that  the second term of the r.h.s. of \eqref{ups}
 is less than $C(\log N)^{10+8\al}(M\eta)^{-1}$
(and with \eqref{lbeta}, we know it is much less than 1), i.e.,  in $\widehat\Omega^o_z$
\be\label{temp3.30}
\left|\sigma^2_{ii}G_{ii}+\sum_{j\neq i}\sigma^2_{ij}G_{ij}G_{ji}G_{ii}^{-1}\right|
\leq C(\log N)^{10+8\al}(M\eta)^{-1}\,.
\ee
 The last term of  the r.h.s. of \eqref{ups} can be bounded, using \eqref{BI2} with
 ${\bf T}=\emptyset$, with a very large probability. Using \eqref{GijGkij} and 
\eqref{cGjkC}, the $G^{(i)}_{kl}$'s in \eqref{BI2} can be bounded as
\be
|G^{(i)}_{kl}|\leq |G^{}_{kl}|+C|G_{ki}||G_{il}|.
\ee
Therefore, again with  the bound on $G_{ij}$ ($i\neq j$) in \eqref{proplsresult2} 
and the one on $G_{ii}$ from \eqref{cGjkC}, we see that
 \be\label{temp3.31}
|\wH^{(i)}_{ii}-\E_{\ba^i}\wH^{(i)}_{ii}|\leq 2(\log N)^{8+6\al}(M\eta)^{-1/2}
\ee
holds in $\widehat\Omega^o_z$ with a probability larger
 than $\P(\widehat\Omega^o_z)-CN^{-c(\log \log  N)}$ for sufficiently large $N$. 
Inserting \eqref{temp3.30} and \eqref{temp3.31} into \eqref{ups} and together
   $\eta\geq3$, we have proved   \eqref{selfequation2}.

\bigskip

\bigskip

\par 
Now we prove \eqref{EOmz} for $\eta\geq 3$. By the definition of $\Upsilon_i$ in \eqref{ups},
we have 
\be\label{temp3.36}
\left| \E\left[{\bf 1}(\widehat\Omega^o_z) \Upsilon_i(z) \right]\right|
 \leq
\E{\bf 1}(\widehat\Omega^o_z)
\left|\sigma^2_{ii}G_{ii}+
\sum_{j\neq i}\sigma^2_{ij}G_{ij}G_{ji}G_{ii}^{-1}\right|
+\left|\E\,{\bf 1}([ \widehat\Omega^o_z ]^c)\left(\wH^{(i)}_{ii}-\E_{\ba^i}\wH^{(i)}_{ii}\right)\right|,
\ee
since $\E\left(\wH^{(i)}_{ii}-\E_{\ba^i}\wH^{(i)}_{ii}\right)=0$. Using \eqref{cGjkC},  in $\Omega^0$  we have that   $|\sigma^2_{ii}G_{ii}+
\sum_{j\neq i}\sigma^2_{ij}G_{ij}G_{ji}G_{ii}^{-1}|$ 
is always less than  a constant ${C}$ for some $C>0$. Inserting this and
 \eqref{temp3.30} into \eqref{temp3.36}, we obtain that 
\be\label{temp3.38}
\left| \E\left[{\bf 1}(\widehat\Omega^o_z) \Upsilon_i(z) \right]\right|
\leq C(\log N)^{10+8\al}(M\eta)^{-1}
+\left|\E\left[{\bf 1}([ \widehat\Omega^o_z ]^c)\big| \wH^{(i)}_{ii}-\E_{\ba^i}\wH^{(i)}_{ii}\big|\right]\right|
+CN^{-c\log\log N}.
\ee
We now claim that for some large enough $C>0$ there exists $c>0$
such that
\be\label{temp3.39}
    \P (|Z_{ii}^{(i)}|\ge N^C) \le e^{-N^c} \qquad \mbox{and}
   \qquad \P (|K_{ii}^{(i)}|\ge N^C) \le e^{-N^c}.
\ee
The first estimate follows from the definition of $Z_{ii}^{(i)}$
given in Definition \ref{basicd} 
by using the sub-exponential decay of the matrix elements
and by using the trivial bound $|G_{kl}^{(i)}|\le \eta^{-1}\ne N$.
The second estimate is a trivial consequence of the
first one and the definition of $K_{ii}^{(i)}$. Together with \eqref{temp3.38}, we obtain \eqref{EOmz} in the case that $3\leq \eta\leq 10$.

We now  prove  \eqref{main<3} and \eqref{selfequation3}
for  the case  $\eta\le 3$  satisfying \eqref{relkaeta}. We will work in the
event $\Omega_{z'}^d\cap\Omega_{z'}^d$ where $z'=z+i N^{-5}$.
Similarly as we proved 
 \eqref{temp2.27}, from the bound below \eqref{lbeta} and the Lipschitz continuity of
 $g(z)$ 
 , we obtain that 
\be\label{syj440}
|G_{ii}(z)-m_{sc}(z)|\leq 2(\log N)^{11+6\al} \frac{(\kappa+\eta)^{1/4}}{\sqrt{M\eta}\,\,g(z)}
\ee
and
\be
|G_{ij}(z)|\leq 2(\log N)^{5+4\al} \frac{(\kappa+\eta)^{1/4}}{\sqrt{M\eta}}
\ee
hold in $\Omega_{z'}^d\cap\Omega_{z'}^d$.  We note the r.h.s of these inequalities are much less than $(\log N)^{-1}$ by \eqref{relkaeta}.
{F}rom the explicit formula \eqref{temp2.8} we obtain that $c\leq |m_{sc }(z)|\leq C$ for any $|z|\leq 10$ with some positive constants. Using this observation and the fact that the r.h.s. of \eqref{syj440} is much less than $(\log N)^{-1}$, we have 
$$
c\leq|G_{ii}(z)|\leq C\,.
$$
  Hence, using \eqref{lbeta}, \eqref{GiiGjii}, \eqref{GijGkij} and the lower bound of $|G_{ii}|$,  one can easily obtain that
\be\label{cGjkC2}
|G^{}_{ii}(z)-m_{sc}(z)|,\,\,\, |G^{(j)}_{ii}(z)-m_{sc}(z)|,\,\,\, |G^{(jk)}_{ii}(z)-m_{sc}(z)|
\leq C(\log N)^{11+6\al} \frac{(\kappa+\eta)^{1/4}}{\sqrt{M\eta}\,\,g(z)}
\ee
hold in $\Omega_{z'}^o\cap \Omega_{z'}^d$, (for the third term in l.h.s., we have also used the lower bounds of $G^{(j)}_{ii}$'s as above). 
Then we also have
\be\label{cGjkC3}
c\leq|G_{ii}(z)|,\,\,\, |G^{(j)}_{ii}(z)|,\,\,\, |G^{(jk)}_{ii}(z)|\leq C
\ee
with some positive constants.

  The definition of $m_{sc}(z)$ implies  $\im \,   \,\,m_{sc}(z)\leq C\sqrt{\kappa+\eta}$. 
Then  with \eqref{cGjkC2}, \eqref{relkaeta} and $g(z)\leq \sqrt{\kappa+\eta}$, we have that 
\be
\im \,   G^{(j k)}_{ii}(z)\leq C\sqrt{\kappa+\eta}
\ee
holds in $\Omega_{z'}^o\cap \Omega_{z'}^d$ for some constant $C>0$. Inserting it into 
\eqref{BI3}, we obtain that 
\be
|K^{(ij)}_{ij}(z)|\leq C(\log N)^{4+4\al} \frac{(\kappa+\eta)^{1/4}}{\sqrt{M\eta}}
\ee
hold in $\Omega_{z'}^o\cap \Omega_{z'}^d$ with a probability larger than
  $\P(\Omega_{z'}^o\cap \Omega_{z'}^d)-CN^{-c(\log \log  N)}$ for sufficiently 
large $N$. Again,  with \eqref{GijHij} and  \eqref{cGjkC3}, we obtain 
 \eqref{main<3} for sufficiently large $N$.
\par Then, as we proved in \eqref{temp3.30} and \eqref{temp3.31}, we get that
\be
|\sigma^2_{ii}G_{ii}(z)+\sum_{j\neq i}
\sigma^2_{ij}G_{ij}G_{ji}G_{ii}^{-1}(z)|\leq C(\log N)^{10+8\al}(M\eta)^{-1}
\ee
and 
\be
|\wH^{(j)}_{jj}(z)-\E_{\ba^j}\wH^{(j)}_{jj}(z)|\leq C(\log N)^{8+6\al}(M\eta)^{-1/2}
\ee
hold in $\widehat\Omega_{z}^o\cap\Omega_{z'}^o\cap \Omega_{z'}^d$ with a probability 
larger than  $\P(\widehat\Omega_{z}^o\cap\Omega_{z'}^o\cap \Omega_{z'}^d)-CN^{-c(\log \log  N)}$, 
which implies  \eqref{selfequation3}. 
\par  Finally, similarly as using  \eqref{temp3.36}- \eqref{temp3.38} to prove 
\eqref{EOmz}, we can obtain \eqref{EOmz2} in the case that $\eta<3$.  
\qed
\bigskip

\section{Stability of the self-consistent equation:
proof of  Lemma  \ref{solvesleq}}\label{proofsolvesleq}

In this section,  we prove Lemma  \ref{solvesleq}, i.e.,  we will prove 
 the stability of the self-consistent equation with a precise error estimate given
 in  \eqref{resultsolvesleq}.  We set $m_{sc}=m_{sc}(z)$ and $\Upsilon=\max_i|\Upsilon_i(z)|$ 
for simplicity
of notation and we will omit all $z$ dependences in all the symbols. With the 
definition of $m_{sc}(z)$ in \eqref{defmsc} and \eqref{temp2.8}, the following properties 
of $m_{sc}(z)$ can be easily  established: 
\begin{lemma} \label{msc}
Let  $z=E + i \eta$ with $\eta > 0$ and $|z|\le 20$. Then  we have
\be\label{zmsc2}
|z+m_{sc}|^{-2}=|m_{sc}|^2\leq 1
\ee
and
\be\label{temp5.4}
\left|(z+m_{sc}(z))^{-2}-1\right|\geq C\sqrt{\kappa+\eta}
\ee
for some constant $C$. Furthermore, 
suppose that either $2\leq|E|\leq10$ or $\kappa\leq\eta$. Then 
\be\label{templemmare}
|z+m_{sc}|^{-2}=|m_{sc}|^2\leq 1-C\sqrt{\kappa+\eta}.
\ee
For small values of $|z^2-4|\asymp \kappa +\eta$, $m_{sc}(z)$ has the asymptotic expansion
\be\label{taylor}
m_{sc}=\mp 1+\frac12\sqrt{z^2-4}+O(|z^2-4|), \quad \mbox{near} \quad z \asymp \pm 2.  
\ee
\qed
\end{lemma}

We first prove \eqref{resultsolvesleq} for the case that $3\leq \eta\leq 10$. 
In this case, we can easily check that $g(z)=\sqrt{\kappa+\eta}$.  
Denote the difference between $G_{ii}$ and $m_{sc}$ by
\[
v_i = G_{ii}- m_{sc}, \qquad 1\leq i\leq N .
\]
By   the self consistent equation \eqref{selfequation1}, \eqref{sum} and   \eqref{defmsc}, we have
\be\label{temp1.47}
v_i=\frac{\sum_{i}\sigma^2_{ij}v_j+ \Upsilon_i}{(z+m_{sc}+\sum_{j}\sigma^2_{ij}
v_j+ \Upsilon_i)(z+m_{sc})}, \,\,\,1\leq i\leq N.
\ee
  For $\eta\ge 3$,    $|z+m_{sc}(z)|>2$ by \eqref{temp2.8}. Using
 $|G_{ii}|\leq \eta^{-1}$ and $|m_{sc}|\leq \eta^{-1}$,  we obtain 
 \be\label{vileq2}
 |v_i|\leq 2/\eta\leq 2/3, \qquad 1\leq i\leq N.
 \ee  From  the assumption \eqref{temp2.28} and \eqref{relkaeta}, 
we have   $\Upsilon=\max_i|\Upsilon_i|\ll1$ in this region.  Together with $|z+m_{sc}(z)|>2$
 and  \eqref{vileq2}, we obtain that   the absolute value of the r.h.s.  
of \eqref{temp1.47} is less than 
\be
\frac{\sup_i|v_i|}{|z+m_{sc}(z)|-\sup_i|v_i|}+O(\Upsilon).
\ee
Taking the absolute value of \eqref{temp1.47} and maximizing over $n$, we have  
\be\label{temp1.49}
\sup_n |v_n|\leq \frac{\sup_i|v_i|}{|z+m_{sc}|-\sup_i|v_i|}+O(\Upsilon).
\ee
The denominator satisfies $|z+m_{sc}(z)|-\sup_i|v_i|\geq 2-2/3=4/3$, therefore 
 we obtain  $\sup |v_i|=$ $\sup_i|G_{ii}-m_{sc}(z)|\leq O(\Upsilon)$, which shows
  \eqref{resultsolvesleq} for $3\leq \eta\leq 10$.

\bigskip 
\par Next, we prove \eqref{resultsolvesleq} in the case that $\eta\le 3$ with 
 $\eta$  satisfying  \eqref{relkaeta}
and under the condition \eqref{condmaxi}. 
 Define
\be\label{defG0}
 { m}=m(z)  :=\frac1N\sum_{i} G_{ii}\quad {\rm and }\quad u_i:=G_{ii}-   m.
\ee

Combining \eqref{temp2.28}, \eqref{relkaeta}, \eqref{condmaxi} with the
 fact that $g(z)\leq C$, we can see that 
\be\label{upclogn4}
\Upsilon\leq  (\log N)^{-4}g^2(z) \leq C(\log N)^{-4},\,\,\,\,
 |G_{ii}-m_{sc}(z)|\leq C(\log N)^{-2}.
\ee
 Together with \eqref{zmsc2}, we have 
$$|z+m_{sc}(z)|-|G_{ii}-m_{sc}(z)|-|\Upsilon|\geq C$$
for some $C>0$. Furthermore  \eqref{condmaxi} implies 
\be\label{mmsc}
|m(z)-m_{sc}|\leq 2(\log N)^{-2} g(z)
\ee
 thus  there exists $C>0$ such that 
\be\label{zmzgiimz}
|z+  m(z)  |-|G_{ii}- m(z)  |-|\Upsilon|\geq C.
\ee
Therefore, expanding  the self consistent equation \eqref{selfequation1}
 around $z+m(z)$, we obtain that   
\be\label{temp1.472}
0=G_{ii}+\frac{1}{z+\sum_j \sigma_{ij}^2 G_{jj}+\Upsilon_i}=G_{ii}+\frac{1}{z+  m(z) }
+ \Omega_i\ee
where $\Omega_i$ is defined by the second equality and it satisfies
\be
 \Omega_i = -\frac{\sum_{j}\sigma_{ij}^2 u_j}{(z+  m(z)  )^2}+
O(\|{\bf u}\|_\infty^2)+O(\Upsilon)
\ee
with error bounds uniform in  $i$. Here $\|{\bf u}\|_\infty=\max_i|u_i|$.
 Taking the average of the r.h.s of \eqref{temp1.472} with respect to $i$, we obtain that 
\be\label{temp4.7}
m(z)+\frac{1}{z+  m(z)  }=-\Omega
\ee
where 
\be
\Omega:= \frac{1}{N} \sum_i \Omega_i,
\ee
and it satisfies
\be\label{temp5.152}
|\Omega| \le  O(\|{\bf u}\|_\infty^2)+O(\Upsilon).
\ee
Here we used $\sum_{i}\sum_{j}\sigma_{ij}^2u_j= \sum_{j}u_j=0$.  The bound \eqref{condmaxi}
, \eqref{mmsc} and   $g(z)\leq\sqrt{\kappa+\eta}$ (from \eqref{defgz}) implies  that 
\be\label{vinfrou}
\|\bu\|_\infty\leq 4(\log N)^{-2}g(z)\leq C(\log N)^{-2}\sqrt{\kappa+\eta}.
\ee  
Together with  \eqref{upclogn4} and \eqref{temp5.152}, we obtain 
\be\nonumber
|\Omega |\le   C(\log N)^{-4} (\kappa+\eta).
\ee
To bound $m(z)$, we use the following lemma.

\begin{lemma}\label{newonmz}
Let $z=E+i\eta\in\C$,  $|z|\leq 10$ and let $\delta>0$ be
a sufficiently small constant. Let $t\in \C$ such that 
\be\label{condont}
|t|\leq \delta(\kappa+\eta).
\ee
 Suppose  there is a function $s_z(t)\in \C$  that solves
the equation
\be\label{szt}
s_z(t)+\frac{1}{z+s_z(t)}= t, \,\,\,
\ee 
with  $\im \,  s_z(t)>0$
and the estimate
\be\label{szmscz}
|s_z(t)-m_{sc}(z)|\leq \delta \sqrt{\kappa+\eta}
\ee 
holds.
 Then 
\be\label{resultszmscz}
|s_z(t)-m_{sc}(z)|\leq C\frac{|t|}{\sqrt{\kappa+\eta}}
\ee
for some constant $C>0$. 
\end{lemma}

{\it Proof of Lemma \ref{newonmz}.} 
It follows from \eqref{szt} that
\be\label{szt-t2}
s_z(t)=t+\frac{-z-t}{2}\pm\frac{\sqrt{(z+t)^2-4}}{2}.
\ee
We  denote by $s^1_z(t)$ and $s^2_z(t)$ the two solutions of this equation,
 which are continuous with respect to $t$ locally in the neighborhood \eqref{condont}.
 When $t=0$, one of them is equal to 
$m_{sc}(z)$, we choose $s^1_z(0)=m_{sc}(z)$.  
{F}rom \eqref{szt-t2}, we have 
\be\label{5.25}
 |s^1-s^2|=|( z+t)^2-4|^{1/2}.
\ee
 Then, for small enough $\delta$, if $|t|\le \delta (\kappa+\eta)$, then
 $|s^1-s^2|\ge \frac{1}{2} \min \{ |z-2|, |z+2|\}$ by using \eqref{5.25}
and that
 $\kappa+\eta \asymp \min \{ |z-2|, |z+2|\}$.
 We thus see that only one out of $s^1$ and $s^2$ can satisfy  \eqref{szmscz}.
 With the assumption that $s^1_z(0)= m_{sc}(z)$, it is $s^1$
 that satisfies \eqref{szmscz}. Then 
 \be
 s_z(t)-m_{sc}(z)=s^1_z(t)-s^1_z(0)
 \ee
 and \eqref{resultszmscz} follows from the fact that
 \be
 |\partial_t s_z(t)|\leq O\left(\frac{1}{ |\sqrt{(z+t)^2-4}|}\right)\leq O\left(\frac{1}{ \sqrt{\kappa+\eta}}\right),
 \ee
where for the second inequality, we used $|t|\leq \delta(\kappa+\eta)$. 
\qed

\bigskip

Using Lemma \ref{newonmz}, for $s_z(t)=m(z)$ and $t=-\Omega$,  we have
\be\label{G0-m}
|  m(z)  -m_{sc}(z)|\leq \frac{C| \Omega |}{\sqrt{\kappa+\eta}}\leq
 C(\log N)^{-2}\|{\bf u}\|_\infty+\frac{C\Upsilon}{\sqrt{\kappa+\eta}},
\ee
where in the second inequality we used 
\eqref{temp5.152} and \eqref{vinfrou}.
 Subtracting \eqref{temp5.152} from \eqref{temp1.472},  we have the equation for  $u_i$
\be\label{visum}
u_i = G_{ii} - m(z) = \frac{\sum_{j}\sigma_{ij}^2 u_j}{(z+  m(z) )^2} + \Omega
+ O(\|{\bf u}\|_\infty^2)+O(\Upsilon) =w_i+\frac{\sum_{j}\sigma_{ij}^2 u_j}{(z+m_{sc}(z))^2},
\ee
where $w_i$ is defined as $u_i-(\sum_{j}\sigma_{ij}^2 u_j)(z+m_{sc})^{-2}$. By
 \eqref{temp5.152}, it is bounded by 
\be\label{winfty2}
\|{\bf w}\|_\infty=O(\|{\bf u}\|_\infty^2)+O\left(\|{\bf u}\|_\infty 
|(z+m)^{-2}-(z+m_{sc})^{-2}|\right)+O(\Upsilon).
\ee 
Then, using \eqref{mmsc} and \eqref{zmsc2}, we obtain that 
\be
|(z+m)^{-2}-(z+m_{sc})^{-2}|\leq C|m(z)-m_{sc}(z)|.
\ee
Inserting this into \eqref{winfty2}, using the bounds on $\|{\bf u}\|_\infty$
 in \eqref{vinfrou} and \eqref{G0-m}, we have
\be\label{winfty}
\|{\bf w}\|_\infty=O(\|{\bf u}\|_\infty^2)+O(\Upsilon).
\ee  
\noindent 

{F}rom \eqref{templemmare} in Lemma \ref{msc}, whenever $|E|\geq 2$ or  $\kappa\leq \eta$,  in which case $g(z)\asymp \sqrt{\kappa+\eta}$, we have  
\be
|z+m_{sc}|^{-2}\leq 1-C\sqrt{\kappa+\eta},
\ee
for some $C>0$. Therefore  \eqref{visum} 
 imply  in this region that 
\be
\|{\bf u}\|_\infty\leq C(\kappa+\eta)^{-1/2}\|{\bf w}\|_\infty.
\ee
Using \eqref{winfty}, we get 
\be
\|{\bf u}\|_\infty\leq \frac{C}{\sqrt{\kappa+\eta}} \|{\bf u}\|_\infty^2+
\frac C{\sqrt{\kappa+\eta}}\Upsilon,
\ee 
and using $\|{\bf u}\|_\infty\ll \sqrt {\kappa+\eta}$ from   \eqref{vinfrou}, we conclude that 
\be\label{uinftyu}
\|{\bf u}\|_\infty\leq O\Big(\frac{\Upsilon}{\sqrt{\kappa+\eta}}\Big).
\ee
 Combining this with the bound on $  m-m_{sc} $ \eqref{G0-m} and $G_{ii}-m_{sc}=
m-m_{sc}+u_i$, we obtain \eqref{resultsolvesleq}.

\bigskip
Finally, we consider the main interesting regime: $ |E|\leq 2$ and   $\kappa\geq \eta$. 
 We claim that the following inequality about $m_{sc}(z)$ holds. 

\begin{lemma}\label{lm:arit} 
Let $1>\delta_->0$ be a given constant. Then there exist small real numbers 
$\tau\geq 0$ and $c_1>0$, depending only on $\delta_-$, such that
 we have 
\be
    \max_{x\in [-1+\delta_-,1-\delta_+]}\left\{\Big| \tau + x\, m_{sc}^2\Big|^2 
\right\}\le \left(1-c_1\,\widehat g(z)\right)(1+\tau)^2
\label{al}
\ee
with 
\be
\widehat g(z)=\max\{\delta_+,\,\, |1- {\rm Re} \, m_{sc}^{2}(z)| \}
\ee
for any positive number  $\delta_+$ such that $-1+\delta_-\leq1-\delta_+$.
\end{lemma}

We postpone the proof of this lemma to the end of this subsection
and we first complete the main argument. 
Recall that $B= \{\sigma_{ij}^2\}_{i,j=1}^N$ is the
matrix of variances which is symmetric. We also recall
$\delta_\pm$ from \eqref{de-de+} and we will apply Lemma \ref{lm:arit}
with these $\delta_-$ and $\delta_+$.  Fix $z$,  set  $\zeta:= m_{sc}^2(z) = (m_{sc}(z) + z)^{-2}$ and rewrite \eqref{visum} as
\be\label{bv}
{\bf u}=(I-\zeta B)^{-1}{\bf w}= \frac{1}{1+\tau}  \Big[ I- \frac{\zeta B +\tau}{1+\tau}\Big]^{-1}{\bf w}
\ee
with $\tau$ given in Lemma \ref{lm:arit}.
Define $Q:= I - |{\bf e}\rangle\langle {\bf e}|$ to be the projection onto
the orthogonal complement of the normalized eigenvector
${\bf e} = N^{-1/2}(1,1,\ldots, 1)$ belonging
to the simple eigenvalue 1 of $B$. Note that $B$ and $Q$ commute
and that the spectrum of $BQ$ lies in $[-1+\delta_-, 1-\delta_+]$.   Denote by $\| A\|$  the usual $\ell^2\to\ell^2$ norm  of a matrix $A$. 
Since 
$$
   \Big\| \frac{\zeta B +\tau}{1+\tau}Q \Big\| \le \sup_{x\in[-1+\delta_-,1-\delta_+]}
  \Big| \frac{\zeta x+\tau}{1+\tau} \Big|
  \le (1-c_1\,\,\widehat g(z))^{1/2}<1
$$
by the Lemma \ref{lm:arit} and  $\bw\perp  (1, . . . , 1) $,  the Neumann expansion of  \eqref{bv} converges on span $((1, . . . , 1))^\perp $ and
\be
    \bu =(I-\zeta B)^{-1}{\bf w}= \frac{1}{1+\tau}\sum_{n=0}^\infty 
\Big( \frac{\zeta B +\tau}{1+\tau} \Big)^n\bw.
\label{neu}
\ee

We will compute the $\ell^\infty\to\ell^\infty$ norm 
 of this matrix.
First note that
\be
   \Big\|  \frac{\zeta B +\tau}{1+\tau} \Big\|_{\infty\to\infty} = \max_i \sum_j 
   \Big| \Big( \frac{\zeta B +\tau}{1+\tau} \Big)_{ij} \Big| 
   \le \frac{1}{1+\tau} \max_i \sum_j |\zeta B_{ij} + \tau \delta_{ij}|
  \le \frac{ |\zeta |+\tau}{1+\tau}\le 1,
\label{triv}
\ee
since $|\zeta| =  |m_{sc}|^{2}\le 1$  and $\sum_j |B_{ij}| =\sum_j B_{ij}=\sum_{j}\sigma_{ij}^2=1$.
 Then we have          
$$
    \Big\| \Big( \frac{\zeta B +\tau}{1+\tau} \Big)^n {\bf u}\Big \| = \Big\| \Big( \frac{\zeta B +\tau}{1+\tau} \Big)^n Q{\bf u}\Big \| \le 
  \sup_{x\in[-1+\delta_-, 1-\delta_+]}  \Big| \frac{\zeta x+\tau}{1+\tau} \Big|^n {\| \bf u\|}
  \le (1-c_1\,\,\widehat g(z))^{n/2}{\| \bf u\|}
$$
by Lemma \ref{lm:arit}.  Since  for any $N\times N$ matrix we have
$$
    \| A \|_{\infty\to\infty} \le \sqrt{N}\|A\|,
$$
we obtain
\be\label{nontr}
    \Big\| \Big( \frac{\zeta B +\tau}{1+\tau} \Big)^nQ \Big \|_{\infty\to\infty} 
  \le \sqrt{N}(1-c_1\, \widehat g(z))^{n/2}.
\ee
Thus, estimating the first $n\le n_0:= (\log N)(c_1\widehat g(z))^{-1}$ terms in
  \eqref{neu} by \eqref{triv}, and the rest by \eqref{nontr}, we get
$$
   \| \bu\|_\infty \le \Bigg( 
 \frac{\log N}{c_1\widehat g(z)}+ \sum_{n =n_0}^\infty \sqrt{N}(1-c_1\,\widehat g(z))^{n/2}
 \Bigg)\|\bw\|_\infty
  \le C\frac{\log N}{\widehat g(z)}\|\bw\|_\infty.
$$
Using the bound  \eqref{winfty}  on $\|{\bf w}\|_\infty$ and the bound  \eqref{vinfrou}
  on $\|{\bf u}\|_\infty$, we have 
\be
 \| \bu\|_\infty\leq (\log N)^{-1} \| \bu\|_\infty+C\frac{\log N}{\widehat g(z)}\Upsilon
\ee 
which implies
\be
\|\bu \|_\infty\leq C\frac{\log N}{ \widehat g(z)}\Upsilon
\ee
for some $C>0$
. Combining this with  \eqref{G0-m}, we find
$$
|G_{ii}-m_{sc}|\leq |m(z)-m_{sc}(z)|+|u_i|\leq C\left[\|\bu\|_\infty+\frac{\Upsilon}{\sqrt{\kappa+\eta}}\right]\leq C\left[\frac{\log N}{\hat g(z)}+\frac{1}{\sqrt{\kappa+\eta}}\right]\Upsilon
$$ 
which implies  \eqref{resultsolvesleq}, since $g(z)=\min\{\sqrt{\kappa+\eta}, \widehat g(z)\}$.
\qed
\bigskip

\bigskip
\noindent 
{\it Proof of Lemma \ref{lm:arit}.} 
First,  if $\widehat g(z)=\delta_+$, then we choose $\tau=0$. 
 With $|m_{sc}|\le 1$ in \eqref{zmsc2}, one can see that \eqref{al} holds. 
\par In the case of $\widehat g(z)= |1-{\rm Re}\, m_{sc}^2|$,  we have $\widehat g\leq 2$ by using $|m_{sc}|\le 1$. We  choose $\tau= \delta_-/10$, then
\be\label{maxxx}
    \max_{x\in [-1+\delta_-, 1-\delta_+]}\Big| \tau + x m_{sc}^2 \Big|^2 \leq\max\left\{
    \left| \frac {\delta_-} {10}+m_{sc}^2 \right|^2, 
\left|  \frac {\delta_-}  {10} -(1-{\delta_-} ) m_{sc}^2 \right|^2 \right\},
\ee
and \be\label{leftfracdelta}
\left|  \frac {\delta_-}  {10} -(1-{\delta_-} ) m_{sc}^2 \right|^2  
\leq \left|1-\frac{9\delta_-}{10}\right|^2\leq (1-\tau \widehat g(z)) (1+{\delta_-} /10)^2\,.
\ee 
For the other term in r.h.s. of \eqref{maxxx}, we have 
\be\label{temp5.36}
 \left| \frac {\delta_-}  {10}+m_{sc}^2 \right|^2= |m_{sc}|^4+{0.2{\delta_-} } {\rm Re}\,(m_{sc}^{2}) 
\, +(\delta_-/10)^2\,.
\ee
With $|m_{sc}|\le 1$ in \eqref{zmsc2} and $\widehat g(z)= |1-{\rm Re}\, (m_{sc}^{2})|$ 
in this case, \eqref{temp5.36} is bounded as
\be\label{lastmz}
 \left| \frac {\delta_-}  {10}+m_{sc}^2 \right|^2\leq \left(1+ 
\frac {\delta_-}  {10}\right)^2-0.2{\delta_-} 
 \widehat g(z)\leq \left(1+ \frac {\delta_-}  {10}\right)^2(1-C\widehat g(z))
\ee
for some $C$ depending on $\delta_-$. At last, we  complete the proof by combining 
\eqref{lastmz} and \eqref{leftfracdelta}. 
\qed

\section{Proof of the universality of local statistics} \label{sec:sk}

We now outline the main steps to prove Theorem \ref{mainsk}. 
\bigskip

\noindent
{\it Step 1. Local relaxation flow.} 
 Following \cite{ESYY}, we first prove that the local eigenvalue statistics
 of Dyson Brownian motion (DBM)
at a fixed time $t$ are the same as those of GUE if $t\asymp N^{-\e_0}$ for some $\e_0$.
  The DBM is generated by the flow
\be\label{matrixdbm}
H_t = e^{-t/2} H_0 + (1-e^{-t})^{1/2}\, V,
\ee
where $ H_0$ is the initial matrix  and $V$ is an independent  GUE matrix
whose matrix elements are centered Gaussian random variables with variance $1/N$.
Strictly speaking, for each matrix element we have used
the Ornstein-Uhlenbeck (OU) process on $\bC$ instead of the
Brownian motion which was used in the original definition
of DBM in \cite{ESYY}. It is easy to check that the
eigenvalues of $H_t$ follow a process, very similar
to the original DBM in \cite{ESYY}, but  with a drift.
With a slight abuse of terminology, we will still
call this process DBM.
More precisely, let 
\be\label{H}
\mu=\mu_N(\rd{\bf x})=
\frac{e^{-\cH({\bf x})}}{Z_\beta}\rd{\bf x},\qquad \cH({\bf x}) =
N \left [ \beta \sum_{i=1}^N \frac{x_{i}^{2}}{4} -  \frac{\beta}{N} \sum_{i< j}
\log |x_{j} - x_{i}| \right ]
\ee
($\beta=2$ for  GUE) be the probability measure of the eigenvalues of the general $\beta$ ensemble,
$\beta\ge 1$ (in this section, we often use the notation $x_j$ for 
the eigenvalues to follow the notations of \cite{ESYY}).
In this paper we consider the $\beta=2$  case for
simplicity, but we stress that our proof applies to  the case of symmetric matrices
as well. Denote the distribution of 
the eigenvalues  at  time $t$
by $f_t ({\bf x})\mu(\rd {\bf x})$.
Then $f_t$ satisfies
\be\label{dy}
\partial_{t} f_t =  \cL f_t.
\ee
where (see (2.2) in \cite{ESYY})
\be
\cL=   \sum_{i=1}^N \frac{1}{2N}\partial_{i}^{2}  +\sum_{i=1}^N
\Bigg(- \frac{\beta}{4} x_{i} +  \frac{\beta}{2N}\sum_{j\ne i}
\frac{1}{x_i - x_j}\Bigg) \partial_{i}.
\label{L}
\ee

\begin{theorem}\label{gengenJT}
Suppose that the probability law for the initial matrix $H_0$ satisfies the
  assumptions of Theorem \ref{mainsk}. 
 Then there exists $\e_0>0$ such that for any  
\be
t\geq N^{-\e_0},
\ee
the probability law for the eigenvalues of $H_t$ satisfies  \eqref{matrixthm}, i.e., 
for any $k\ge 1$ and for any compactly supported continuous test function
$O:\bR^k\to \bR$, we have
\be
\begin{split}
 \lim_{\kappa\to0}\lim_{N\to \infty} \frac{1}{2\kappa}
\int_{E-\kappa}^{E+\kappa}\rd E' \int_{\R^k} & \rd\alpha_1 
\ldots \rd\alpha_k \; O(\alpha_1,\ldots,\alpha_k)  \non \\
&\times
 \frac{1}{\varrho_{sc}(E)^k} \Big ( p_{N}^{(k)}  - p_{GU\! E, N} ^{(k)} \Big )
  \Big (E'+\frac{\alpha_1}{N\varrho_{sc}(E)}, 
\ldots, E'+\frac{\alpha_k}{N \varrho_{sc}(E)}\Big) =0,
\non  
\end{split}
\ee
where  $p_{GU\! E, N} ^{(k)}$ is the $k$-point correlation
function of the GUE ensemble. 
\end{theorem}

\bigskip

\noindent 
{\it Proof of Theorem \ref{gengenJT}.}
We first recall the following general theorem concerning the Dyson Brownian motion from \cite{ESYY}
that asserts that under four general assumptions, the local eigenvalue statistics
of the time evolved matrix $H_t$ coincide with GUE.
The first assumption (called Assumption I in \cite{ESYY}) is a convexity
bound on $\cH$ which is automatically satisfied in our case and we 
 only have to verify  the following three assumptions. 

\medskip

{\bf Assumption II.} There exists a continuous, compactly supported
density function $\varrho(x)\ge 0$, $\int_\bR \varrho =1$, on the
real line, independent of $N$, such that for any fixed $a,b\in \bR$
\be
\lim_{N\to\infty}   \sup_{t\ge 0}  \Bigg|
\int \frac{1}{N}\sum_{j=1}^N {\bf 1} ( x_j \in [a, b]) f_t(\bx)\rd\mu(\bx)
- \int_a^b \varrho(x) \rd x \Bigg| =0.
\label{assum1}
\ee

\bigskip
For the next assumption, we introduce a notation.
Let $\gamma_j =\gamma_{j,N}$ denote the location of the $j$-th point
under the limiting density, i.e., $\gamma_j$ is defined by
\be\label{def:gamma}
 N \int_{-\infty}^{\gamma_j} \varrho(x) \rd x = j, \qquad 1\leq j\le N, \quad
\gamma_j\in \mbox{supp} \varrho.
\ee
We will call $\gamma_j$ the {\it classical location} of the $j$-th point.

\bigskip
{\bf Assumption III.} There exists an $\e>0$ such that
\be
 \sup_{t\ge 0} \int  \frac{1}{N}\sum_{j=1}^N(x_j-\gamma_j)^2
 f_t(\rd \bx)\mu(\rd \bx) \le CN^{-1-2\e}
\label{assum3}
\ee
with a constant $C$ uniformly in $N$.

\bigskip

The final assumption is an upper bound on the local density. 
For any $I\in \R$, let
$$
    \cN_I: = \sum_{i=1}^N {\bf 1}( x_i \in I)
$$
denote the number of points in $I$.

\bigskip

{\bf Assumption IV.} For any compact subinterval $I_0\subset \{ E\;: \; \varrho(E)>0\}$ independent of $N$, 
and for  any $\delta>0$,  $\sigma>0$ and $r>0$, 
there are constants $c$  depending on $I_0$,
$\delta$, $\sigma$ and $r$ such that for any interval $I\subset I_0$ with
$|I|\ge N^{-1+\sigma}$, we have
\be
   \sup_{\tau \ge N^{-2\e+\delta}}
    \int {\bf 1}\big\{ \cN_I \ge K N |I| \big\}f_\tau \rd\mu
\le N^{-c\log\log N},  \quad K=N^r
\label{ass4}
\ee
where $\e$ is the exponent from Assumption III.
\bigskip

\begin{theorem}\label{thm:main} \cite[Theorem 2.1]{ESYY}
Let  $\e>0$ be the exponent from Assumption III. 
 Suppose that there is a time $\tau < N^{-2 \e}$ 
such that the following entropy bound holds
\be
S_{\mu}(f_\tau):=\int f_\tau \log f_\tau \rd\mu \le CN^m
\label{entro}
\ee
for  some fixed exponent $m$. Suppose that  the 
Assumptions II, III and IV  hold for the solution $f_t$ of the forward equation
\eqref{dy} for all time $ t \ge \tau$.  Let $E\in \bR$ be a point
where $\varrho(E)>0$. Then 
for any $k\ge 1$ and for any compactly supported continuous test function
$O:\bR^k\to \bR$, we have
\be
\begin{split}
\lim_{b\to 0}\lim_{N\to \infty} \sup_{t\ge N^{-2\e+\delta}}  \;
\frac{1}{2 b }\int_{E-b}^{E+b}\rd E'
\int_{\R^k} &  \rd\alpha_1
\ldots \rd\alpha_k \; O(\alpha_1,\ldots,\alpha_k) \\
&\times \frac{1}{\varrho(E)^k} \Big ( p_{t,N}^{(k)}  - p_{\mu, N} ^{(k)} \Big )
\Big (E'+\frac{\alpha_1}{N\varrho(E)},
\ldots, E'+\frac{\alpha_k}{ N\varrho(E)}\Big) =0 \, .
\label{abstrthm}
\end{split}
\ee
\end{theorem}

Theorem \ref{thm:main} was exactly  Theorem 2.1 of \cite{ESYY} except
 that the assumption \eqref{entro} on the entropy 
in \cite{ESYY} was stated for the initial probability density $f_0$. 
 Clearly, we can start the flow \eqref{L} 
from a fixed time $\tau \ll 
N^{-2\e+\delta}$ since the statement of Theorem \ref{thm:main} concerns 
only the time $t \ge N^{-2\e+\delta}$. 
In the case that the flow  \eqref{L} is generated from  the matrix evolution
 \eqref{matrixdbm},  the  entropy assumption
\eqref{entro} is satisfied automatically. To see this,  
let $\nu_t^{ij}$ denote the probability measure of the $ij$-th 
element of the matrix $H_t$, $i\le j$,
and $ \bar \nu_t$  the  probability measure of the matrix $H_t$. 
Let $\bar\mu$ denote the probability measure 
of the GUE and $\mu^{ij}$ the probability measure of 
its $ij$-th element which is a Gaussian measure 
with mean zero and variance $1/N$. Since the dynamics of  matrix 
elements are independent (subject to the Hermitian condition), we have the identity 
\be
\int  \log \left ( \frac {\rd \bar \nu_t} { \rd \bar \mu}  \right ) \rd \bar \nu_t 
= \sum_{ij} \int  \log \left ( \frac {\rd  \nu^{ij}_t} 
{ \rd \mu^{ij} }  \right ) \rd \nu^{ij}_t.
\ee
The process $t \to \nu^{ij}_t$ is an Ornstein-Uhlenbeck  process and each entropy term 
on the right hand side of the last equation is bounded by $ C N$ provided that $t \ge 1/N$
and $ \nu^{ij}_0$ has a  subexponential decay. It is easy to check
from the explicit OU kernel. Since the entropy of the marginal distribution 
on the eigenvalues is bounded by the entropy of the total  measure
 on the matrix, we have proved that 
\be
\int f_{1/N}\log f_{1/N} \rd\mu\leq CN^3,
\ee
and this verifies \eqref{entro}. 
Therefore, in order to  apply Theorem \ref{thm:main}, we only have to 
verify the Assumptions II, III and IV. 
Clearly, Assumption II follows from Theorem \ref{mainls}
(note that in the case of generalized Wigner
matrix, $M\asymp N$ and $g(z)\asymp\sqrt{\kappa+\eta}$).
Assumption IV also follows from Theorem \ref{mainls} by
noting that $\cN_I \le C\, \im  (E+i\eta)$
if $I$ is an interval of length $\eta$ about $E$.
We also note that Assumption IV in \cite{ESYY} was stated
in a slightly stronger form, requiring a large
deviation bound \eqref{ass4} for all $K\ge 1$, but
inspecting the proof of Theorem 2.1 of \cite{ESYY}
reveals that Assumption IV is used only for 
$K$ larger than some positive power of $N$
and smaller than $N$ (the main observation is
that the upper limit of the summation in (7.16) of \cite{ESYY}
is effectively $N$ and not $\infty$).

Having verified all other assumptions, it remains to prove
\eqref{assum3}, which we state as the next theorem.

\begin{theorem}\label{lagaN-1}
Suppose $H$  satisfies the  assumptions of Theorem \ref{mainsk},
in particular, it is a
 generalized  Wigner matrix with positive constants $C_{inf}$, $C_{sup}$ in
\eqref{VV}. Let $\wt\nu_{ij}(x)\rd x := \sigma_{ij}\nu_{ij}(\sigma_{ij}x)\rd x$
be the rescaling of the distributions $\nu_{ij}$  of the matrix elements and suppose that
they satisfy the logarithmic Sobolev inequality (LSI) with a constant $C_S$
independent of $N, i, j$, i.e.,
\be
   \int u\log u \rd\wt\nu_{ij}\le C_{S} \int |\nabla \sqrt{u}|^2\rd\wt\nu_{ij}
\label{lsi}
\ee
holds
for any smooth probability density $u$, $\int u \rd\wt \nu_{ij}=1$.
Denote
$ \lambda_i$ the $i$-th eigenvalue of $H$ in increasing order,
 $\lambda_1\leq\lambda_2\le\ldots\leq \lambda_N$.
Then there exists $\eps>0$  depending on $\alpha, \beta$ in \eqref{subexp} but
 independent of $C_{inf}$, $C_{sup}$ and $C_S$ such that 
\be
	\frac1N\sum_{i=1}^N\E(\lambda_i-\gamma_i)^2\leq CN^{-1-2\eps},
\label{lag}
\ee 
if $N$ is sufficiently large (depending on $C_{inf}$, $C_{sup}$, $C_S$, $\al$ and $\beta$).
\end{theorem}
The proof of Theorem \ref{lagaN-1} will be given in Section \ref{sectlem}.
It is easy to check that if an initial matrix $H=H_0$ satisfies
the conditions of Theorem \ref{lagaN-1}, then its 
evolution $H_t$ under the Ornstein-Uhlenbeck flow will also
{  satisfy these conditions with  constants changed at most by a factor two. 
The main condition to check is  that the logarithmic Sobolev inequality  \eqref{lsi} holds for $0 \le t \ll 1$.  But this was proved in 
the argument following Lemma 5.3 of \cite{ESY4} using an estimate on the logarithmic Sobolev constant 
 for convolution of two measures, i.e., Lemma B.1 of \cite{ESY4}. }
Therefore Theorem  \ref{lagaN-1} guarantees  \eqref{lag}
for all positive times $t>0$  and this proves
 Assumption III provided that the initial distribution satisfies 
 the LSI \eqref{lsi}.  We have thus proved   Theorem \ref{mainsk}  for matrix ensembles of the form 
\be\label{formm} 
h_{ij}=e^{-t/2} \widehat h_{ij}+(1-e^{-t})^{1/2} N^{-1/2}\xi^G_{ij},  \quad t\ge N^{-2\e+\delta}
\ee
where $\xi^G_{ij}$ are i.i.d. complex random variables with  
Gaussian distribution with mean $0$
and variance $1$, and $\widehat h_{ij}$'s are independent random variables such that the rescaled  
variables $\widehat \zeta_{ij} = \widehat h_{ij}/\sigma_{ij}$ satisfy th
 LSI assumption \eqref{lsi}.  In \eqref{formm} $\delta>0$ is arbitrary and
 $\e$ is fixed in Theorem \ref{lagaN-1}.
In particular,  with the choice $\delta=\e$ and $t \asymp  N^{-\e}$,
we have proved   Theorem \ref{mainsk} for matrix ensembles 
 $h_{ij}= \sigma_{ij}\zeta_{ij}$ if $\zeta_{ij}$ is of the form 
\be\label{3.5} 
\zeta_{ij}= (1-\gamma)^{1/2} \widehat \zeta_{ij} + \gamma^{1/2} \xi^G_{ij}, 
 \quad \gamma \asymp N^{-\e}, \qquad \mbox{distribution of $ \wh\zeta_{ij}$ satisfies \eqref{lsi}.}
\ee

\bigskip 
\noindent
{\it Step 2. Eigenvalue  correlation function comparison theorem}.

The next step is to prove that the correlation functions 
of eigenvalues for two matrix ensembles  are identical up to 
scale $1/N$ provided that the first four moments 
of all  matrix elements  of these two ensembles are almost identical.
This theorem is a corollary of  Theorem \ref{comparison} and we state it 
as the following  correlation function comparison theorem.  The proof 
will be given in Section \ref{sec:proofcomp}.
Note that the assumption \eqref{basic} in Theorem \ref{comparison}
is satisfied by Theorem \ref{thm:detailed};
in case of generalized Wigner matrix we have $g(z) = \sqrt{\kappa+\eta}$
and $M\asymp N$,
so in the regime where $|E|$ is separated away from 2, we have
from \eqref{mainlsresult}, that $G_{ii}(z)$ is uniformly bounded 
(modulo logarithmic factors).

\begin{theorem} \label{com} Suppose the assumptions of Theorem \ref{comparison} hold. 
Let $p_{v, N}^{(k)}$ and $p_{w, N}^{(k)}$
be the  $k-$point functions of the eigenvalues w.r.t. the probability law of the matrix $H^{(v)}$
and $H^{(w)}$, respectively. 
Then for any $|E| < 2$,  any
$k\ge 1$ and  any compactly supported continuous test function
$O:\bR^k\to \bR$ we have  
\be \label{6.3}
\int_{\R^k}  \rd\alpha_1 
\ldots \rd\alpha_k \; O(\alpha_1,\ldots,\alpha_k) 
   \Big ( p_{v, N}^{(k)}  - p_{w, N} ^{(k)} \Big )
  \Big (E+\frac{\alpha_1}{N}, 
\ldots, E+\frac{\alpha_k}{N }\Big) =0.
\ee
\end{theorem}

\bigskip 
\noindent
{\it Step 3. Approximation of a measure by Ornstein-Uhlenbeck process for small time}.
\bigskip 

Summarizing,   we have proved  Theorem \ref{mainsk} in Step 1
 for matrix ensembles
whose probability distributions of the  normalized  matrix elements $
\zeta_{ij}$ are
 of the form \eqref {3.5}.
Using the  Green's function comparison theorem, i.e. Theorem \ref{com}, we
  extended the class of distributions
to all random variables whose first four moments can almost be matched
 (more precisely,   match  the first three moments  and almost match the
fourth moments in the sense of   \eqref{4match}) by  random variables  in
the class
\eqref{3.5}.  In order to complete the proof of  Theorem \ref{mainsk}, 
 it remains to prove that for 
all measures in the class given by the assumptions of Theorem \ref{mainsk}, i.e., 
 measures satisfying  the subexponential decay
condition, the uniformly bounded-variance condition \eqref{VV} 
 and the moment restriction  \eqref{no3no4} for the real and imaginary parts,
we can find   random variables  in the class
\eqref{3.5} to almost match the first four moments.
Since the real and imaginary parts are i.i.d., it is sufficient
to match them individually, i.e., we can work with real random variables
normalized to variance one.
This is the content of the following Lemma \ref{fmam}.
Notice that the uniformity in the conditions \eqref{no3no4}
and \eqref{subexp} guarantees that the bounds \eqref{m4m3cc}
hold with  uniform constants $C_1, C_2$. This implies the uniformity 
of the  LSI
constants, needed in Theorem \ref{lagaN-1}, for  the random variables 
constructed 
in Lemma \ref{fmam}. 
The proof of this Lemma will be given in Appendix \ref{sec:LSI}. 
We have thus proved Theorem \ref{mainsk}. \qed

\begin{lemma}\label{fmam} Let  $m_3$ and $m_4$ be two real numbers  such that 
\be\label{m4m3cc}
m_4-m_3^2-1\ge C_1,\,\,\, m_4\leq C_2
\ee
for some positive constants $C_1$  and $C_2$. 
Then for any sufficient small $\gamma>0$ (depending on $C_1$ and $C_2$),
 there exists a real random variable $\xi_\gamma$ whose distribution 
satisfies LSI and 
the first 4 moments of 
\be\label{defxi'}
\xi'=(1-\gamma)^{1/2}\xi_\gamma+\gamma^{1/2}\xi^G
\ee 
are $0$, $1$, $m_3(\xi')=m_3$ and $m_4(\xi')$, and 
\be\label{m4m4}
|m_4(\xi')-m_4|\leq C\gamma
\ee 
for some $C$ depending on $C_1$ and $C_2$, where $\xi^G$ is
real Gaussian random variable 
 with mean $0$ and variance $1$, independent of $\xi_\gamma$.
The LSI constant of $\xi_\gamma$ (and thus $\xi'$)   is bounded
from above by a function of
 $C_1$ and $C_2$.

\end{lemma}

\section{Proof of Theorem \ref{lagaN-1}}\label{sectlem}

Theorem \ref{lagaN-1} states that the eigenvalues are at a distance $N^{-1/2-\e}$
from their classical locations in a quadratic average sense. We will deduce this
conclusion from the information on the closeness of the local density 
to the semicircle law
. We note the constants appearing in this section may also depend on $\al$ and $\beta$ in \eqref{subexp}, but we will not mention the dependence in the proof.

First we reformulate a result, which  we have proved in \cite{ESYY}, 
in a somewhat more general setup.  It states that  random points, $\lambda_j$,
are close to a fixed set of locations, $\gamma_j$, if the local
fluctuation is controlled, if the averaged counting function 
is close to the counting function of the $\gamma_j$'s in $L^1$-sense
and if some tightness holds.  For simplicity, the result is stated for the case when
$\gamma_j$'s are the classical locations given by the semicircle law $\varrho=\varrho_{sc}$,
\eqref{def:gamma}, but the statement (and its proof) holds for any
density function with support being a compact interval and
with square root singularity at the edges. In particular, we applied this result in
\cite{ESYY} for the Marchenko-Pastur (MP)
distribution instead of the semicircle law. The counting function
of $\gamma_j$ can be replaced by its continuous version, i.e., by the distribution
function of the semicircle law which defined by
\be
   n_{sc}(E):= \int_{-\infty}^E \varrho_{sc}(x)\rd x.
\label{nsc}\ee

\begin{lemma}\label{N-1ESYY}
Let $\lambda_1\leq\lambda_2\le \ldots\leq \lambda_N$ be
 an  ordered collection of random points in $\R$. 
Denote the   averaged counting function of $\lambda_j$'s
\be\label{defnlambda}
n^\lambda(E)=\frac1N\E \#[{\lambda_j\leq E}].
\ee
   Suppose  the following four assumptions hold. 
\begin{enumerate}
	\item {[Tightness at the edge]}  There exist $m<7$ and  $\eps> 0$ such that
\be 
   n^\lambda(-2-N^{-1/m})\leq Ce^{-N^{\eps}}\quad \mbox{and} \quad n^\lambda(2+N^{-1/m})\geq
 1-Ce^{-N^{\eps}}
\label{neartight}
\ee
and for any $K\ge 3$,
\be
n^\lambda(K)\geq
 1-e^{-N^\e\log K} \quad \mbox{and} \quad n^\lambda(-K)\leq e^{-N^\e\log K}.
\label{largetight}
\ee

\item {[$L^1$-closeness of the counting functions]}
 \be\label{fakeresN-1}
 \int_{-\infty}^{\infty}\left|n^\lambda(E)-n_{sc}(E)\right|\rd E\leq CN^{-6/7}.
 \ee

\item {[Fluctuation of moving averages]}  For any small $\delta>0$ there is a constant $C$
such that
for any $j, K\in \mathbb N$ with $j+K\leq N+1$, 
the local averages
$\lambda_{j,K}:=K^{-1}\sum_{i=0}^{K-1}\lambda_{j+i}$ satisfy
\be\label{fakerangexjk}
\mathbb P\left(|\lambda_{j,K}-\E\, \lambda_{j,K}|\geq N^{-1/2+\delta}K^{-1/2}\right)
\leq Ce^{-N^{\delta/2}}.
\ee
\item  {[Positivity of the bulk density]} There exists a small enough $\delta>0$ such that:
 for any interval $I$ with $|I|=N^{-5/8}$ and $I\subset[-2+N^{-\delta}, 2-N^{-\delta}]$,
 the number of the $\lambda$'s  in $I$ is bounded from below as follows  
\be\label{lIllN}
\P\left(\#\{\lambda_j\in I\}\geq N^{-\delta}N|I|\right)\geq 1-CN^{c\log\log N}.
\ee
\end{enumerate}
Then there exists $\eps>0$ (independent of the constants in these four assumptions) such that 
\be\label{alg}
\frac1N	\sum_{i=1}^N\E(\lambda_i-\gamma_i)^2\leq CN^{-1-\eps}
\ee 
when $N$ is large enough (depending on  the constants in these four assumptions).
\end{lemma}
\bigskip

 {\it Proof of Lemma \ref{N-1ESYY}.} In Theorem 9.1 of \cite{ESYY} we have proved
 the analogous result on the singular values of the covariance matrix,
where the role of the semicircle law
was played by the MP law and the spectral edges, $\pm 2$,
were replaced by  $\lambda_\pm$, the
two edges of the support of the MP distribution.
In that paper we first proved the analogues of these four assumptions,
then we presented the proof of \eqref{alg} via a general argument that
 used only these assumptions.
Inspecting the proofs of Lemma  9.5, 9.6 and 9.7 in \cite{ESYY}, leading
to \eqref{alg}, we observe that only equations (9.6),
 (9.8), (9.9) and (9.13) from \cite{ESYY} were used, in addition
to the lower bound on the density of the points in the scale $N^{-5/8}$, which is used below (9.51) of \cite{ESYY}.
The lower bound on the density is granted by the last assumption \eqref{lIllN}
(even with a better control on the probability than we required in \cite{ESYY}).
Repeating  the argument from \cite{ESYY}, for the proof of Lemma \ref{N-1ESYY}
it is sufficient to check that the first three assumptions in   Lemma \ref{N-1ESYY}
 imply 
 equations (9.6),
 (9.8), (9.9) and (9.13) in \cite{ESYY}.
We now explain how to obtain these necessary  bounds
 from our assumptions.

 The first  condition \eqref{neartight} corresponds to the input for Lemma 9.2  in
\cite{ESYY}, in particular, the analogue of (9.6) of \cite{ESYY},
$$
    -2 - N^{-1/m} \le \E \lambda_j \le 2 + N^{1/m},
$$
follows immediately from \eqref{neartight} and \eqref{largetight}.
 We note that  (9.6) in \cite{ESYY}
contains a threshold  $N^{-1/5}$ but actually in the proof we only needed
it to be much less than $N^{-1/7}$ (see (9.36)--(9.37) of \cite{ESYY} for
the application of (9.6)).

The second condition \eqref{fakeresN-1} corresponds to Eq. (9.8) in \cite{ESYY}.
 As we showed in the  proof of Lemma 9.3 of \cite{ESYY},
Eq. (9.9) directly follows from (9.8). Here the analogous bound
$$
   \sup_E \big| n^\lambda(E)- n_{sc}(E)\big|\le CN^{3/7}
$$
follows directly from \eqref{fakeresN-1} in the same way.

Finally, the third condition \eqref{fakerangexjk} is exactly the same as (9.13)  in \cite{ESYY}.
Simply repeating now the proof of Theorem 9.1 from 
\cite{ESYY}, we proved  Lemma \ref{N-1ESYY}.
\qed 

\bigskip

Theorem \ref{lagaN-1} will now follow from  Lemma \ref{N-1ESYY}  if
we prove that  the four conditions in  the Lemma \ref{N-1ESYY} hold in the case of 
generalized Wigner matrices \eqref{VV}.
The last condition \eqref{lIllN}
 follows from the local semicircle law (Theorem \ref{mainls}) and from the
fact that $\varrho_{sc}(x)\ge c\sqrt{\kappa}$ for $x\in (-2+\kappa, 2-\kappa)$.
Here we list the first three conditions as three separate lemmas that will be proven
 in the next three subsections.  This will complete the proof of Theorem \ref{lagaN-1}. \qed

\begin{lemma}\label{N-1/6}

(1) Let $H$ be a generalized Wigner matrix with 
subexponential decay, in fact it is
sufficient to assume that \eqref{subexp} and the upper bound $C_{sup}<\infty$
in \eqref{VV}  hold.
Define $n^\lambda(E)$ as in \eqref{defnlambda}. Then 
\be\label{resnlambda1}
n^\lambda(-2-N^{-1/6+\e})\leq Ce^{-N^{\eps'}}\,\,\,{\rm and }
\,\,\,n^\lambda(2+N^{-1/6+\e})\geq 1-Ce^{-N^{\eps'}}
\ee
for  any small $\eps>0$ with an $\e'>0$ depending on $\e$.
 Furthermore, for $K\geq 3$, 
\be\label{resnlambda2}
n^\lambda(-K)\leq e^{-N^\e\log K}\,\,\,{\rm and }\,\,\,n^\lambda(K)\geq 1-e^{-N^\e\log K}
\ee
for some $\eps> 0$.

(2) In fact, the last tightness bound holds in a more general situation, namely,
let the universal Wigner matrix
 $H$ satisfy \eqref{sum}, \eqref{subexp} and  $M\geq (\log N)^9$ where $M$ is defined in \eqref{defM}. Then we have
\be\label{resnlambda3}
n^\lambda(-3)\leq CN^{-c\log\log N}\,\,\,{\rm and }\,\,\,n^\lambda(3)\geq 1-CN^{-c\log\log N}.
\ee
\end{lemma}
\bigskip
\begin{lemma}\label{N-1}
Let $H$ satisfy the conditions of Theorem \ref{lagaN-1}. 
 Then  for any $\e>0$ we have
 \be\label{resN-1}
 \int_{-\infty}^{\infty}\left|n^\lambda(E)-n^\lambda_{sc}(E)\right|\rd E\leq CN^{-1+\e}.
 \ee
\end{lemma}
\bigskip
\begin{lemma}\label{N-1/2}
Let $H$ satisfy the conditions
in Theorem \ref{lagaN-1}, in particular, let the distribution of the matrix elements satisfy the
uniform LSI \eqref{lsi}.
For $j, K\in \mathbb N$, $j+K\leq N+1$, define 
$\lambda_{j,K}=K^{-1}\sum_{i=0}^{K-1}\lambda_{j+i}$. Then for any $\delta>0$ small enough, 
\be\label{rangexjk}
\mathbb P\left(|\lambda_{j,K}-\E(\lambda_{j,K})|\geq N^{-1/2+\delta}K^{-1/2}\right)
\leq Ce^{-N^{\delta}},
\ee
with $C$ depending on $C_{sup}$ in \eqref{VV} and $C_S$ in \eqref{lsi}.
\end{lemma}

\subsection{Proof of Lemma \ref{N-1/6}.}

Extreme eigenvalues are typically controlled by the moment method, evaluating
$\E \, \tr H^k$ for  large $k$ using  some graphical representation.
Our proof follows the standard path, but since we were unable to find 
a reference that would apply precisely to our case, we include
the proof for completeness. The main technical estimate
\eqref{W} is borrowed from \cite{Vu}.
We remark that if we use the strongest result in \cite{Vu}, one
can improve the exponent $1/6$ to $1/4$ in \eqref{resnlambda1}.

We start with the proof of \eqref{resnlambda1}
and \eqref{resnlambda2} in the case of generalized  Wigner matrices (see \eqref{VV}).
First we truncate the random variables.
With the assumption of subexponential decay
  of $h_{ij}$, for any small $\delta>0$, one can find a 
$\widehat h_{ij}$ such that 
\be
\P(\widehat h_{ij}=h_{ij})\geq 1-e^{-N^{\e'}}
\ee
and 
\be\label{hijc0}
|\widehat h_{ij}|\leq N^{-1/2+\delta},\,\,\,\,\,\,\E(\widehat h_{ij})=0,
\,\,\,\,\,\,\,\,\E(|\widehat h_{ij}|^2)\leq \E(|h_{ij}|^2)
\ee
for some small number $\e'$, depending on $\delta$.  Then we only need to bound the spectral norm 
of the new matrix $\widehat H=(\widehat h_{ij})$.  To prove 
 \eqref{resnlambda1},
it only remains to  prove that, for some small $\e'>0$,
\be\label{hatH2}
\P(\|\widehat H\|\geq 2+ N^{-1/6+\e})\leq e^{-N^{\e'}}.
\ee 
With
$$
\P(\|\widehat H\|\geq 2+ N^{-1/6+\e})\leq \frac{\E\tr \widehat H^k}{(2+ N^{-1/6+\e})^k},
$$
for any even $k$, \eqref{hatH2} follows from
 \be\label{highm}
 \E \, \tr \widehat H^{k_0}\leq 2^{k_0+O(\log N)},
 \ee 
with the choice of  $k_0=N^{1/6-\delta/3}$ and $\delta =3\e/2$, since $\|\wh H\|^k \le \tr \widehat H^{k}$ for even powers.
The proof of \eqref{resnlambda2} is analogous.

\bigskip

To estimate $\E(\tr \widehat H^{k})$ for $k\in\N$, we start with introducing 
some notations and concepts on graphs.   

Let $p$ and $k$ be given integers. We define the concept of
{\it ordered closed walk} of $k$ edges on an abstract ordered set 
$A_p:=\{ a_1, a_2, \ldots, a_p\}$
of $p$ elements with the natural ordering $a_1 < a_2 < \ldots < a_p$.
An ordered closed walk on $p$ vertices with $k$ edge is determined by a sequence $\underline w 
=(w_1, w_2, \ldots, w_k)$ of the elements of $A_p$ with the
following properties:
\begin{itemize}

\item[i)] Along the walk, the fresh vertices from $A_p$ are adjoined in  
increasing order,
i.e., $\max_{j\le m} w_j \le \max_{j\le m-1} w_j +1$.

\item[ii)] $\{w_1, w_2, \ldots w_k\} = A_p$, i.e., all points of $A_p$
are visited.

\item[iii)]  Let $\Gamma(\underline w)$ denote the undirected 
graph associated with $\underline{w}$,
i.e., the vertex set of $\Gamma(\underline w)$ is $A_p$, the edges are
given by $(w_1, w_2), (w_2, w_3), \ldots (w_k, w_1)$; with
multiple edges as well as self-loops ($w_i=w_{i+1}$ for some $i$)  allowed.
Then 
 every edge of $\Gamma$ appears at least twice.

\end{itemize}

Let ${\cal W}(k,p)$ denote the set of ordered closed walks on $p$ vertices
with $k$ edges. Their number  was estimated in Lemma 2.1 of \cite{Vu} 
\be
  W(k,p) : = |{\cal W}(k,p)| \le {k\choose 2p-2} p^{2(k-2p+2)} 2^{2p-2}.
\label{W}
\ee
This bound will be sufficient for the proof of \eqref{resnlambda1}
with exponent $1/6+\e$. We remark that Lemma 4.1 of \cite{Vu} gives a different bound
on \eqref{W} that is better by essentially a factor $[(k-2p)/p]^{k-2p}$.
Applying this bound, one could improve the exponent in \eqref{resnlambda1} to $1/4+\e$
but we will not pursue this improvement here.

We also need the concept of {\it labelling} the elements of $A_p$ by the set
$\{1,2,\ldots, N\}$. A labelling is given by a function $\ell: A_p\to \{1, 2,\ldots,
N\}$
and we require that $\ell$ be injective. The set of such labelling functions
is denoted by ${\cal L}(p,N)$.

With these notations, we have the formula
\begin{align}\label{mainid}
  \E \tr \widehat H^k = & \sum_{i_1, i_2, \ldots i_k=1}^N \E \widehat h_{i_1i_2}
 \widehat h_{i_2i_3} \ldots
\widehat h_{i_ki_1} \non\\
  = & \sum_{p=1}^{k/2+1} \;\; \sum_{\underline w \in {\cal W}(k,p)} \;\;\sum_{\ell
\in {\cal L}(p,N)}
   \E \widehat h_{\ell(w_1) \ell(w_2)}\widehat h_{\ell(w_2) \ell(w_3)}\ldots
\widehat  h_{\ell(w_k) \ell(w_1)}.
\end{align}
To verify this formula, for any given sequence $i_1, i_2, \ldots, i_k$
on the l.h.s., let $p$ denote  the number of different
elements in this sequence and let the set $A_p$
be identified with these different elements
in the order of their appearance 
(i.e. for any $m$ we let $a_m: = i_s$ for some $s$ if 
$i_s \neq i_t$, $t<s$, and $i_s$ is the $m$-th freshest
element among $i_1, i_2, \ldots, i_s$, i.e., $|\{ i_1, i_2, \ldots , i_{s-1}\}|=m-1$).
Let $w_1, w_2, \ldots, w_k$ encode the sequence $i_1, i_2, \ldots, i_k$
with the new labels $a_1, a_2, \ldots a_p$. One may think of
the walk, $w_1, w_2, \ldots, w_k$, as the topological structure
of the sequence $(i_1, i_2, \ldots , i_k)$ where 
the original labels from the set $\{1, 2, \ldots , N\}$
have been replaced by abstract labels, defined intrinsically
from the repetition structure of $(i_1, i_2, \ldots , i_k)$.
Formula \eqref{mainid} is a resummation of all sequences
$(i_1, i_2, \ldots, i_k)$ in terms of topological walks (first and second sum)
and then reintroducing the original labelling with $\{1, 2, \ldots, N\}$
(third sum).
Since the first moment of $\wh h_{ij}$ vanishes and different matrix elements
are independent, all terms on the right hand side have zero expectation
in which at least one factor $\wh h_{ij}$ appears only once. This justifies
the requirement iii) in the definition of the ordered closed walks.
The restriction $p\le k/2+1$ in the summation then comes from iii). 
This proves \eqref{mainid}.

To compute the expectation on the r.h.s. of the \eqref{mainid}, we need to introduce the
concept of the {\it skeleton of the walk}. Given $\underline{w}\in {\cal W}(k,p)$,
its skeleton $S(\underline{w})$ is the undirected graph on $A_p$
that is obtained from $\Gamma(\underline{w})$ after replacing each multiple (parallel)
edge by a single undirected  edge. Here $S(\underline{w})$ allows self-loops (as long 
as every edge has multiplicity 1). Thus the edge set $E(S(\underline{w}))$
of the skeleton coincides with the edge set $E(\Gamma(\underline{w}))$
after neglecting multiplicity and direction. The skeleton is a subgraph
of the complete graph on $A_p$. We will also define the {\it tree of the walk},
$T(\underline{w})$, which is just a spanning tree of the skeleton $S(\underline{w})$
built up successively  along the walk by a greedy algorithm: include an edge to
the  $T(\underline{w})$ if it does not create a loop together with the previously
adjoined edges. Since $\Gamma(\underline{w})$ is connected, and then so is $S(\underline{w})$,
thus $T(\underline{w})$ is indeed a tree on $p$ vertices, in particular
the number of its edges is
\be
   |E(T(\underline{w}))| = p-1.
\label{Ep}
\ee
and  $S(\underline{w}) \setminus  T(\underline{w})$ has total edge 
multiplicity less than $k-2(p-1)$. 

For any edge $e\in E(S(\underline{w}))$
of the skeleton, let $\nu(e)$ denote the multiplicity of $e$ in $\Gamma(\underline{w})$
(edges with both orientations are taken into account). Clearly 
\be
   \sum_{e\in E(S)} \nu(e) =k
\label{sum2}
\ee
for any skeleton graph $S=S(\underline{w})$ for $w\in {\cal W}(k,p)$.
Finally, for a given edge $e=(a_\al, a_\beta)$ in a subgraph of $A_p$
and for any labelling $\ell \in {\cal L}(p,N)$, we define
the induced labelling of the edge $e$ by $\ell (e) = (\ell(a_\al), \ell(a_\beta))$.

With these notations we have
$$
  \Big| \E \widehat  h_{\ell(w_1) \ell(w_2)}\widehat h_{\ell(w_2) \ell(w_3)}
  \ldots \widehat  h_{\ell(w_k)
\ell(w_1)}\Big|
  \le \prod_{e\in E(S(\underline{w}))} \E \; |\widehat h_{\ell (e)}|^{\nu(e)}.
$$
Note that $|\widehat h_{ij}|=|\widehat h_{ji}|$, therefore there is no ambiguity
in the notation $|\widehat h_{\ell (e)}|$.
Since $|\widehat h|\le N^{-1/2+\delta}$ and $\nu(e)\ge 2$, we have
\be
   \E \; |\widehat h_{\ell(e)}|^{\nu(e)} \le N^{(-1/2+\delta)(\nu(e)-2)} \sigma^2_{\ell(e)},
\label{htree}
\ee
or, alternatively, 
\be
   \E \; |\widehat h_{\ell(e)}|^{\nu(e)} \le N^{(-1/2+\delta)\nu(e)} .
\label{hloop}
\ee

We will use \eqref{htree} for the edges of the tree, $e\in E(T(\underline{w}))$,
and we use \eqref{hloop} for the remaining edges
 $e\in E(S(\underline{w}))\setminus E(T(\underline{w}))$.
We can now estimate \eqref{mainid} using \eqref{sum2} and \eqref{Ep}:
\begin{align}
|\E \;\tr \widehat H^k|\le & 
\sum_{p=1}^{k/2+1} \;\; \sum_{\underline w \in {\cal W}(k,p)} \;\;\sum_{\ell \in
{\cal L}(p,N)}
\prod_{e\in E(S(\underline{w}))} \E \; |\widehat h_{\ell (e)}|^{\nu(e)} \non\\
\le & \sum_{p=1}^{k/2+1} \;\; \sum_{\underline w \in {\cal W}(k,p)} \;\;
N^{(-1/2+\delta)(k-2(p-1))}
\sum_{\ell \in {\cal L}(p,N)}
\prod_{e\in E(T(\underline{w}))}  \sigma_{\ell (e)}^2 \non\\
\le & \sum_{p=1}^{k/2+1} \;\; \sum_{\underline w \in {\cal W}(k,p)} \;\;
N^{1+(-1/2+\delta)(k-2(p-1))}.
\end{align}
In the last step we used that
$$
 \sum_{\ell \in {\cal L}(p,N)}
\prod_{e\in E(T)}  \sigma_{\ell (e)}^2\leq N.
$$
holds for any tree $T$. 
This   identity follows from successively
summing up the labels for  vertices with
degree one in $T$ by using
the identity $\sum_i \sigma_{ij}^2=1$.

Using \eqref{W}, we obtain the bound
\be\label{etrw}
  |\E \;\tr \widehat H^k|\le \sum_{p=1}^{k/2+1} S(k,p) ,
  \ee
  with 
 \be\label{defskp}
  S(k,p):= {k\choose 2p-2} p^{2(k-2p+2)} 2^{2p-2}
N^{1+(-1/2+\delta)(k-2(p-1))}.
\ee

It is easy to show that
\be\label{defskp-1}
S(k,p-1)\leq \frac{N^{2\delta} k^6}{4N}S(k,p).
\ee
Choosing $k=N^{1/6-\delta/3}$, we have $S(k,p-1)\leq S(k,p)$.
 Inserting this into \eqref{etrw}, we obtain \eqref{highm} and complete the proof. 

\bigskip 

Now we prove \eqref{resnlambda3} with the same method. 
Similarly, with the assumption on the distribution  of $h_{ij}$,
 one can find a $\widehat h_{ij}$ such that 
\be
\P(\widehat h_{ij}=h_{ij})\geq 1-CN^{-c\log\log N}
\ee
and 
\be\label{hijc02}
|\widehat h_{ij}|\leq M^{-1/2} n,\,\,\,\,\,\,\E(\widehat h_{ij})=0,
\,\,\,\,\,\,\,\,\E(|\widehat h_{ij}|^2)\leq \E(|h_{ij}|^2)+N^{-c\log\log N}
\ee
for  $n=(\log N)(\log\log N)$. Here $\wh h_{ij}$ can be obtained by considering
the cutoff random variables
$h_{ij}  {\bf 1}\big(|h_{ij}|\le M^{-1/2} (\log N)(\log \log N) \big)$
and then  slightly modifying them to recover  their zero expectation 
value. 

We can again bound  $|\E \;\tr \widehat H^k|$ as in  \eqref{etrw} but with
a slightly different $S(k,p)$; instead of the factor
 $N^{1+(-1/2+\delta)(k-2(p-1))}$
we will have $N\cdot M^{(-1/2+\delta)(k-2(p-1))}$ in the definition  \eqref{defskp}.
These modified $S(k,p)$ numbers satisfy
\be
S(k,p-1)\leq \frac{ n^2 k^6}{4M}S(k,p)
\ee
and 
\be
S(k,k/2+1)=2^k\cdot N.
\ee
Choosing $k=n$, we have $\frac{ n^2 k^6}{4M}<1$. Thus we obtain
\be
 |\E \;\tr \widehat H^k|\le 2^k\cdot 2n N,
\ee
which implies \eqref{resnlambda3} .
\qed

\subsection{ Proof of Lemma \ref{N-1}.}
 
First we show that the estimate on the expectation of $m-m_{sc}$ is better
 than the estimate  \eqref{fakemainlsresult} on $m-m_{sc}$  itself.

\begin{lemma}\label{Emzmscz2}  Assume that the $N\times N$ 
generalized Wigner matrix $H$ (see \eqref{VV}) satisfies
 \eqref{sum}, \eqref{de-de+}, 
\eqref{speccond} and \eqref{subexp}, $\E\, h_{ij}=0$, for any $1\leq i,j\leq N$
(i.e. the assumptions 
 of Theorem \ref{thm:detailed} apart from \eqref{relkaeta} hold). Then we have, with some $C>0$,
\be\label{resEmzmscz2}
|  \E m(z)-   m_{sc}(z)|\leq \frac{ (\log N)^C}{(N\eta)(\kappa+\eta)}
\ee
 for any $z=E+i\eta$, $\eta>0$.
\end{lemma}

As a preparation to the proof, we
 need the following technical lemma that we state
under more general conditions so that it is
applicable for universal Wigner matrices.

\begin{lemma}\label{Emzmscz} With the assumption of Theorem \ref{thm:detailed},
suppose \eqref{relkaeta} holds, we have the estimate
\be\label{resEmzmscz}
\left|\E m(z)+\frac{1}{\E m(z)+z}\right|\leq \frac{(\log N)^{C_0}\sqrt{\kappa+\eta}}{(M\eta)g^2(z)}
\ee
for some sufficiently large positive constants  $C_0$ (depending on $\al, \beta$ in \eqref{subexp}).
\end{lemma}

\par
{\it Proof of Lemma \ref{Emzmscz}.}
 Recall the definitions of $\Omega^o_z $, $\Omega^d_z $ and $\widehat\Omega^\Upsilon_z$ in
 \eqref{temp2.28}, \eqref{proplsresult1}, \eqref{proplsresult2} and \eqref{defOmzdo} and we define 
\be\label{defomegaz}
\Omega_z\equiv \Omega_z^o\cap \Omega_z^d\cap \widehat\Omega^\Upsilon_z.
\ee
With \eqref{allomega}, we have 
\be\label{temp5.7}
\P(\Omega_z)\geq 1-CN^{-c\log\log N}.
\ee
 The r.h.s. of \eqref{resEmzmscz} is larger than $N^{-2}$. Then with \eqref{temp5.7} 
and $|m(z)|\leq\eta^{-1}\leq M$ (see \eqref{lbeta}), we only need to prove
\be\label{resEmzmsc2}
\left|\E {\bf 1}(\Omega_z) m(z)- \frac{1}{\E {\bf 1}(\Omega_z) m(z) + z}\right|\leq \frac{C (\log N)^C  \sqrt{\kappa+\eta}}{(M\eta)g(z)^2}.
\ee 
Taking the expectation of   the self consistent equation \eqref{mainseeq} with 
\eqref{seeqerror}, we obtain that  
\be\label{temp5.9}
\E \big[  {\bf 1}(\Omega_z) \cdot G_{ii}\big]+\E\Big[ {\bf 1}(\Omega_z) \Big(z+ \sum_{j}\sigma^2_{ij}
G_{jj}+\Upsilon_i\Big)^{-1} \Big]=0.
\ee
For simplicity, we define 
$$
A_i:=\E\big[ {\bf 1}(\Omega_z) \cdot G_{ii}\big], \qquad A:=\sum_{i}A_i/N\,.
$$ 
Together with \eqref{defomegaz} and \eqref{temp5.7}, we have 
$$
|A-m_{sc}|, \; |A_j - A| \ll1. 
$$
Then, similarly to  \eqref{zmzgiimz}, 
 on the event $\Omega_z$ we have
$$
|z+ A|- |\sum_j\sigma_{ij}^2A_j-A|-|\Upsilon|>C,
$$ 
by using $|z+m_{sc}|\ge 1$ and that on the set $\Omega_z$,  $\Upsilon$ is small.
Therefore, we can  expand \eqref{temp5.9} as 
\begin{eqnarray}
0=&&A_i+\frac1{z+ A}
-\frac{\sum_j\sigma_{ij}^2A_j-A}
{(z+ A)^2}-\frac{\E{\bf 1}(\Omega_z)\Upsilon_i}
{(z+ A)^2}\\\nonumber
&&+O\left(\frac{\E \Big[ {\bf 1}(\Omega_z)\left|\sum_{j}\sigma^2_{ij}G_{jj}-A\right|^2\Big]}
{(z+ A)^3}\right)+O\left(\frac{\E\big[{\bf 1}(\Omega_z)\left|\Upsilon_i\right|^2\big]}
{(z+ A)^3}\right).
\end{eqnarray}

Then summing up $1\leq i\leq N$
,  we obtain that 
\begin{eqnarray}\label{equEomz}
\left|A+\frac{1}{z+A}\right|\leq C\max_i \Big| \E \big[ {\bf 1}(\Omega_z)\Upsilon_i\big]\Big|+
C\max_i \E \Big[ {\bf 1}(\Omega_z)\Big|\sum_{j}\sigma^2_{ij}G_{jj}-A \Big|^2\Big]
+C\E\big[{\bf 1}(\Omega_z)\left|\Upsilon\right|^2 \big].
\end{eqnarray}
Applying \eqref{proplsresult1}
 and the definition of $\Omega_z$, we can bound the second and third terms in the 
r.h.s. of \eqref{equEomz} with some constant $C$ as follows, 
\begin{eqnarray}\label{equEomz2}
\left|A+\frac{1}{z+A}\right|\leq C\max_i \Big| \E \big[ {\bf 1}(\Omega_z)\Upsilon_i\big]\Big|+
\frac{(\log N)^C\sqrt{\kappa+\eta}}{M\eta g(z)^2}.
\end{eqnarray}
If $\eta>3$, we estimate $ \E \big[ {\bf 1}(\Omega_z)\Upsilon_i\big]$ as 
\be\label{temp5.13}
\Big| \E \big[ {\bf 1}(\Omega_z)\Upsilon_i\big]\Big|
\leq \Big|\E \big[ {\bf 1}(\widehat\Omega^o_z)\Upsilon_i\big]\Big|
+\E \Big[ {\bf 1}([\widehat\Omega^o_z]^c){\bf 1}(\widehat\Omega_z)|\Upsilon_i|\Big].
\ee
With \eqref{GiiGjii}, we have
\be
|G_{ij}G_{ji}/G_{ii} |\leq 2/\eta.
\ee
Then, with the definition of $\Upsilon_i$ \eqref{seeqerror} and \eqref{temp3.39}, we have 
\be
\P(|\max{\Upsilon_i}|\geq N^C)\leq e^{-N^c}
\ee
for some positive constants $c$ and $C$. Inserting this and \eqref{EOmz}
 into \eqref{temp5.13}, we have 
\be\label{temp5.15}
\left|A+\frac{1}{z+A}\right|\leq \frac{(\log N)^C\sqrt{\kappa+\eta}}{M\eta g^2(z)}
\ee
in the case of $\eta>3$. If $\eta<3$, similarly, with \eqref{EOmz2} we have the same result.  
This proves \eqref{resEmzmsc2} and thus completes the proof of Lemma \ref{Emzmscz}. \qed

\bigskip

{\it Proof of Lemma \ref{Emzmscz2}.}
First we will prove the result for large $\eta$, more precisely we show
\eqref{resEmzmscz2}
 under the additional assumption 
that 
\be\label{NetaC01}
N\eta(\kappa+\eta)^{3/2}\geq (\log N)^{C_1},
\ee
with a sufficiently large constant $C_1$.

In the case of the generalized Wigner matrix, \eqref{VV},
 we have $M\ge (\Cs)^{-1}N$ and $\delta_+\geq \Ci$ \eqref{de-de+2}, then 
$$
g(z)\asymp \sqrt{\kappa+\eta}
$$
up to an $O(1)$ factor. Note that with a sufficiently large $C_1$,
\eqref{NetaC01} implies \eqref{relkaeta} and thus
combining Lemma \ref{Emzmscz} with Lemma \ref{newonmz}  we obtain 
\eqref{resEmzmscz2} under the condition that $\eta$ satisfies \eqref{NetaC01}.

To prove \eqref{resEmzmscz2} for any $\eta>0$,
it remains to consider the case when \eqref{NetaC01} does not hold.
For a fixed $E$, let $\eta^*=\eta^*(E)>0$ be the (unique) solution
of $N\eta (\kappa+\eta)^{3/2}=(\log N)^{C_1}$, i.e. when 
\eqref{NetaC01} becomes an equality. In particular, we know that
\be\label{temp7.30}
|  \E m(z^*)-   m_{sc}(z^*)|\leq \frac{ (\log N)^C}{(N\eta^*)(\kappa+\eta^*)}.
\ee
Consider $\eta< \eta^*$,
set $z=E+i\eta$, $z^*=E+i\eta^*$ and
estimate
\be\label{mmmsc}
   | \E m(z)-m_{sc}(z)|\le   | \E m(z^*)-m_{sc}(z^*)|
  + \int_\eta^{\eta^*} \big| \partial_y 
\big(\E m(E+iy)-m_{sc}(E+iy)\big)\big|\rd y.
\ee
Note that
\begin{align}
   |\partial_y m(E+iy)| = & \Big|\frac{1}{N}\sum_j \partial_y G_{jj}(E+iy)\Big|
 \\ \le & \frac{1}{N}\sum_{jk} |G_{jk}(E+iy)|^2 =
 \frac{1}{Ny}\sum_j \im G_{jj}(E+iy) = \frac{1}{y}\im m(E+iy),
\end{align}
and similarly
$$
  |\partial_y m_{sc}(E+iy)| = 
\Big|\int \frac{\varrho_{sc}(x)}{(x-E-iy)^2}\rd x\Big|
\le \int \frac{\varrho_{sc}(x)}{|x-E-iy|^2}\rd x
 = \frac{1}{y} \im m_{sc}(E+iy).
$$
Now we use the fact that the functions $y\to y\im m(E+iy)$ and
$y\to y\im m_{sc}(E+iy)$ are monotone increasing for any $y>0$
since both are Stieltjes transforms of a positive measure.
Therefore the integral in \eqref{mmmsc} can be bounded by
\be\label{intbb}
   \int_\eta^{\eta^*} \frac{\rd y}{y} \big[ \im \E m(E+iy) +  
\im m_{sc}(E+iy)\big] \le  \eta^*\big[ \im \E m(E+i\eta^*) +  
\im m_{sc}(E+i\eta^*)\big] \int_\eta^{\eta^*} \frac{\rd y}{y^2}
\ee
By the choice of $\eta^*$ and using that
$\im \,   m_{sc}(z^*)\leq C\sqrt{\kappa+\eta^*} $, we have  
\be\label{temp5.10}
\im \,   m_{sc}(z^*)\leq \frac{ (\log N)^C}{(N\eta^*)(\kappa+\eta^*)}.
\ee
and then $ \im \,  \E m(z^*)$ can be estimated from
 \eqref{temp7.30}.
Inserting these estimates into \eqref{mmmsc} and \eqref{intbb},
and using \eqref{temp7.30},
we get 
$$
 | \E m(z)-m_{sc}(z)|\le   | \E m(z^*)-m_{sc}(z^*)|
  +  \frac{ 2(\log N)^C}{N\eta^*(\kappa+\eta^*)} \frac{\eta^*}{\eta}
 \le  \frac{ (\log N)^C}{N\eta(\kappa+\eta)}
$$
with a possible larger $C$ in the r.h.s.
   This completes the proof of 
 Lemma \ref{Emzmscz2}. \qed

\bigskip
With Lemma \ref{Emzmscz2}, it follows that for any $E$ and $\eta>0$, 
\be\label{denss}
\left|n^\lambda(E+\eta)-n^\lambda(E-\eta)\right|+
\left|n^\lambda_{sc}(E+\eta)-n^\lambda_{sc}(E-\eta)\right|\leq \eta 
(\log N)^C\left(1+\frac{1}{N\eta(|E-2|+\eta)}\right).
\ee
Now we return to the main argument to prove \eqref{resN-1} in Lemma \ref{N-1}.
Given \eqref{resnlambda2}, we only need to prove 
\be\label{resN-133}
\int_{-3}^{3}\left|n^\lambda(E)-n^\lambda_{sc}(E)\right|\rd E\leq CN^{-1+\e}.
\ee
This inequality follows from the next lemma by choosing the signed measure
\be
\varrho^\Delta(\rd x)=\varrho_{sc}(\rd x)-\frac{\rd n^\lambda(E)}{\rd E},
\ee
whose Stieltjes transform is given by
\be
m^\Delta(z)=m_{sc}(z)-\E m(z)
\ee 
and the conditions \eqref{temp7.43} and \eqref{secc} are provided by
\eqref{resEmzmscz2} and \eqref{denss}.
 This will complete the proof of Lemma  \ref{N-1}.

\begin{lemma}\label{new67}
Let $\varrho^\Delta(\rd x)$ be a finite signed measure with support in $[-K,K]$ for
some $K > 0$. Let
\be
m^\Delta(z):=\int_\R\frac{\varrho^\Delta (\rd x)}{x-z},\qquad
n^\Delta(E):=\int_{-\infty}^E \varrho^\Delta(\rd x)
\ee 
be the Stieltjes transform and the distribution function of $\varrho^\Delta(\rd x)$,
 respectively. Let $\kappa_x$, $\kappa_E$ denote $||x|-2|$ and $||E|-2|$.  
We assume that $m^\Delta$ satisfies the following bound with some constant $C$:
\begin{eqnarray}
\left|   m^\Delta(x+iy)\right|\leq \frac{ (\log N)^C}{(Ny)(\kappa_x+y)}&\mbox{for} & 
  y>0, \;\; |x|\leq  K + 1, \label{temp7.43} 
\end{eqnarray}
and for any $a>0$ 
\be\label{secc}
\int_{E-a}^{E+a}|\varrho^\Delta|(\rd x)\leq a (\log N)^C\left(1+\frac{1}{Na(\kappa_E+a)}\right) .
\ee
Then
\be\label{resultnew67}
\int_{-K}^{K}\rd E\left|n^\Delta(E)\right|\leq CN^{-1}(\log N)^{C'}
\ee
for some constant $C'>0$ when $N$ is sufficiently large. 
\end{lemma}

This lemma  is  similar to    Lemma B.1 in \cite{ERSY}, but with different
 assumptions. Since the assumptions here are stronger than (B.3) and (B.4) in \cite{ERSY},
we  actually  obtain  a better
bound  \eqref{resultnew67} than in  \cite{ERSY}, where the
l.h.s. of \eqref{resultnew67} was bounded by $N^{-6/7}$.

\medskip

{\it Proof of Lemma \ref{new67}.}
 \,\,\,For simplicity, we omit the $\Delta$ superscript in the proof.
For a fixed $E\in [-K,K]$, $\eta>0$, define a function $f= f_{E,\eta}$: $\R\to \R$:
 such that $f(x) = 1$ for $x\in [-K, E-\eta]$, $f(x)$ vanishes  
for  $x\in (-\infty, -K-1)\cap [E+\eta, \infty)$, moreover
 $|f'(x)|\leq C\eta^{-1}$ and $|f''(x)|\leq C\eta^{-2}$. Then
 \be
 \left|n(E)-\int_{\R}f_{E,\eta}(\lambda)\varrho(\lambda)\rd\lambda\right|\leq 
\int_{E-\eta}^{E+\eta}| \varrho|(\rd x)\leq 
\eta (\log N)^C\left(1+\frac{1}{N\eta(\kappa_E+\eta)}\right).
 \ee
We will choose $\eta= N^{-1}$ and set $f_E:= f_{E,\eta}$ with $\eta=1/N$.
 Then to prove \eqref{resultnew67}, we only need to prove that 
 \be\label{boudnfel}
 \left|\int_{|E|\le K+1}
\int_{\R}f_E(\lambda)\varrho(\lambda)\rd\lambda\rd E \right|\leq N^{-1}(\log N)^{C'}
 \ee
 for some $C'>0$. 
\par To express $f_E(\lambda)$ in terms of the Stieltjes transform, we use 
the Helffer-Sj\"{o}strand functional calculus, as (B.12) in \cite{ERSY}. 
We formulate this result in a more general form.

\begin{lemma}\label{lm:HS} Let $f_{E,\eta}$ be given as above with some $E\in [-K, K]$,
$K\ge 3$,
and $0 < \eta \le 1/2$. Suppose that the Stieltjes transform $m$ of
the signed measure $\varrho$ satisfies
\begin{eqnarray}
\left|  m(x+iy)\right|\leq
 \frac{ L}{(Ny)^\tau(\kappa_x+y)^\sigma}&\mbox{for} & 
  y>0, \;\; |x|\leq  K + 1, \label{temp7.431} 
\end{eqnarray}
  with some exponents $0\le \tau, \sigma\le1 $ and some constant $L$.
Then
\be
   \left|\int f_E(\lambda)\varrho(\lambda)\rd\lambda \right|  \le
   \frac{C L |\log \eta|}{N^\tau (\kappa_E+\eta)^\sigma},
\label{genHS}
\ee
with some constant $C$ depending on $K$.
\end{lemma}
The condition of this lemma with $\tau=\sigma=1$ and $L=(\log N)^C$
 coincides with 
\eqref{temp7.43}, therefore, after integrating in $E$ and using $\eta=1/N$,
we obtain \eqref{boudnfel} which completes the proof of Lemma \ref{new67}. \qed

\medskip

{\it Proof of Lemma \ref{lm:HS}.}
Analogously to  (B.13), (B.14)  and (B.15) in \cite{ERSY}
we obtain that 
\begin{eqnarray}\nonumber
\left|\int f_E(\lambda)\varrho(\lambda)\rd\lambda \right|
\leq && C\int_{\R^2} ( |f_E(x)| +|y| |f'_E(x)|) |\chi'(y)| 
|m(x+iy)| \rd x\rd y\non \\ 
&& +C\left|\int_{|y|\leq \eta}\int y f''_E(x) \chi(y)
\im \,  m(x+iy)\rd x\rd y\right|\label{intr2fe} \\
&&+C\left|\int_{|y|\geq \eta}\int_\R y f''_E(x)\chi(y) \im \,  m(x+iy)\rd x
\rd y\right|, \non
\end{eqnarray}
where $\chi(y)$ is  a smooth cutoff function with support in $[-1,1]$, with $\chi(y) = 1$ for
$|y|\leq  1/2$ and with bounded derivatives.
The first term is estimated by
\be 
 \int_{\R^2} ( |f_E(x)| +|y| |f'_E(x)|) |\chi'(y)| 
|m(x+iy)| \rd x\rd y
 \le \frac{CL}{N^\tau},
\label{firss}
\ee
 using \eqref{temp7.431} and the  support of $\chi'$.

 With \eqref{temp7.431} 
and $|f''_E|\leq C\eta^{-2}$ and
$${\rm supp} f'_E(x)\subset\{|x-E|\leq \eta\},$$
  the second term in r.h.s. of \eqref{intr2fe} is bounded by  
\begin{eqnarray}\nonumber
CL\left|\int_{0\leq y\leq \eta}\int_{|x-E|\leq \eta} 
 \frac{ y |f''_E(x)|}{(Ny)^\tau(\kappa_x+y)^\sigma}\rd x\rd y\right|&&\leq 
\frac{CL}{N^\tau \eta^2}\left|\int_{0\leq y\leq \eta}\int_{ |x-E|\leq \eta} 
 \frac{y^{1-\tau}}{(\kappa_x+y)^\sigma}\rd x\rd y\right|\\\label{temp7.49}
&&\leq \frac{CL\eta^{1-\tau}|\log \eta|}{N^\tau (\kappa_E +\eta)^\sigma}.
\end{eqnarray}
Here we used that for $y\le 1/2$ we have
$$
   \int_{ |x-E|\leq \eta} 
 \frac{1}{(\kappa_x+y)^\sigma}\rd x \le \frac{C\eta|\log y|}{(\kappa_E +\eta)^\sigma}.
$$

As the (B.17) and (B.19) in \cite{ERSY}, we integrate the third term in
 \eqref{intr2fe} by parts first in $x$, then in $y$. Then  bound it with absolute value by 
\be\label{temp7.50}
C\int_{|x|\leq K+1}\eta |f'_E(x)| |\re \,  m(x+i\eta)|\rd x 
+
C \int _{\R^2}\, |f_E'(x)  \chi'(y)   \re \, m (x +iy) |
+
\frac C\eta \int_{\eta\le y\leq 1}\int_{ |x-E|\leq \eta} \!\!|\re \,  m(x+iy)|\rd x\rd y.
\ee
The middle term is bounded as \eqref{firss}.  With \eqref{temp7.431} again, we have
\begin{eqnarray}\nonumber
\eqref{temp7.50}\leq&& \frac{CL}{(N\eta)^\tau}\int_{|x-E|\leq \eta}
 \frac{1}{(\kappa_x+\eta)^\sigma}
 \rd x
 +\frac{CL}{(N\eta)^\tau} 
 + \frac{CL}{(N\eta)^\tau}\int_{\eta\le y\le 1}
\int_{|x-E|\leq \eta}\frac{1}{(\kappa_x+y)^\sigma}
 \rd x\rd y\\\label{temp7.51}
\medskip
&& \leq \frac{CL\eta^{1-\tau}|\log \eta|}{ N^\tau(\kappa_E+\eta)^\sigma}.
\end{eqnarray}

 Then combining \eqref{intr2fe}, \eqref{firss},
 \eqref{temp7.49}, \eqref{temp7.50} and \eqref{temp7.51} 
 we obtain    \eqref{genHS} and complete the proof of Lemma \ref{lm:HS}.
\qed
\bigskip

\subsection{Proof of Lemma \ref{N-1/2}}

\par Define the variables $v_{ij}$ as
\be 
h_{ij}=\sigma_{ij}v_{ij}.
\ee
Denote by $u_\al$ and $\lambda_\al$  the eigenvectors and eigenvalues of $H$.
 For any collection of real numbers, $C_\al\in \R$,  we  have 
\be\label{temp5.11}
\sum_{ij}\left|\sum_\al C_\al \frac{\partial \lambda_\al}{\partial v_{ij}}\right|^2
= \sum_{ij}\left|\sum_\al C_\al \sigma_{ij}\bar u_\al(i)u_\al(j)\right|^2=
\sum_{ij}\sigma^{2}_{ij}\left|\sum_\al C_\al \bar u_\al(i)u_\al(j)\right|^2
\leq \Cs N^{-1}\sum_{\al}|C_\al|^2.
\ee
With the choice $C_\al = K^{-1}$, $\al=j, j+1, \ldots, j+K-1$, and $C_\al=0$ otherwise,
we get $|\nabla \lambda_{j,K}|^2\le C_{sup} (NK)^{-1}$.
Using the Bobkov-G\"otze concentration inequality \cite{BG} and the uniform
bound on the LSI constant \eqref{lsi}, we get
$$
   \P \big( |\lambda_{j,K} - \E \, \lambda_{j,K}|\ge \gamma\big) \le e^{-\gamma T} 
  \E \, e^{ C_S T^2 |\nabla \lambda_{j,K}|^2} \le e^{-\gamma T+ C_S C_{sup} T^2/(NK) }
$$
for any $T$ and $\gamma$. Choosing $\gamma= N^{-1/2+\delta}K^{-1/2}$
and $T= (NK)^{1/2}$, we obtain
\eqref{rangexjk}. \qed

\section{Proof of the Green's function comparison theorem }\label{sec:proofcomp}

\noindent 
{\it Proof of Theorem \ref{comparison}.}  {F}rom  
the trivial bound 
\[
 \im \,   \left ( \frac 1 {H-E- i\eta} \right )_{jj} \le  \left ( \frac {y} {  \eta} \right )\,  \im \, 
  \left ( \frac 1 {H-E-iy} \right )_{jj} , \qquad \eta\le y,
 \] 
and from  \eqref{basic}
we have the following a priori bound
\be\label{basic3}
\P\left(\max_{0 \le \gamma \le \gamma(N)} \max_{1 \le k \le N}  \max_{|E|\le 2-\kappa}
\sup_{\eta\ge N^{-1-\e}} \left |  \im \,  \left (\frac 1 { H_{\gamma}-E\pm i\eta} \right )_{k k } 
\right | \le N^{3\tau+ \e} \right)\geq 1-CN^{-c\log\log N}.
\ee
Note that the supremum over $\eta$ can be included by establishing the estimate first
for a fine grid of $\eta$'s with spacing $N^{-10}$ and then extend the bound for all $\eta$ by
using that the Green's functions are Lipschitz continuous in $\eta$
with a Lipschitz constant $\eta^{-2}$.

Let $\lambda_m$ and $u_m$ denote the eigenvalues and eigenvectors
of $H_\gamma$, then by  the definition of the Green's function,   we have 
\[
\left | \left ( \frac 1 {H_\gamma-z} \right )_{jk} \right | 
\le  \sum_{m = 1}^N  \frac {| u_{m}(j)| |  u_m(k)| }{| \lambda_m -z |} 
\le \left [ \sum_{m = 1}^N  \frac {| u_{m}(j)|^2  }{| \lambda_m-z |}  \right ]^{1/2} 
\left [ \sum_{m = 1}^N  \frac {| u_{m}(k)|^2  }{| \lambda_m -z  |}  \right ]^{1/2} .
\]
Define a dyadic decomposition
\be
U_n = \{m:  2^{n-1} \eta \le |\lambda_m - E|<  2^{n} \eta \}, \qquad n=1,2,\ldots, n_0:=C\log N,
\label{dya}
\ee
\[ 
   U_0 = \{m:    |\lambda_m - E|< \eta \},\qquad U_\infty:=
\{m:  2^{n_0} \eta \le |\lambda_m - E| \},
\]
and divide the summation over $m$ into $\cup_n  U_n$
\[
\sum_{m = 1}^N  \frac {| u_{m}(j)|^2  }{| \lambda_m-z  |} 
= \sum_n  \sum_{m \in U_n} \frac {| u_{m}(j)|^2  }{| \lambda_m-z |}   
\le C  \sum_n  \sum_{m \in U_n}  \im \,  \frac {| u_{m}(j)|^2  }{ \lambda_m -E - i 2^n \eta }  
\le C  \sum_n   \im \,  \left ( \frac 1 {H_\gamma -E - i 2^n \eta} \right )_{jj}. 
\]
Using the estimate \eqref{basic} for $n=0,1, \ldots, n_0$ and a trivial bound of $O(1)$
for $n=\infty$, we have proved that
\be\label{basic4}
\P\left(\sup_{0 \le \gamma \le \gamma(N)} \sup_{1 \le k, \ell \le N}  \max_{|E|\le 2-\kappa}
\sup_{\eta\ge N^{-1-\e}}
  \left |  \left (\frac 1 { H_{\gamma}-E\pm i\eta} \right )_{k \ell } 
\right | \le N^{4\tau+ \e} \right)\geq 1-CN^{-c\log\log N}.
\ee

For simplicity, we will consider the case when the test function $F$ has only $n=1$
variable and $k_1=1$, i.e., we consider the trace of a first order monomial;
the general case follows analogously.  Consider the telescoping sum of differences of expectations 
\begin{align}\label{tel}
\E \, F \left ( \frac{1}{N}\tr  \frac 1 {H^{(v)}-z} \right )   - 
 & \E \, F \left  (  \frac{1}{N}\tr  \frac  1 {H^{(w)}-z} 
\right )  \\
= & \sum_{\gamma=1}^{\gamma(N)}\left[  \E \, F \left (  \frac{1}{N}\tr \frac 1 { H_\gamma-z} \right ) 
-  \E \, F \left (  \frac{1}{N}\tr \frac  1 { H_{\gamma-1}-z} \right ) \right] . \non
\end{align}
Let $E^{(ij)}$ denote the matrix whose matrix elements are zero everywhere except
at the $(i,j)$ position, where it is 1, i.e.,  $E^{(ij)}_{k\ell}=\delta_{ik}\delta_{j\ell}$.
Fix an $\gamma\ge 1$ and let $(i,j)$ be determined by  $\phi (i, j) = \gamma$.
We will compare $H_{\gamma-1}$ with $H_\gamma$.
Note that these two matrices differ only in the $(i,j)$ and $(j,i)$ matrix elements 
and they can be written as
$$
    H_{\gamma-1} = Q + \frac{1}{\sqrt{N}}V, \qquad V:= v_{ij}E^{(ij)}
+ v_{ji}  E^{(ji)}
$$
$$
    H_\gamma = Q + \frac{1}{\sqrt{N}} W, \qquad W:= w_{ij}E^{(ij)} +
   w_{ji} E^{(ji)},
$$
with a matrix $Q$ that has zero matrix element at the $(i,j)$ and $(j,i)$ positions and
where we set $v_{ji}:= \ov v_{ij}$ for $i<j$ and similarly for $w$.
Define the  Green's functions
$$
      R = \frac{1}{Q-z}, \qquad S= \frac{1}{H_\gamma-z}.
$$

We first claim that the estimate \eqref{basic4} holds for the Green's function $R$ as well. 
To see this, we have, from the 
resolvent expansion, 
\[
 R = S  +   N^{-1/2}SV S + \ldots + N^{-9/5} (SV)^9S+
N^{-5} (SV)^{10} R.
\] 
Since $V$ has only at most two nonzero element, when
computing the $(k,\ell)$ matrix element of this matrix identity,
each term is a finite sum involving
matrix elements of $S$ or $R$ and $v_{ij}$, e.g.  $(SVS)_{k\ell} =S_{ki} v_{ij} S_{j\ell}
+ S_{kj} v_{ji} S_{i\ell}$.  Using the bound \eqref{basic4} for the $S$ matrix elements,
the subexponential decay for $v_{ij}$ and 
 the trivial bound $|R_{ij}| \le  \eta^{-1}$, we obtain that 
the estimate \eqref{basic4} holds for $R$. 

\medskip

We can now start proving the main result. 
By the resolvent expansion, 
\[
   S = R - N^{-1/2} RVR+ N^{-1} (RV)^2R - N^{-3/2} (RV)^3R + N^{-2} (RV)^4R - N^{-5/2} (RV)^5S,
\]
so we can write
\[
\frac{1}{N} \tr S =  \wh R + \xi, \qquad \xi =  \sum_{m=1}^4 N^{-m/2}\wh R^{(m)}+N^{-5/2}\Omega
\]
with
\[
   \wh R = \frac{1}{N} \tr R, \qquad \wh R^{(m)} = (-1)^m\frac{1}{N}\tr (RV)^mR, \qquad
 \Om = - \frac{1}{N}\tr (RV)^5S.
\]
For each diagonal element in the computation
of these traces, the contribution to $\wh R$,  $\wh R^{(m)}$ and $\Omega$ is a 
sum of a few terms. E.g.
\[
  \wh R^{(2)} = \frac{1}{N}\sum_k \Big[ R_{ki} v_{ij} R_{jj} v_{ji} R_{ik} + 
 R_{ki} v_{ij} R_{ji} v_{ij} R_{jk}
+ R_{kj} v_{ji} R_{ii} v_{ij} R_{jk} + R_{kj} v_{ji} R_{ij} v_{ji} R_{ik} \Big]
\]
and similar formulas hold for the other terms.

Then we have 
\begin{align}
\E F  \left(\frac{1}{N}\tr \frac 1 { H_\gamma-z} \right )
= & \E F \left ( \wh R + \xi \right )  
\label{temp6.6}\\
= & \E \left [ F(  \wh R) + F'( \wh R ) \xi + F^{\prime \prime} ( \wh R) \xi^2
+ \ldots+ F^{(5)} ( \wh R+\xi') \xi^5  \right ]  \non\\
= & \sum_{m=0}^5 N^{-m/2} \E    A^{(m)},\non
\end{align}
where $\xi'$ is a number between $0$ and $\xi$ 
and it depends on $ \wh R$ and $\xi$;  the $A^{(m)}$'s are defined as 
\[
A^{(0)} = F(   \wh R), \quad 
A^{(1)}=  F'(\wh R)   \wh R^{(1)},\quad 
A^{(2)}=  F''(\wh R)(  \wh R^{(1)})^2 + F'(\wh R)  \wh R^{(2)},\quad
\]
and similarly for $A^{(3)}$ and $A^{(4)}$. Finally, 
$$
  A^{(5)} = F'(\wh R)\Omega + F^{(5)} (\wh R+ \xi') (\wh R^{(1)})^5 +\ldots.
$$

 The expectation values
of the terms $A^{(m)}$, $m\le 4$, 
with respect to $v_{ij}$ are determined by the first 
four moments of $v_{ij}$, for example
$$
  \E \, A^{(2)} = F'(\wh R)
\Big[ \frac{1}{N}\sum_k R_{ki}R_{jj}R_{ik} + \ldots
   \Big] \E \, |v_{ij}|^2  + F''(\wh R) \Big[
  \frac{1}{N^2}\sum_{k,\ell} R_{ki}R_{j\ell}R_{\ell j} R_{ik} +\ldots\Big]
 \E \, |v_{ij}|^2 
$$
$$
\qquad  +  F'(\wh R) \Big[ \frac{1}{N}\sum_k R_{ki}R_{ji}R_{jk} + \ldots
   \Big] \E \, v_{ij}^2  + F''(\wh R) \Big[
  \frac{1}{N^2}\sum_{k,\ell} R_{ki}R_{j\ell}R_{\ell i} R_{jk} +\ldots\Big]
 \E \, v_{ij}^2 .
$$
Note that the coefficients involve up to four derivatives of $F$ and normalized sums
of matrix elements of $R$.  Using the estimate \eqref{basic4} for $R$ and the derivative
bounds \eqref{lowder}  for the typical values of $\wh R$,
 we see that all these coefficients are bounded by $N^{C(\tau +\e)}$
with a very large probability, where $C$ is an explicit constant. 
We use the bound \eqref{highder} for the extreme values of $\wh R$
but this event  has a very small probability by \eqref{basic4}.
Therefore, the coefficients of the moments $\E\, \bar v_{ij}^s v_{ij}^u$, $u+s\le 4$,
in the quantities $A^{(0)}, \ldots , A^{(4)}$ are essentially bounded,
modulo a factor $N^{C(\tau +\e)}$.
Notice that the fourth moment of $v_{ij}$ appears only in the $m=4$
term that already has a prefactor $N^{-2}$   in \eqref{temp6.6}. Therefore, to
compute the $m\le 4$ terms in \eqref{temp6.6} up to a precision $o(N^{-2})$,
it is sufficient to know the first three moments of $v_{ij}$ exactly and 
the fourth moment only with a precision $N^{-\delta}$; if
$\tau$ and $\e$ are chosen such that $C(\tau +\e) < \delta$, then
the discrepancy in the fourth moment is irrelevant.

Finally, we have to estimate the error term $A^{(5)}$.
All terms without $\Omega$ can be dealt with as before;
after estimating the derivatives of $F$ by $N^{C(\tau + \e)}$,
one can perform the expectation with respect to $v_{ij}$ 
that is independent of $\wh R^{(m)}$. For the terms
involving $\Omega$ one can argue similarly, by
appealing to the fact that the matrix elements of $S$
are also essentially bounded by $N^{C(\tau + \e)}$, see \eqref{basic4},
and that $v_{ij}$ has subexponential decay.
Alternatively, one can use H\"older inequality to decouple
$S$ from the rest and use \eqref{basic4} directly,
for example:
\[
\E  | F'(\wh R)  \Omega | = \frac{1}{N}\E  | F'(\wh R) \tr (RV)^5 S | 
\le  \frac{1}{N}   \left [ \E  ( F'(\wh R))^2 \tr S^2 \right ]^{1/2}    \left [ \E  \, 
   \tr (RV)^5 (VR^*)^5    \right ]^{1/2}
\le C N^{C(\tau+ \e)}.
\]

Note that exactly the same perturbation expansion holds for
the resolvent of $H_{\gamma-1}$, just  $v_{ij}$ is replaced
with $w_{ij}$ everywhere. By the moment matching condition,
the expectation values $\E A^{(m)}$ of terms for $m\le 3$  in \eqref{temp6.6}
are identical and the $m=4$ term differs by
$N^{-\delta + C(\tau +\e)}$. Choosing $\tau =\e$,  we have
\[
\E \, F \left ( \frac{1}{N}\tr \frac 1 { H_\gamma-z} \right ) 
-  \E \, F \left ( \frac{1}{N}\tr \frac  1 { H_{\gamma-1}-z} \right )
\le  C N^{-5/2 + C \e}+C N^{-2 -\delta + C \e}.
\]
After summing up in \eqref{tel}  we have thus proved that 
\[
\E \, F \left ( \frac{1}{N}\tr  \frac  1 { H^{(v)} - z}\right ) 
-  \E \, F \left ( \frac{1}{N}\tr \frac  1 { H^{(w)} - z} \right )\le 
 C N^{-1/2+ C \e}+C N^{-\delta+ C \e}.
\]
The proof can be easily generalized to functions of several variables. This concludes the proof
of Theorem  \ref{comparison}.
\qed

\bigskip 
\noindent 
{\it Proof of Theorem \ref{com}.}  
Define an approximate  delta function (times $\pi$) at the scale $\eta$ by 
\[
\theta_\eta(x)  =  \im \,  \frac 1 {x - i \eta}.
\] 
For notational simplicity, we will prove only the case of three point correlation functions;
the proof is analogous for the general case.
 By definition of the correlation function, for any fixed $E$, $\al_1, \al_2, \al_3$,
\begin{align}\label{6.5}
    \E_\bw & \frac 1 {N(N-1)(N-2)} \sum_{i \not = j \not = k}
  \theta_\eta\Big(\lambda_i - E-\frac{\alpha_1}{N} \Big) 
 \theta_\eta\Big(\lambda_j-  E-\frac{\alpha_2}{N}\Big)  
\theta_\eta\Big(\lambda_k- E-\frac{\alpha_3}{N}\Big) 
 \nonumber \\
 & = \int \rd x_1 \rd x_2  \rd x_3 
p_{w, N}^{(3)}(x_1, x_2, x_3)  \theta_\eta(x_1 - E_1)  \theta_\eta(x_2- E_2) 
 \theta_\eta(x_3- E_3),   \qquad E_j :=  E+\frac{\alpha_j}{N}.
\end{align}
By the exclusion-inclusion principle,
\be\label{6.6}
\E_\bw \frac 1 {N(N-1)(N-2)} \sum_{i \not = j \not = k}   \theta_\eta(x_1 - E_1)
  \theta_\eta(x_2- E_2) 
 \theta_\eta(x_3- E_3)  = \E_\bw  A_1 +  \E_\bw  A_2 +  \E_\bw A_3, 
\ee
where 
\[
A_1: = \frac 1 {N(N-1)(N-2)}  \prod_{j=1}^3 
\left [  \frac 1 {N} \sum_{i }  \theta_\eta(\lambda_i - E_j) \right ],
\]
\[
A_3 :=   \frac 2 {N(N-1)(N-2)}   \sum_{i  }  \theta_\eta(\lambda_i - E_1) 
 \theta_\eta(\lambda_i - E_2) 
  \theta_\eta(\lambda_i - E_3) + \ldots 
\]
and
\[
A_2 :=  B_1+ B_2 + B_3, \quad\mbox{with}
\quad 
B_3=  -  \frac 1 {N(N-1)(N-2)}   \sum_{i  } 
  \theta_\eta(\lambda_i - E_1)  \theta_\eta(\lambda_i - E_2)  \sum_k  \theta_\eta(\lambda_k - E_3),
\]
and similarly, $B_1$ consists of terms with $j=k$, while  $B_2$ consists of terms with $i=k$.

Notice that, modulo a trivial change in the prefactor,
$\E_\bw A_1$ can be approximated by  
\[
\E_\bw  F \left ( \frac{1}{N}  \im \,  \tr\frac 1 { H^{(v)}-z_1}, \ldots, 
\frac{1}{N} \im \,  \tr \frac 1 { H^{(v)}-z_3} \right ),
\]
where the function $F$ is chosen to be
 $F(x_1, x_2, x_3) := x_1 x_2 x_3$ if $\max_j |x_j| \le N^\e$ and it is smoothly
cutoff to go to zero in the regime $\max_j |x_j|\ge N^{2\e}$.
The difference between the expectation of $F$ and $A_1$ is negligible, since it 
comes from  the regime where $ N^\e\le \max_j \frac{1}{N}  | \im \,  \tr (H^{(v)}-z_j)^{-1}| 
\le N^2$, which has an exponentially small probability by \eqref{basic4}
(the upper bound on the Green's function always holds since $\eta\ge N^{-2}$).  
Here the arguments of $F$ are imaginary parts of the trace of the Green's function, but
this type of function is allowed when applying Theorem \ref{comparison},
since 
$$
   \im \,  \tr G(z) = \frac{1}{2}\big[ \tr G(z) - \tr G(\bar z)\big].
$$
We remark that the main assumption  \eqref{basic} for Theorem \ref{comparison}  is satisfied by 
using \eqref{Gii} of Theorem \ref{mainls} with the choice of $M \asymp N$.

Similarly, we can approximate  $\E_\bw B_3$  by  
\[
\E_\bw  G \left ( \frac{1}{N^2}  \tr \left \{ \im \,  \frac 1 { H^{(v)}-z_1}
 \im \,  \frac 1 { H^{(v)}-z_2}  \right \}, \; 
\frac{1}{N} \im \,  \tr \frac 1 { H^{(v)}-z_3} \right ), 
\]
where $G(x_1, x_2) = x_1 x_2$ with an appropriate cutoff for large arguments, 
and there are  similar expressions for  $B_1, B_2$ and also for $A_3$, the latter
involving the trace of the product of three resolvents. 
By Theorem \ref{comparison}, these expectations w.r.t. $\bw$ in the approximations of $\E_\bw A_i$ 
can be replaced by expectations w.r.t. $\bv$ with only negligible errors provided that 
$\eta \ge N^{-1-\e}$.  We have thus proved that 
\begin{align}\label{6.8}
\lim_{N \to \infty}   \int \rd x_1 \rd x_2  \rd x_3 
\big[ p_{w, N}^{(3)}(x_1, x_2, x_3) -  p_{v, N}^{(3)}(x_1, x_2, x_3)\big]
  \theta_\eta(x_1 - E_1)  \theta_\eta(x_2- E_2) 
 \theta_\eta(x_3- E_3) = 0.
\end{align}

Set $\eta = N^{-1-\e}$ for the rest of the proof.
We now show that the validity of \eqref{6.8} for any choice of $E$, $\al_1, \al_2, \al_3$
(recall $E_j = E + \al_j/N$)
implies that the rescaled correlation functions,
$p_{w, N}^{(3)}(E+\beta_1/N,\ldots, E+ \beta_3/N)$
and $p_{v, N}^{(3)} (E+\beta_1/N,\ldots, E+ \beta_3/N)$, as functions
of the variables $\beta_1, \beta_2, \beta_3$, have the same weak limit.

Let $O$ be a smooth, compactly supported test function and let 
\[
O_\eta(\beta_1, \beta_2, \beta_3): =  \frac{1}{(\pi N)^3} \int_{\R^3}  \rd\alpha_1 \rd\alpha_2
 \rd\alpha_3  O(\alpha_1,\alpha_2,\alpha_3) \theta_\eta\left(\frac{\beta_1-\alpha_1}{N}\right )
\ldots \theta_\eta\left(\frac{\beta_3-\alpha_3}{N}\right ) 
\]
be its smoothing on scale $N\eta$. 
Then  we can write
\begin{align}\label{6.8.1}
  \int_{\R^3}  \rd\beta_1 \rd\beta_2
 \rd\beta_3  & \; O(\beta_1,\beta_2,\beta_3) 
p_{w, N}^{(3)}\left( E+ \frac{\beta_1}{N}, \ldots, E+ \frac{\beta_3}{N}\right)  \nonumber \\
 = &\int_{\R^3}  \rd \beta_1 \rd \beta_2
 \rd\beta_3   \; O_\eta (\beta_1,\beta_2,\beta_3) 
p_{w, N}^{(3)}\left( E+ \frac{\beta_1}{N}, \ldots, E+ \frac{\beta_3}{N}\right)
\nonumber  \\& +  \int_{\R^3}  \rd \beta_1 \rd \beta_2
 \rd\beta_3   \; (O-O_\eta) (\beta_1,\beta_2,\beta_3) 
p_{w, N}^{(3)}\left( E+ \frac{\beta_1}{N}, \ldots, E+ \frac{\beta_3}{N}\right) .
\end{align}
The first term on the right side, after the change of variables $x_j = E + \beta_j /N$, is  equal to 
\begin{align}\label{6.8.2}
&  \int_{\R^3}  \rd\alpha_1 \rd\alpha_2
 \rd\alpha_3   \; O(\alpha_1,\alpha_2,\alpha_3)  \int_{\R^3}\rd x_1 \rd x_2
 \rd x_3  
p_{w, N}^{(3)}(x_1, x_2, x_3)  \theta_\eta(x_1 - E_1)  \theta_\eta(x_2- E_2) 
 \theta_\eta(x_3- E_3)   ,
\end{align}
i.e., it can be written as an integral of expressions of the form \eqref{6.8}
for which limits with $p_{w,N}$ and $p_{v,N}$ coincide.

Finally, the second term on the right hand side of \eqref{6.8.1} is negligible. To see this,
notice that for any  test function $Q$, we have
\begin{align}
 \int_{\R^3}  \rd \beta_1 \rd \beta_2
 \rd\beta_3 &   \; Q (\beta_1,\beta_2,\beta_3) 
p_{w, N}^{(3)}\left( E+ \frac{\beta_1}{N}, \ldots, E+ \frac{\beta_3}{N}\right)
\non\\
& =  N^3 \int_{\R^3}  \rd x_1 \rd x_2
 \rd x_3    \; Q\big( N (x_1-E) , N (x_2-E),  N (x_3-E)\big) 
p_{w, N}^{(3)}(x_1 , x_2, x_3)  \non \\
& =   \Big(1-\frac{1}{N}\Big)\Big(1-\frac{2}{N}\Big)
 \E_\bw \sum_{i\ne j\ne k} Q\big( N (\lambda_i-E), N (\lambda_j-E), N (\lambda_k-E ) \big). 
\label{idQ}
\end{align}
If the test function $Q$ were supported on a ball of size $N^{\e'}$, $\e'>0$, then
this last term were bounded by 
\be
\| Q\|_\infty  \E_\bw \cN_{CN^{-1+\e'}}^3(E) 
\le C\| Q\|_\infty N^{4\e'}.
\label{Q}
\ee
Here
$\cN_\tau(E)$ denotes the number of eigenvalues in the interval $[E-\tau, E+\tau]$
and in the estimate we used the local semicircle law
 on intervals of size $\tau \ge N^{-1+\e'}$. 

Set now $Q:= O-O_\eta$. {F}rom the definition of $O_\eta$, it is
 easy to see that the function 
\[
Q_1 (\beta_1,\beta_2,\beta_3) = O (\beta_1,\beta_2,\beta_3) -
O_\eta  (\beta_1,\beta_2,\beta_3) \prod_{j=1} ^3  1 ( |\beta_j|\le N^{ \e'})
\]
satisfies the bound $\| Q_1\|_\infty \le \|Q\|_\infty= \| O-O_\eta\|_\infty\le C  {N\eta} = CN^{-\e}$. 
So choosing $\e' < \e/4$, the contribution of $Q_1$ is negligible.
Finally, $Q_2 = Q - Q_1$ is given by 
\[
Q_2 (\beta_1,\beta_2,\beta_3) =  -O_\eta  (\beta_1,\beta_2,\beta_3) 
 \left [1-  \prod_{j=1} ^3 1 ( |\beta_j|\le N^{ \e'}) \right ]
\]
and 
\begin{align} 
|Q_2| \le  &  C  \left [  \frac {1} { 1   +  \beta_1^2 } \right ] 
   \left [  \frac {1} { 1   +  \beta_2^2 } \right ] 
   \left [  \frac {1} { 1   +  \beta_3^2 } \right ] 
 \Big \{ 1 ( |\beta_1|\ge N^{ \e'}) +\ldots\Big\}  \non  \\
   \le  & C  \left \{  N^{-\e'} \left [  \frac {N^{\e'} } { N^{2\e'}   +  \beta_1^2 } \right ] 
   \left [  \frac {1} { 1   +  \beta_2^2 } \right ] 
   \left [  \frac {1} { 1   +  \beta_3^2 } \right ]   +\ldots\right \}. 
\end{align} 
Hence the contribution of   $Q_2$ in the last term of \eqref{idQ}  is bounded by 
\[
 C  N^{-3-\e'} \E_\bw \sum_{i,j,k}  \left \{   \left [  
\frac {N^{-1+ \e'} } {   N^{-2+ 2\e'} +  (\lambda_i - E)^2 } \right ] 
 \left [  \frac {N^{-1} } { N^{-2}    +   (\lambda_j - E)^2 } \right ] 
\left [  \frac {N^{-1} } {  N^{-2} +   (\lambda_k- E)^2 } \right ]
+\ldots\right \} 
\label{triple}
\]
From Theorem \ref{mainls}, the  last term is bounded by $N^{-\e'}$ up to some logarithmic factor. 
This completes the proof of
 Theorem   \ref{com}.  \qed

\appendix
\section{Spectral condition for band matrices}\label{sec:spec}

\begin{lemma} Let $B=(\sigma_{ij})$ satisfying  \eqref{sum}
and \eqref{BM} with $W\ge 1$  and with $f$ being  a nonnegative  symmetric 
function with $\int f =1$ and  $f \in  L^\infty (\bR)$. Then we have 
\be
       B\ge -1 + \delta
\label{gap}
\ee
for some $\delta>0$ and $W$ large enough, depending on $f$.
\end{lemma}

{\it Proof.}
Recall that the discrete Fourier transform in $d=1$
dimensions is defined as follows. 
Let  $\e:= 1/N $ and 
$$
    \Lambda_{\e} := \Lambda = \e \bZ/ \bZ 
$$
 be the periodic one dimensional lattice (torus) of
size $1$ and spacing $\e$ with its dual lattice being
$$
   \Lambda_{\e}^*:=
     \Lambda^*:= \Big( 2\pi \bZ \Big)
    / \Big(\frac{2\pi}{\e} \bZ\Big).
$$
\newcommand{\F}{{\cal F}_N}
\newcommand{\Fi}{{\cal F}_N^{-1} }
Let $\psi$ be a function on $\Lambda$. Then its Fourier transform
$ {\cal F}_N \psi$ is a function on $\Lambda^*$ defined as
$$
    \F \psi(p) 
    =  \e \sum_{x\in \Lambda} \psi(x) e^{-i p \cdot x}
$$
and it is an isometry
$$
   \e\sum_{x\in \Lambda} \ov{\psi(x)} \phi(x) = \sum_{p\in \Lambda^*}
  \ov{\F \psi(p)} \F \phi(p).
$$
In our case, $x = k/N$ and define
\[
F_W(x): =  N W^{-1} f(xN/W). 
\]
Then, for $p \in    \Lambda^*$, we have
$$
    (\F F_W) (p) 
    =  \  \sum_{x\in \Lambda} F_W (x) e^{-i p \cdot x}
    =    \sum_{k = 1}^N W^{-1} f (k/W ) e^{-i (Wp/N) \cdot (k/W)}= \wh f(q) + o(1) , 
  \quad q= Wp/N ,
$$
where   the error term vanishes as $W\to \infty$
and
$\wh f$ denotes   the usual Fourier transform in $L^1(\bR)$
$$
\wh f(q)    =  \int f(y) e^{-i q y} \rd y .
$$
 With this formula, and with the notation $\psi_N(j):=\psi(j/N)$
for any $\psi$ defined on $\Lambda$, we have 
\[
(B\psi_N)(k)=\e \sum_{\ell = 1}^N  N W^{-1} f ((k-\ell)/W ) \psi_N(\ell)
 = \e \sum_{y\in \Lambda}  F_W\Big( \frac{k}{N}-y \Big) \psi (y)
= \sum_{p\in \Lambda^*} e^{ipk/N} 
 \F F_W (p) \F \psi(p) 
\]
and 
\[
 \sum_{p\in \Lambda^*} | \F \psi(p) |^2 = \e \sum_{j = 1}^N  |\psi_N(j)|^2 
\]
which is normalized to be $1$. 
Hence $B$ on the Fourier side acts as a multiplication by the function
$\F F_W$, so
$$
   \mbox{Spec} B  = \mbox{Range}\; \F F_W\subset \mbox{supp}\; \wh f  + o(1).
$$
Since $f$ is nonnegative, symmetric function and $\int f = 1$, we have  $\wh f$ is real and
$$
   \inf \wh f > -1 +\delta
$$
for some $\delta> 0$, which completes the proof.

\section{Large deviation estimates}\label{sec:lde}

In this Appendix we prove two large deviations results. They are weaker
than the corresponding results of Hanson and Wright \cite{HW}, used
in \cite{ESY3}, but they require only independent, not necessarily
identically distributed random variables, moreover the proofs are much simpler.

\begin{lemma}\label{generalCLT}
Let $a_i$ ($1\leq i\leq N$) be $N$ independent complex random variables with mean zero,
 variance $\sigma^2$ and  uniform subexponential  decay, i.e.,  there exist  $\al$,
 $\beta>0$ that for any $x>0$  
\be\label{assumdelta0}
\P(|a_i|\geq x^{\alpha})\leq  \beta e^{- x}.
\ee
Then for any $A_i\in \C$ ($1\leq i\leq N$) and $D\ge 1$ we have,
\be\label{resgenCLT1}
\P\left\{\left|\sum_{i}a_iA_i\right|\geq D \sigma
 \Big(\sum_{i}|A_i|^2\Big)^{1/2}\right\}  \leq C\exp{\big( -cD^{\frac{2}{2+\al}}\big)}
\ee
for some positive constants $C$ and $c$ depending on $\al$ and $\beta$ in \eqref{assumdelta0}.   
\end{lemma}
{\it Proof of Lemma \ref{generalCLT}.} 
Without loss of generality, we may assume that $\sigma=1$. 
The assumption \eqref{assumdelta0} implies that the $k-$th moment of $a_i$ is bounded by: 
\be\label{Eaik}
\E|a_i|^k\leq (Ck)^{\al k}
\ee
for some $C>0$ depending on $\al$ and $\beta$. 
\par First,  for $p\in \N$, we estimate
\be
\E\left|\sum_{i=1}^Na_iA_i\right|^p.
\ee 
With the Marcinkiewicz--Zygmund inequality, for an integer $p\geq 2$,  we have  
\be\label{iiEaA}
\E\left|\sum_{i}a_iA_i\right|^p\leq (Cp)^{p/2}\E\left[\left(\sum_{i}|a_iA_i|^2\right)^{p/2}\right]
\ee
(for the estimate of the constant, see e.g. Exercise 2.2.30 of \cite{Str}).
Using \eqref{Eaik},  we have $  \E |a_{i_1}a_{i_2}\cdots a_{i_{p/2}}|^2\leq (Cp)^{\al p}$. Inserting it into \eqref{iiEaA}, we obtain
\be\label{temp9.6}
\E\left|\sum_{i}a_iA_i\right|^p\leq (Cp^{\frac{1}{2}+\al})^p\left(\sum_{i}|A_i|^2\right)^{p/2},
\ee
which implies \eqref{resgenCLT1} by choosing an even integer $p$
of the order $(D/Ce)^{\frac{2}{2+\al}}$ 
and applying a high moment Markov inequality. 
\qed
\bigskip

\begin{lemma}\label{generalHWT1}
Let $a_i$ ($1\leq i\leq N$) be $N$ independent random complex  variables with mean zero, 
variance $\sigma^2$  and having the uniform  subexponential decay \eqref{assumdelta0}. 
Let  $B_{ij}\in \C$ ($1\leq i,j\leq N$). 
 Then we have that 
\be\label{resgenHWTD1}
\P\left\{\left|\sum_{i=1}^N\overline a_iB_{ii}a_i-\sum_{i=1}^N\sigma^2 B_{ii}\right|\geq 
D \sigma^2
 \Big( \sum_{i=1}^N|B_{ii}|^2\Big)^{1/2}\right\}\leq  C\exp{\big( -cD^{\frac{1}{1+\al}}\big)}
\ee
and
\be\label{resgenHWTO1}
\P\left\{\left|\sum_{i\neq j}\overline a_iB_{ij}a_j\right|\geq D \sigma^2 
\Big(\sum_{i\ne j} |B_{ij}|^2 \Big)^{1/2}\right\}\leq C\exp{\big( -cD^{\frac{1}{2(1+\al)}}\big)}
\ee
for some positive constants $C$ and $c$ depending on $\al$ and $\beta$ in \eqref{assumdelta0}.  
\end{lemma}
{\it Proof of Lemma \ref{generalHWT1}.}
Without loss of generality, we may again assume that $\sigma=1$.
 First, we prove  \eqref{resgenHWTD1}.
 Notice that $|a_i|^2-1$ ($1\leq i\leq N$) are independent random variables
 with mean $0$ and variance less than some constants $C$. Furthermore, the $k$-th moment of
 $|a_i|^2-1$ is bounded as 
\be
\E(|a_i|^2-1)^k\leq (Ck)^{2\alpha k} .
\ee 
Then following the proof of the  Lemma \ref{generalCLT} with $|a_i|^2-1$ replacing $a_i$, 
we obtain   \eqref{resgenHWTD1}. 

\medskip

Next, we prove \eqref{resgenHWTO1}. 
For  any $p\in \N$, $p\ge 2$, we estimate
\be
\E\left|\sum_{i }\overline a_i\xi_i\right|^p\equiv 
\E\left|\sum_{i>j}\overline a_iB_{ij}a_j\right|^p
\ee 
where $\xi_i:=\sum_{j<i}B_{ij}a_j$. Note that $a_i$ and $\xi_i$ are independent for any fixed $i$. 
 By the definition, 
\be
X_n\equiv \sum_{i=1}^n \overline a_i\xi_i
\ee
is martingale. Using  the Burkholder inequality, we have that
\be\label{iiEaA2}
\E\left|\sum_{i}\overline a_i\xi_i\right|^p\leq
 (Cp)^{3p/2}\E\left[\Big(\sum_{i}|\overline a_i\xi_i|^2\Big)^{p/2}\right]
\ee
(for the constant, see Section VII.3 of \cite{Shy}).
By the generalized Minkowski inequality, by the independence of $a_i$ and $\xi_i$
and using \eqref{Eaik}, we have
\[
 \left [ \E\Big(\sum_{i}|\overline a_i\xi_i|^2\Big)^{p/2}\right]^{2/p}
 \le  \sum_{i}  \bigg [ \, \E |\overline a_i\xi_i|^{p} \, \bigg ]^{2/p} 
 =  \sum_{i}  \bigg [ \, \E(|\overline a_i|^p)\E(|\xi_i|^p) \, \bigg ]^{2/p}
\le
(C p)^{ 2 \alpha} \sum_{i}  \bigg [ \, \E(|\xi_i|^p) \, \bigg ]^{2/p}.
\]
Using \eqref{temp9.6}, we have
$$
    \E(|\xi_i|^p) \le (Cp^{\frac{1}{2}+\al})^p \Big(\sum_{j} |B_{ij}|^2\Big)^{p/2}.
$$
Combining this with \eqref{iiEaA2} we obtain
\be\label{iiEaA3}
\E\left|\sum_{i}\overline a_i\xi_i\right|^p\leq
(Cp)^{2p(1+\alpha)}\Big( \sum_{i} \sum_j |B_{ij}|^2 \Big)^{p/2}.
\ee
 Then choosing $(D/Ce)^{\frac{1}{2(1+\al)}}$ and applying Markov inequality, 
we obtain \eqref{resgenHWTO1} . 
\qed

\bigskip

In our applications we will need these two lemmas when
$D$ is a power of $\log N$. For simplicity, we do not want to keep
track of the precise powers in the estimate and
we are interested only in error bounds that decay faster than
any fixed power of $N$, say $CN^{-\log\log N}$.  Therefore, 
in this paper we will use the following weaker form of these two lemmas,
the stronger form will be useful in future applications.
 
\begin{corollary}\label{generalHWT}
Let $a_i$ ($1\leq i\leq N$) be $N$ independent random complex  variables with mean zero, 
variance $\sigma^2$  and having the uniform  subexponential decay \eqref{assumdelta0}.
Let $A_i$, $B_{ij}\in \C$ ($1\leq i,j\leq N$). 
 Then we have that 
 \begin{align}
\P\left\{\left|\sum_{i=1}^N a_iA_i\right|\geq 
(\log N)^{\frac32+\al} \sigma \,\Big(\sum_{i}|A_i|^2\Big)^{1/2}\right\}\leq & CN^{-\log\log N},
\label{resgenHWTD} \\
\P\left\{\left|\sum_{i=1}^N\overline a_iB_{ii}a_i-\sum_{i=1}^N\sigma^2 B_{ii}\right|\geq 
(\log N)^{\frac32+2\al} \sigma^2 \Big( \sum_{i=1}^N|B_{ii}|^2\Big)^{1/2}\right\}\leq &
 CN^{-\log\log N},\label{diaglde}\\
\P\left\{\left|\sum_{i\neq j}\overline a_iB_{ij}a_j\right|\geq (\log N)^{3+2\al} \sigma^2 
\Big(\sum_{i\ne j} |B_{ij}|^2 \Big)^{1/2}\right\}\leq & CN^{-\log\log N}, \label{resgenHWTO}
\end{align}
for some constants $C$ depending on $\al$ and $\beta$ in \eqref{assumdelta0}.
\end{corollary}

\section{Proof of Lemma \ref{fmam}}\label{sec:LSI}

We first prove a version of this lemma when the fourth moment exactly matches, i.e., $\gamma=0$,
then we explain how to deal with the approximation. More precisely, we first show the following:

\begin{lemma}\label{exact}
Under the condition
\be\label{m4m3cc1}
m_4-m_3^2-1\ge C_1,\,\,\, m_4\leq C_2
\ee
for some positive constants $C_1$  and $C_2$,
there exists a real random variable $\xi$ such that the first four moments of
 $\xi $ are $0$, $1$, $m_3$ and $m_4$ and the distribution $\nu$ of $\xi$ satisfies 
logarithmic Sobolev inequality and the LSI constant
is bounded from above by a function
of $C_1$ and $C_2$.  Moreover, $\nu$ can be chosen to be absolutely continuous
with a smooth positive  density, $\nu(\rd x) = e^{-U(x)}\rd x$,
such that the derivatives of $U$ satisfy
\be
  | U^{(k)}(x)| \le C_k (1+ x^2)^{Ck}
\label{Uk}
\ee
with some fixed constant $C$ and $k$-dependent constants $C_k$.
\end{lemma}

{\it Remark.} The last statement about the smoothness of $U$ 
will not be needed in this paper, but we state it for further reference.

\bigskip

{\it Proof.}
We start with the case  $|m_3|>\delta$, where $\delta$ is small enough number
to depend only on $C_1$, see below.
Let $\xi$ be the sum of two Gaussians, with density function of the form
\be\label{deffxi}
f_\xi(x)=\frac{b}{(a+b)}\frac1{\sqrt{2\pi \sigma} }e^{-{(x-a)^2/(2\sigma)}}
+\frac{a}{(a+b)}\frac1{\sqrt{2\pi \sigma} }e^{-{(x+b)^2/(2\sigma)}}
\ee
 with some  parameters $a>0$, $b>0$, $\sigma>0$. If  the first 4 moments of
 $f_\xi(x)$ are $0$, $1$, $m_3$ and $m_4$, then we have the relations
 \be\label{temp6.20}
 m_4=1+\frac{m_3^2}{1-\sigma}+4\sigma-2\sigma^2,
 \ee
\be\label{ab}
a b=1-\sigma\,\,\,\mbox{and}\,\,\,a-b=\frac{m_3}{1-\sigma}.
\ee
With $m_3$, $m_4$  in \eqref{m4m3cc1} and $|m_3|\geq \delta$, one can always
 find a solution of \eqref{temp6.20} such that $0<\sigma<1$. Actually, one 
can see that $c<\sigma<C$, where $c$ and $C$ only depend on $C_1$, $C_2$ 
in \eqref{m4m3cc} and $\delta$.

Once $\sigma$ is found, it is easy to check that 
one can always find real solutions $a,b$ for \eqref{ab} as long as
 $m_3$, $m_4$ satisfy  \eqref{m4m3cc1} 
and $|m_3|\geq \delta$. Since the solutions $a,b,\sigma$ are continuous with respect to 
$m_3$ and $m_4$, then they are uniformly bounded. 
Distributions of the form \eqref{deffxi} satisfy the LSI, since 
the are log concave away from a compact set.
Since the parameters $a,b,\sigma$ are in a compact set, the
LSI constant will remain uniformly bounded with a
bound depending on $C_1$, $C_2$ and $\delta$.
 It is clear that the density function \eqref{deffxi} is positive
and its logarithm satisfies \eqref{Uk}.

\medskip

Now we consider the case that  $|m_3|<\delta$ with a small
 $\delta= \frac{1}{100}\min \{1,C_1\}$,
where $C_1$ is the constant in \eqref{m4m3cc1}.
Without loss of generality, we may assume $m_3>0$. 
We consider the following three parameter family of probability densities
$$
  f_{d, \beta, \e}(x) = (1-\e) g_{d,\beta} (x) + \e h(x)
$$
with 
$$
   g_{d,\beta} (x) = \frac{\beta+1}{2d^{\beta+1}} \cdot |x|^\beta \cdot {\bf 1}(|x|\le d), 
 \qquad h(x) = \frac{b}{(a+b)}\frac1{\sqrt{2\pi} }e^{-{(x-a)^2/2}}
+\frac{a}{(a+b)}\frac1{\sqrt{2\pi } }e^{-{(x+b)^2/2}},
$$
where the parameters are in the range $-1 < \beta <\infty$, $0<d<\infty$, $0\le \e \ll 1$
and $a,b$ will be chosen explicitly.
Simple calculation shows that the moments of $f_{d, \beta, \e}$ are $m_1=0$,
\begin{align}\label{moments}
   m_2 = & (1-\e)\frac{\beta+1}{\beta+3} d^2 + \e (1+ ab), \\
   m_3 = & \e ab(a-b),\\
   m_4 = & (1-\e) \frac{\beta+1}{\beta+5} d^4 + \e\Big[ 3 + ab(6 + a^2+b^2 -ab)\Big] .
\end{align}
Choosing, say, $a=2$, $b=1$, and setting $m_2=1$, we obtain 
$d^2= \frac{1-3\e}{1-\e}\frac{\beta+3}{\beta+1}$
from the first equation, $\e = m_3/2$ from the second equation and finally the last equation
becomes
\be
  m_4= \frac{(1-3m_3/2)^2}{1-m_3/2} \frac{(\beta+3)^2}{(\beta+1)(\beta+5)} +  \frac{23}{2} m_3.
\label{m44}
\ee
Recall that we are in the regime where  $|m_3|\le \delta \le C_1/100$.
For any fixed $0\le m_3 \le \delta$, the right hand side of \eqref{m44}
is a monotonically decreasing
function in $\beta \in (-1,\infty)$ whose value goes down from $\infty$ to
$\frac{(1-3m_3/2)^2}{1-m_3/2} +  \frac{23}{2} m_3 \le 1+20\delta$. But we know
from \eqref{m4m3cc1} that $C_2\ge m_4\ge 1+100\delta$, thus there is a value $\beta$
such that \eqref{m44} holds, moreover, $\beta$ is in a compact
subinterval of $(-1, \infty)$ 
that depends only on $\delta$ and $C_2$. It is then easy to check that
the support and the supremum norm of the density $g_{d, \beta}$ also remains in a compact
set, depending only on $\delta$. Therefore we constructed
a probability measure with the given moments, that is a linear
combination of two Gaussians plus a compactly supported piece
with a  nonnegative bounded  density.  To ensure smoothness,
we replace $g_{d,\beta}$ with $\wt g_{d,\beta,\tau}: = \vartheta_\tau \ast g_{d,\beta}$,
where $\vartheta_\tau(x) = \tau^{-1}\vartheta(x/\tau)$ and 
$\vartheta$ is a compactly supported nonnegative smooth symmetric function with $\int\vartheta=1$.
The first moment $m_1$ is unchanged and
the formulas \eqref{moments} for the higher moments will get modified by an error term of order $\tau$.
Let $\tau$ be much smaller than all other parameters in this proof.  It is easy to see that, by 
a simple calculation  treating $\tau$ as a small perturbation,    one can still choose $a, b, \e$ 
and $\beta$ in the previous argument  to match   $m_2= 1, m_3 $ and $m_4$.

Finally, note that the sum of two Gaussians satisfy
the LSI, as well as its compact perturbation
 and the new LSI constant depends only on
the supremum norm of the density of the perturbation.
Since all these parameters remain uniformly controlled by $C_1$ and $C_2$,
we proved Lemma \ref{exact}, i.e.,  Lemma \ref{fmam} for $\gamma=0$. \qed

\bigskip
\par Now consider the case $\gamma>0$.  For any real random variable
$\zeta$, independent of $\xi^G$, and with the first 4 moments being $0$, $1$, $m_3(\zeta)$
 and $m_4(\zeta)<\infty$, the first 4 moments of 
\be
\zeta'=(1-\gamma)^{1/2}\zeta+\gamma^{1/2}\xi^G
\ee
are $0$, $1$, 
\be\label{relm3}
m_3(\zeta')=(1-\gamma)^{3/2}m_3(\zeta)
\ee
 and 
 \be\label{relm4}
m_4(\zeta')=(1-\gamma)^{2}m_4(\zeta) +6\gamma-3\gamma^2.
 \ee

Given $m_3$ and $m_4$, satisfying \eqref{m4m3cc1} and
 using   Lemma \ref{exact}, we obtain that for any $\gamma$
small enough, there exists a real random variable $\xi_\gamma$ such that the 
first four moments are $0$, $1$, 
\be\label{m3xi}
m_3(\xi_\gamma)=(1-\gamma)^{-3/2} m_3
\ee
 and
 $$
 m_4(\xi_\gamma)=m_3(\xi_\gamma)^2+(m_4-m^2_3).
 $$ 
With $m_4\leq C_2$, we have $m_3^2\leq C_2$, thus   
 \be
 |m_4(\xi_\gamma)-m_4 |\leq C\gamma
 \ee
for some $C$ depending on $C_2$. 

\par Hence with \eqref{relm3} and \eqref{relm4}, we obtain that $\xi'
=(1-\gamma)^{1/2}\xi_\gamma+\gamma^{1/2}\xi^G$
 satisfies $m_3(\xi')=m_3$ and \eqref{m4m4}. With  Lemma \ref{exact},
 we obtain that the LSI constant of $\xi_\gamma$ is bounded by a constant only depends on 
$C_1$ and $C_2$, which completes the proof of Lemma \ref{fmam}. 
\qed


\thebibliography{hhhhh}
   

\bibitem{AZ} Anderson, G.; Zeitouni, O. : 
 A CLT for a band matrix model. Probab. Theory Related Fields
 {\bf 134} (2006), no. 2, 283--338.



\bibitem{BP} Ben Arous, G., P\'ech\'e, S.: Universality of local
eigenvalue statistics for some sample covariance matrices.
{\it Comm. Pure Appl. Math.} {\bf LVIII.} (2005), 1--42.


\bibitem{BI} Bleher, P.,  Its, A.: Semiclassical asymptotics of 
orthogonal polynomials, Riemann-Hilbert problem, and universality
 in the matrix model. {\it Ann. of Math.} {\bf 150} (1999): 185--266.


\bibitem{BG} Bobkov, S. G., G\"otze, F.: Exponential integrability
and transportation cost related to logarithmic
Sobolev inequalities. {\it J. Funct. Anal.} {\bf 163} (1999), no. 1,
1--28.


\bibitem{BH} Br\'ezin, E., Hikami, S.: Correlations of nearby levels induced
by a random potential. {\it Nucl. Phys. B} {\bf 479} (1996), 697--706, and
Spectral form factor in a random matrix theory. {\it Phys. Rev. E}
{\bf 55} (1997), 4067--4083.

%

\bibitem{De1} Deift, P.: Orthogonal polynomials and
random matrices: a Riemann-Hilbert approach.
{\it Courant Lecture Notes in Mathematics} {\bf 3},
American Mathematical Society, Providence, RI, 1999


\bibitem{De2} Deift, P., Gioev, D.: Random Matrix Theory: Invariant
Ensembles and Universality. {\it Courant Lecture Notes in Mathematics} {\bf 18},
American Mathematical Society, Providence, RI, 2009


\bibitem{DKMVZ1} Deift, P., Kriecherbauer, T., McLaughlin, K.T-R,
 Venakides, S., Zhou, X.: Uniform asymptotics for polynomials 
orthogonal with respect to varying exponential weights and applications
 to universality questions in random matrix theory. 
{\it  Comm. Pure Appl. Math.} {\bf 52} (1999):1335--1425.

\bibitem{DKMVZ2} Deift, P., Kriecherbauer, T., McLaughlin, K.T-R,
 Venakides, S., Zhou, X.: Strong asymptotics of orthogonal polynomials 
with respect to exponential weights. 
{\it  Comm. Pure Appl. Math.} {\bf 52} (1999): 1491--1552.






\bibitem{DPS} Disertori, M., Pinson, H., Spencer, T.: Density of
states for random band matrices. Commun. Math. Phys. {\bf 232},
83--124 (2002)


\bibitem{D}  Dyson, F.J.: Correlations between eigenvalues of a random
matrix. {\it Commun. Math. Phys.} {\bf 19}, 235-250 (1970).


\bibitem{ESY1} Erd{\H o}s, L., Schlein, B., Yau, H.-T.:
Semicircle law on short scales and delocalization
of eigenvectors for Wigner random matrices.
{\it Ann. Probab.} {\bf 37}, No. 3, 815--852 (2008)


\bibitem{ESY2} Erd{\H o}s, L., Schlein, B., Yau, H.-T.:
Local semicircle law  and complete delocalization
for Wigner random matrices. {\it Commun.
Math. Phys.} {\bf 287}, 641--655 (2009)


\bibitem{ESY3} Erd{\H o}s, L., Schlein, B., Yau, H.-T.:
Wegner estimate and level repulsion for Wigner random matrices.
{Int Math Res Notices } {\bf 2010} (3): 436-479 (2010) 

\bibitem{ESY4} Erd{\H o}s, L., Schlein, B., Yau, H.-T.: Universality
of random matrices and local relaxation flow.
To appear in {\it Invent. Math.}  
arxiv.org/abs/0907.5605

\bibitem{ERSY}  Erd{\H o}s, L., Ramirez, J., Schlein, B., Yau, H.-T.:
{\it Universality of sine-kernel for Wigner matrices with a small Gaussian
 perturbation.}  Electr. J. Prob. {\bf 15},  Paper 18, 526--604 (2010)

\bibitem{EPRSY}
Erd\H{o}s, L.,  P\'ech\'e, G.,  Ram\'irez, J.,  Schlein,  B.,
and Yau, H.-T., Bulk universality 
for Wigner matrices.  Comm. Pure Appl. Math. {\bf 63}, No. 7, 895-925 (2010)

\bibitem{ERSTVY}
Erd\H{o}s, L.,  Ram\'irez, J.,  Schlein,  B., Tao, T., Vu, V. and Yau, H.-T.,
Bulk universality for Wigner hermitian matrices with subexponential decay.
{\it Math. Res. Lett.} {\bf 17} (4), 667--674 (2010).

\bibitem{ESYY} Erd{\H o}s, L., Schlein, B., Yau, H.-T., Yin, J.:
The local relaxation flow approach to universality of the local
statistics for random matrices.  To appear in Annales Inst. H. Poincar\'e, Prob. and Stat. 
Preprint arXiv:0911.3687




\bibitem{EYY}  Erd{\H o}s, L.,  Yau, H.-T., Yin, J.:
Universality for generalized Wigner matrices with
Bernoulli distribution.  To appear in Journal of Combinatorics. 
Preprint arXiv:1003.3813




\bibitem{gui}  Guionnet, A.:
Large deviation upper bounds
and central limit theorems for band matrices,
{\it Ann. Inst. H. Poincar\'e Probab. Statist }
{\bf 38 }, (2002), pp.  341-384.





\bibitem{HW} Hanson, D.L., Wright, F.T.: A bound on
tail probabilities for quadratic forms in independent random
variables. {\it The Annals of Math. Stat.} {\bf 42} (1971), no.3,
1079-1083.


\bibitem{J} Johansson, K.: Universality of the local spacing
distribution in certain ensembles of Hermitian Wigner matrices.
{\it Comm. Math. Phys.} {\bf 215} (2001), no.3. 683--705.

\bibitem{J1} Johansson, K.: Universality for certain hermitian Wigner
matrices under weak moment conditions. Preprint 
{arxiv.org/abs/0910.4467}



\bibitem{M} Mehta, M.L.: Random Matrices. Academic Press, New York, 1991.


\bibitem{MG} Mehta, M.L., Gaudin, M.: On the density of eigenvalues
of a random matrix. {\it Nuclear Phys.} {\bf 18}, 420-427 (1960).


\bibitem{PS} Pastur, L., Shcherbina M.:
Bulk universality and related properties of Hermitian matrix models.
{\it J. Stat. Phys.} {\bf 130} (2008), no.2., 205-250.


\bibitem{Shy} Shiryayev, A. N.: Probability. Graduate Text in Mathematics. {\bf 54}.
Springer, 1984.


\bibitem{Spe}  Spencer, T.: {\it Random banded and sparse matrices (Chapter 23)}
to appear in ``Oxford Handbook of Random Matrix Theory'', edited by
G. Akemann, J. Baik and P. Di Francesco.



\bibitem{Str} Stroock, D.W.: Probability theory, an analytic view. Cambridge
University Press, 1993.
 
\bibitem{TV} Tao, T. and Vu, V.: Random matrices: Universality of the 
local eigenvalue statistics.
 Preprint arXiv:0906.0510. 


\bibitem{TV2} Tao, T. and Vu, V.: Random matrices: Universality 
of local eigenvalue statistics up to the edge. Preprint. arXiv:0908.1982

\bibitem{TV3} Tao, T. and Vu, V.: Random covariance matrices:
 Universality of local statistics of eigenvalues. Preprint. arXiv:0912.0966


\bibitem{Vu} Vu, V.: Spectral norm of random matrices. {\it Combinatorica},
{\bf 27} (6) (2007), 721-736.
\end{document}